\documentclass[a4paper,11pt]{article}
\pdfoutput=1

\usepackage[utf8]{inputenc}
\usepackage{jheppub}
\usepackage{amsmath,amssymb,mathtools}
\usepackage{graphicx}
\usepackage{xspace}
\usepackage{subcaption}

\renewcommand{\tabcolsep}{2em}

\allowdisplaybreaks

\DeclareRobustCommand{\ensuremathrm}[1]{\ensuremath{\mathrm{#1}}\xspace}

\DeclareRobustCommand{\rd}{\ensuremathrm{d}} 

\DeclareRobustCommand{\GeV}{\ensuremathrm{GeV}\xspace}
\DeclareRobustCommand{\TeV}{\ensuremathrm{TeV}\xspace}

\DeclareRobustCommand{\pb}{\ensuremathrm{pb}\xspace}

\DeclareRobustCommand{\nnlojet}{NNLO\scalebox{0.8}{JET}\xspace}
\preprint{{\raggedleft ZU-TH 16/22\\ KA-TP-12-2022\\ P3H-22-044\\ IPPP/22/30\\ CERN-TH-2022-072\\ LMU-ASC 17/22\\ }}

\title{Single Photon Production at Hadron Colliders at NNLO QCD with Realistic Photon Isolation}

\author{X.\,Chen$^{a,b}$, T.\,Gehrmann$^{c}$,  E.W.N.\,Glover$^{d}$, M.\,H\"ofer$^{e}$,  A.\,Huss$^{f}$, R.\,Sch\"urmann$^{c}$}

\affiliation{
        $^a$Institute for Theoretical Physics, Karlsruhe Institute of Technology, 76131 Karlsruhe, Germany\\
        $^b$Institute for Astroparticle Physics, Karlsruhe Institute of Technology,\\ 76344 Eggenstein-Leopoldshafen, Germany\\
        $^c$Physik-Institut, Universit\"at Z\"urich, Winterthurerstrasse 190, 8057 Z\"urich, Switzerland\\
        $^d$Institute for Particle Physics Phenomenology, Department of Physics, University of Durham, Durham, DH1 3LE, UK\\
        $^e$Ludwig-Maximilians-Universit\"at M\"unchen, Fakult\"at f\"ur Physik, \\  Arnold Sommerfeld Center for Theoretical Physics, D-80333 M\"unchen, Germany,\\
        $^f$Theoretical Physics Department, CERN, 1211 Geneva 23, Switzerland
        }

\emailAdd{xuan.chen@kit.edu}
\emailAdd{thomas.gehrmann@uzh.ch}
\emailAdd{e.w.n.glover@durham.ac.uk}
\emailAdd{m.hoefer@physik.uni-muenchen.de}
\emailAdd{alexander.huss@cern.ch}
\emailAdd{robins@physik.uzh.ch}

\abstract{Isolated photons at hadron colliders are defined by permitting only a limited amount of hadronic energy inside a fixed-size cone around the 
candidate photon direction. This isolation criterion admits contributions from collinear photon radiation off QCD partons and from parton-to-photon 
fragmentation processes. We compute the NNLO QCD corrections to isolated photon and photon-plus-jet production, including these two contributions. Our newly derived 
results allow us to reproduce the isolation prescription used in the experimental measurements, performing detailed comparisons with data from the LHC experiments. 
We quantify the impact of different photon isolation prescriptions, including no isolation at all, on photon-plus-jet cross sections and discuss possible measurements 
of the photon fragmentation functions at hadron colliders. 
}
\keywords{Hadron Colliders, QCD Phenomenology, Photons, Jets, NNLO Corrections}

\begin{document}
\maketitle

\section{Introduction}\label{sec:intro}

High-transverse momentum photons are produced copiously at hadron colliders and provide a distinctive final-state signature in the 
form of an electromagnetic energy deposit. Their production cross sections have been studied at early hadron colliders~\cite{Athens-Athens-Brookhaven-CERN:1979oxx, Anassontzis:1982gm, CMOR:1989qzc, UA1:1988zam, UA2:1991wce} as well as in fixed-target experiments~\cite{WA70:1987vvj,FermilabE706:2004emk}, followed by first precision measurements at the Fermilab Tevatron~\cite{D0:2005ofv,D0:2008chx,D0:2013lra,CDF:2017cuc}.
Production cross sections for photons at high transverse momentum have been measured at the LHC by the ATLAS~\cite{ATLAS:2017nah,ATLAS:2019buk,ATLAS:2017xqp}, CMS~\cite{CMS:2018qao,CMS:2019jlq} and ALICE~\cite{ALICE:2019rtd} collaborations, either fully inclusively
or in association with hadronic jets. 
Besides their importance for precision QCD phenomenology, these  cross sections are also key ingredients in data-driven background estimations in new physics 
searches~\cite{Lindert:2017olm}.

With the Born-level production process~\cite{Halzen:1978rx,Ruckl:1978mx} being photon radiation off a quark in 
quark--antiquark annihilation or quark--gluon scattering, these measurements offer direct sensitivity to the gluon distribution in the proton. 
In global determinations of parton distribution functions, the inclusion of photon production data has long been a matter of 
debate~\cite{Harriman:1990hi,Vogelsang:1995bg,dEnterria:2012kvo,Carminati:2012mm,Campbell:2018wfu}, relating especially to the quantification of 
uncertainties associated with the identification of photons in the experimental measurements and the implementation of photon identification in the corresponding 
theory calculations. 
Photons that are observed in a hadronic environment can arise from different sources: in addition to the direct Born-level process that is calculable at parton level, 
they can also be emitted in the course of the parton-to-hadron transition (fragmentation process),
described by photon fragmentation functions~\cite{Koller:1978kq,Laermann:1982jr}, or be produced as secondary photons in 
the decay of hadrons. In order to single out the direct production process over fragmentation and 
photons from hadron decays, observed photons are required to be sufficiently isolated from other hadronic activity in the detector.
This isolation criterion is typically formulated 
in the experimental studies by
admitting only a limited amount of hadronic energy in a fixed-size cone around the direction of the photon candidate. For a fixed-size cone, 
a non-zero amount of hadronic energy must be admitted to ensure the infrared safety of the resulting observable in QCD perturbation theory, where 
the phase space for soft radiation is not allowed to be geometrically restricted. This cone-based isolation is very efficient in suppressing secondary photons, but 
admits contributions from photon fragmentation processes. Consequently, the theory description of photon production observables must 
account for both the direct and fragmentation contributions, and subject both contributions to the photon isolation procedure applied in the 
experimental measurement.  
The fragmentation contribution can be eliminated by imposing a dynamical isolation cone~\cite{Frixione:1998jh}, smoothly lowering
 the hadronic energy cut towards its center where all emissions are vetoed.  
 While theoretically very appealing and practical, in practice, this dynamical cone isolation could not be faithfully implemented in experimental measurements.
  
 With a fixed-cone isolation prescription, direct and fragmentation contributions can not be separated from their final-state signatures and 
 are closely linked to each other in the theory description. 
 In particular, higher order perturbative terms in the direct contribution contain  collinear parton--photon configurations that give rise to 
 infrared singularities that are compensated by the mass factorisation terms~\cite{Koller:1978kq,Laermann:1982jr} of the parton-to-photon fragmentation functions. 
 The next-to-leading order (NLO) predictions for inclusive photon~\cite{Aurenche:1987fs,Baer:1990ra,Aurenche:1992yc,Gordon:1993qc,Gluck:1994iz,Catani:2002ny,Catani:2013oma}, 
and photon-plus-jet~\cite{Aurenche:2006vj} production, which 
share the same parton-level processes, account for these fragmentation contributions. They depend on the photon fragmentation 
functions, which fulfil inhomogeneous DGLAP-type evolution equations with a priori unknown non-perturbative boundary conditions. 
The available parameterisations of the photon fragmentation functions~\cite{Owens:1986mp,Gluck:1992zx,Bourhis:1997yu} rely on model estimates for 
these boundary conditions. Measurements of non-isolated photon production at LEP~\cite{Buskulic:1995au,Ackerstaff:1997nha} are the only data that 
enable a determination of the photon fragmentation functions~\cite{GehrmannDeRidder:1997gf}
and a critical assessment~\cite{GehrmannDeRidder:1998ba} of the model estimates. 
 
 Up to now, next-to-next-to-leading order (NNLO) calculations~\cite{Campbell_2017,Campbell:2017dqk,Chen_2020}
 of isolated photon and photon-plus-jet production were only performed for a dynamical cone isolation (or a hybrid variant thereof~\cite{Siegert:2016bre}), 
 since none of the available infrared subtraction schemes was able to handle fragmentation processes at NNLO. 
 This conceptual limitation has been overcome recently with the incorporation of heavy hadron fragmentation processes into 
 residue subtraction~\cite{Czakon:2021ohs} and photon fragmentation processes in antenna subtraction~\cite{Gehrmann:2022cih}. 
 In this work, we employ the antenna subtraction method~\cite{GehrmannDeRidder:2005cm,Daleo:2006xa,Currie:2013vh} to compute 
 the NNLO corrections to isolated photon and photon-plus-jet observables. 
 
 The paper is structured as follows. Section~\ref{sec:def_photons} reviews the different photon isolation prescriptions used in experiment and theory and 
 discusses their effect on direct and fragmentation contributions. The calculation of the NNLO corrections and their implementation in the \nnlojet 
 parton-level event generator is discussed in Section~\ref{sec:set_up}. Our newly derived predictions are compared with data from the LHC experiments 
 in Section~\ref{sec:exp}, discussing in particular the impact on precision phenomenology applications. 
 The incorporation of fragmentation processes allows us to perform a detailed comparison of different isolation prescriptions in Section~\ref{sec:iso_comp}, 
 where we 
 also quantify the numerical impact of the photon fragmentation functions. In view of a potential hadron collider measurement of the photon fragmentation functions, 
 we introduce a new observable derived from photon-plus-jet final states in Section \ref{sec:z_rec}, and discuss its behaviour for different photon isolation prescriptions. 
Section~\ref{sec:conc} summarises our most important findings and provides an outlook on 
 future precision phenomenology with isolated photon observables.

\section{Definition of Isolated Photons}
\label{sec:def_photons}

Photons can be produced by three different mechanisms in hadronic collisions. Direct photons originate from the point-like QED coupling to quarks and are produced in the hard scattering event. These photons usually have a clean signature, being well separated from additional hadronic radiation. The second class of photons are photons from fragmentation, i.e.\ these photons are radiated during the hadronisation process of an ordinary jet production event.  This fragmentation process is described by non-perturbative parton-to-photon fragmentation functions $D_{p \to \gamma}$. Both direct and fragmentation photons are primary photons. The last class of photons are secondary photons, which result from hadronic decays (e.g.\ $\pi^0 \to \gamma \gamma$). The majority of photons produced in hadronic collisions are secondary photons. In order to suppress this overwhelming background of photons and enhance the contribution from direct photons, a photon isolation criterion is imposed in the experimental measurement. Isolation criteria limit the hadronic energy in the vicinity of the photon by a maximal value $E_T^{\rm max}$. It is noted that lowering the maximal hadronic energy to zero is not feasible from an experimental point of view, nor it is infrared safe since soft gluon emission must be unrestricted across the entire phase space to cancel infrared singularities. Different isolation prescriptions that have been used in experimental analyses and theoretical calculations are summarised as the following:
\begin{itemize}
\item[1.]{\bf Fixed Cone Isolation}: The hadronic energy inside a cone around the photon direction with radius $R =\sqrt{ (\Delta \eta)^2 + (\Delta \phi)^2}$ in pseudorapidity $\eta$ and azimuthal angle $\phi$ is integrated. If the total hadronic energy $E_T^{\rm hadr}$ does not exceed a certain maximal value $E_T^{\rm max}$ the photon is considered to be isolated. The maximal hadronic energy can depend on the photon transverse momentum and is commonly parametrised in the form
\begin{equation}
E_T^{\rm max} = \varepsilon \, p_T^{\gamma} + E_T^{\rm thres} \, .
\end{equation}
The fixed cone isolation is characterised by three parameters $(R,\,\varepsilon, \,E_T^{\rm thres})$ and is used in all experimental measurements of photon production cross sections to date. 
\item[2.]{\bf Smooth Cone Isolation~\cite{Frixione:1998jh}}: All cones with radii $r_d$ smaller than a maximal cone radius $R_d$ are considered. A photon candidate satisfies the smooth cone isolation criterion if
\begin{equation}
E_T^{\rm hadr}(r_d) < E_T^{\rm max}(r_d) \quad \forall \, r_d \leq R_d \, .
\end{equation}
The maximal hadronic energy is smoothly decreasing with smaller cone radii $r_d$ and vanishes for $r_d = 0$. This behaviour can be implemented using the functional form
\begin{equation}
E_T^{\rm max}(r_d) = \varepsilon_d \, p_T^{\gamma} \left( \frac{1 - \cos r_d}{1 - \cos R_d } \right)^n \, ,
\end{equation}
where $\varepsilon_d > 0$ and $n>0$ are free parameters. As exact collinear radiation in the photon direction is vetoed by the smooth cone isolation criterion, the fragmentation contribution is completely eliminated. The smooth cone isolation is characterised by three parameters $(R_d, \, \varepsilon_d, \, n)$. Due to the finite resolution from the cell size of the electromagnetic calorimeter, the smooth cone isolation can only be approximated in an experimental analysis and so far has not been used in any measurement. 
\item[3.]{\bf Hybrid Isolation~\cite{Siegert:2016bre}}: A smooth cone isolation and a fixed cone isolation are combined. First a smooth cone isolation with radius $R_d$ is applied to the event. By doing so, radiation collinear to the photon is completely vetoed and the fragmentation contribution is eliminated. To all events which pass the smooth cone isolation a fixed cone isolation with a cone radius $R$ is applied. The two cone radii are chosen in such a way that $R > R_d$. 
Outside of the inner cone, hadronic radiation is treated in the same manner as in the fixed cone isolation. Therefore, the hybrid isolation is expected to be a better approximation of the fixed cone isolation compared to the smooth cone isolation.
\item[4.]{\bf Democratic Clustering~\cite{Glover:1993xc, GehrmannDeRidder:1997gf}}: The photon candidate is clustered together with the partons by using a 
sequential jet clustering algorithm. Upon completion of the clustering, the cluster containing the photon candidate is identified and the fraction of electromagnetic energy inside the cluster
\begin{equation}
z_{\rm em} = \frac{p_T^{\gamma}}{E_T^{\rm jet}}
\label{eq:z_em}
\end{equation}
is determined. If this energy fraction is larger than a certain minimal value $z_{\rm cut}$ the photon is considered to be isolated. The democratic clustering isolation is described by the jet algorithm and the parameter $z_{\rm cut}$. Refinements to the democratic clustering approach using modern jet substructure techniques 
have been discussed recently~\cite{Hall:2018jub}. 
\end{itemize}

Having implemented fragmentation processes into the theory predictions, it is also possible to   obtain results for the photon production cross section without imposing any photon isolation. These inclusive predictions contain photons from primary production processes. Secondary photons, which would dominate the measured non-isolated photon production cross section, are not included. Therefore, these inclusive predictions can not directly be compared to experimental measurements. It is however noted, that they can still provide important insights on direct and fragmentation photons in inclusive scattering kinematic. A detailed numerical study of the impact of the isolation prescription on the photon-plus-jet cross section is presented in Section~\ref{sec:iso_comp}, where also results for non-isolated photon-plus-jet production are shown.

\section{Computational Set-Up}
\label{sec:set_up}

Isolated photon and photon-plus-jet production at hadron colliders originate from the same parton-level subprocesses, since the  finite transverse momentum of the 
isolated photon requires the presence of a partonic recoil. Isolated photon cross sections are fully inclusive over this extra partonic radiation, while photon-plus-jet cross sections are obtained by requiring this extra radiation to be recombined into a resolved jet within kinematical acceptance. The relevant Born-level subprocesses~\cite{Halzen:1978rx,Ruckl:1978mx} are quark--antiquark 
annihilation $q\bar q \to \gamma g$ and quark--gluon scattering $qg \to \gamma q$. 

Three different parton-level contributions combine to yield the NNLO QCD corrections for isolated photon and photon-plus-jet production: the double virtual contribution (VV) consisting of the two-loop corrections to the Born-level processes, the real-virtual contribution (RV) consisting of one-loop Feynman diagrams with an additional parton radiation and double real (RR) contribution where two additional partons are emitted. All matrix elements in the different types of contributions are known  in compact analytic form~\cite{Anastasiou:2001sv,Bern:2003ck,Bern:1994fz,Signer:1995np,Signer:1995a,DelDuca:1999pa}. 

All three types of NNLO corrections are individually infrared divergent. The RR and RV contributions exhibit singularities from configurations in which additional partons become unresolved, i.e.\ soft and/or collinear. The RV and VV corrections contain explicit singularities from the integration over the corresponding loop momenta. These singularities cancel among each other once all contributions are summed up and the mass factorisation related to the parton distributions and fragmentation functions is performed. In order to extract the singularities in the different types of corrections and combine them into an explicitly finite result, a subtraction method is employed. For our calculation we use the antenna subtraction method~\cite{GehrmannDeRidder:2005cm,Daleo:2006xa,Currie:2013vh}, which is implemented in the \nnlojet parton-level event generator. 

We present in this paper the first NNLO calculation using the fixed cone photon isolation that is applied in all experimental analyses to date. Therefore, it overcomes an essential drawback of previous NNLO calculations for inclusive photon~\cite{Campbell_2017,Chen_2020} and photon-plus-jet production~\cite{Chen_2020, Campbell:2017dqk} that used idealised isolations (smooth cone or hybrid isolation), which rely on an empirical tuning of the isolation parameters in the theory calculation to approximate the effect of the experimental fixed cone procedure. A fixed cone isolation allows singular photon-parton collinear configurations, which have to be subtracted along with the QCD singularities up to NNLO. At variance with the subtraction of QCD singularities for which no information on individual partons in unresolved limits is required, photonic singularities have to be subtracted while retaining the information on the energy fraction of the photon in the collinear cluster. For the subtraction of photonic singularities we use a new class of antenna functions, fragmentation antenna functions~\cite{Gehrmann:2022cih}, whose integrated forms depend on the final-state momentum fraction. Calculations with a fixed cone isolation also receive contributions from fragmentation processes. The mass factorisation terms of the fragmentation functions cancel the explicit poles of the integrated fragmentation antenna functions. Our construction of the antenna subtraction terms relevant to photon-parton collinear singularities and their interplay with the mass factorisation of the fragmentation functions is described in detail elsewhere~\cite{Gehrmann:2022cih}.

The additional subtraction terms needed for the subtraction of the parton--photon collinear limits are largely independent of the genuine QCD subtraction terms, so that they can be combined in an additive manner. For the calculation presented in this paper we build on the QCD subtraction terms which were already used in a previous \nnlojet calculation for photon production cross sections~\cite{Chen_2020} with a hybrid isolation. 

The fragmentation contribution as well as the direct contribution (through the mass factorisation terms of the fragmentation functions) depend on the
fragmentation scale $\mu_A$. 
In the same manner as for the 
 mass factorisation scale in the parton distribution functions, this scale should be chosen close to the characteristic scale of 
the process under consideration. In the presence of photon isolation, the determination of this characteristic scale does however deserve special attention. 
The mass factorisation procedure dresses the bare parton distributions and fragmentation functions with unresolved parton radiation. This unresolved parton radiation is 
inclusive in transverse momentum relative to the primary bare parton, with relative transverse momenta ranging between zero (the exact collinear limit) and 
the factorisation scale. This factorisation scale does therefore reflect up to which relative transverse momentum scale additional parton radiation is admitted into the 
observable under consideration. For the mass factorisation of the parton distributions, this factorisation scale can clearly be chosen to be of the order of the 
transverse momentum of the photon $p_T^{\gamma}$, since the isolated photon and photon-plus-jet cross sections are inclusive in initial-state radiation up to this scale. The
situation for final-state photon fragmentation is however entirely different. Here, the relative transverse 
momentum of the final-state photon with respect to the fragmenting parton--photon cluster defines the characteristic scale of the process. In the absence of any 
isolation criterion, this scale could also be of the order of $p_T^{\gamma}$, since no restrictions on additional radiation are applied. As soon as an isolation requirement is imposed
on the photon, however, any additional radiation associated with the fragmentation process is severely restricted. Only radiation inside the isolation cone and below the 
maximal hadronic energy threshold is admitted, thereby substantially lowering the scale up to which the observable is inclusive in extra radiation. 

For our calculations we set this scale close to the maximal invariant mass inside the isolation cone. For a fixed cone isolation with cone radius $R$ and maximal hadronic energy $E_T^{\rm max}$ it reads
\begin{equation}
m_{\rm cone} = \sqrt{E_T^{\rm max} \, p_T^{\gamma} \, R^2} + \mathcal{O}(R^3)
\label{eq:m_cone}
\end{equation}
where we assumed that the hadronic energy stems from a single parton radiation at the distance $R$ to the photon. 
This assumption is valid as long as the value for the scale remains larger than the hadronisation scale. Due to the small values for $E_T^{\rm max}$ used in the experimental isolation prescriptions the typical numerical values for the fragmentation scale are $\mathcal{O}(1 \, \GeV ) $ to $\mathcal{O} (10 \, \GeV)$ and therefore small compared to the renormalisation and factorisation scale which are set by default to $\mu_R = \mu_F = p_T^{\gamma}$ in our calculation.

For an isolated photon cross section, a choice of fragmentation scale  $\mu_A$ of the order of $p_T^{\gamma}$ would imply that partonic radiation up to this scale 
is removed from the direct contribution and assigned to the fragmentation contribution, which is added back in a resummed form. For photon-parton invariant masses between 
$m_{\rm cone}$ and $p_T^{\gamma}$, this radiation is however either vetoed by the isolation prescription or represents resolved out-of-cone configurations that should remain 
part of the direct contribution. By attributing them to the fragmentation contribution, their kinematics is misrepresented and their overall impact is 
underestimated since the  collinear resummation of the 
photon fragmentation functions attenuates them in the range of momentum fractions relevant to isolated photon cross sections. 
As a matter of fact, cone-isolation cross sections lead to a complicated hierarchy of scales~\cite{Catani:2013oma} and to the emergence of non-global logarithmic corrections~\cite{Balsiger:2018ezi,Ebert:2019zkb} that potentially 
require dedicated resummation. 
  
The scale uncertainties of the predictions are determined by varying the three scales by a factor of 2 and $\tfrac{1}{2}$ around their central values:
\begin{equation}
\mu_R = a \, p_T^{\gamma} \, , \quad \mu_F = b \, p_T^{\gamma} \,  , \quad \mu_A = c \, m_{\rm cone} \, ,
\label{eq:default_scales}
\end{equation}
with $a,b,c \in \bigl\{\tfrac{1}{2}, 1, 2 \bigr\}$ and only retaining combinations with $\tfrac{1}{2} \leq a/b,\,  a/c, \, b/c \leq 2$ (15-point scale variation). For our predictions we use the NNPDF4.0 PDF set~\cite{NNPDF:2021njg} and the BFGII set~\cite{Bourhis:1997yu} for the parton-to-photon fragmentation functions. Moreover, we assign an $\mathcal{O}(\alpha)$ power counting to the quark-to-photon and gluon-to-photon fragmentation functions, which is justified by the low 
mass factorisation scale used for these functions, and their overall small contribution to the cross sections. It should be noted that this power counting differs from 
the usual assignment of $\mathcal{O}(\alpha/\alpha_s)$ to the photon fragmentation functions, which lacks however any formal ground as soon as 
 $\mu_A \neq \mu_R$. 
 
Imposing this power counting, the photon production cross sections at the different levels of accuracy read
\begin{eqnarray}
\rd \hat{\sigma}^{\gamma+ X,{\rm LO}} &= &\rd \hat{\sigma}_{\rm dir}^{\rm LO} \, ,\nonumber \\
\rd \hat{\sigma}^{\gamma+ X,{\rm NLO}} &= & \rd \hat{\sigma}_{\rm dir}^{\rm NLO} +  \rd \hat{\sigma}^{\rm NLO}_{\rm frag}  \, , \nonumber \\
\rd \hat{\sigma}^{\gamma+ X,{\rm NNLO}} &= & \rd \hat{\sigma}_{\rm dir}^{\rm NNLO} + \rd \hat{\sigma}_{\rm frag}^{\rm NNLO} \, ,
\label{eq:sigma}
\end{eqnarray}
where the direct contribution is given by
\begin{eqnarray}
\rd \hat{\sigma}_{\rm dir}^{\rm LO}  &= & \, \rd \hat{\sigma}_{\gamma}^{{\rm LO}},\nonumber \\
\rd \hat{\sigma}_{\rm dir}^{\rm NLO} &= &\rd \hat{\sigma}_{\gamma}^{{\rm NLO}} + \rd \hat{\sigma}_{\rm MF}^{{\rm NLO}}  \, , \nonumber  \\
\rd \hat{\sigma}_{\rm dir}^{\rm NNLO} &= & \rd \hat{\sigma}_{\gamma}^{{\rm NNLO}} + \rd \hat{\sigma}_{\rm MF}^{\rm NNLO} \, ,
\end{eqnarray}
and the fragmentation contribution at the respective order takes the form
\begin{eqnarray}
\rd \hat{\sigma}_{\rm frag}^{\rm NLO} &= &\sum_p \rd \hat{\sigma}^{{\rm LO}}_{p} \otimes D_{p \to \gamma}  \, , \nonumber  \\
\rd \hat{\sigma}_{\rm frag}^{\rm NNLO} &= & \sum_p \rd \hat{\sigma}_{p}^{{\rm NLO}} \otimes D_{p \to \gamma} \, .
\end{eqnarray}
Here, the sum over $p$ runs over all final-state partons and $D_{p\to\gamma}$ are the mass-factorised fragmentation functions. The cross section $\rd \hat{\sigma}_i$ denotes the production of a hard resolved particle $i$ and $\rd \hat{\sigma}_{\rm MF}$ contains the mass factorisation counter-terms of the fragmentation functions.

\subsection{Comparison with JETPHOX}
\label{subsec:comp_JP}

NLO QCD corrections to isolated photon and photon-plus-jet cross sections using a fixed cone isolation have already been calculated and implemented in the JETPHOX code~\cite{Catani:2002ny}. These predictions allow for a validation of the implementation of  fragmentation processes up to NNLO (employing our counting of orders) and the validation of the direct contribution up to NLO in \nnlojet. 

In order to compare with JETPHOX we use a similar set-up to the ATLAS 13\,\TeV photon-plus-jet study~\cite{ATLAS:2017xqp}. The following cuts on the transverse momenta and (pseudo-)rapidities of the photon and the jet are imposed:
\begin{equation}
p_T^{\gamma} > 125 \, \GeV \quad p_T^{{\rm jet}} > 100 \, \GeV, \quad | \eta_{\gamma} | < 2.37, \quad |y_{{\rm jet}} | < 2.37 \, .
\end{equation}
To simplify the set-up we do not apply additional cuts on the pseudorapidity of the photon nor do we require a minimum angular separation between the photon and the jet. A fixed cone photon isolation with the parameters
\begin{equation}
R = 0.4, \quad \varepsilon=0.0042, \quad E_T^{\rm thres} = 10\, \GeV
\end{equation}
is imposed.
Jets are constructed using a $k_T$ jet algorithm with a cone radius $R_{\rm jet}=0.4$.
For comparison only we set all scales (also the fragmentation scale) to the transverse momentum of the photon, i.e.\ $\mu_R = \mu_F = \mu_A = p_T^{\gamma}$ and use the NNPDF31\_nnlo\_as\_0118\_mc PDF set~\cite{NNPDF:2017mvq}. 

\begin{table}
\centering
\begin{tabular}{|l|r|r|}
\hline
                                           & \multicolumn{1}{c|}{\nnlojet} & \multicolumn{1}{c|}{JETPHOX}  \\ \hline
$\sigma_{{\rm dir}}^{{\rm LO}}$   & $(192.487 \pm 0.009) \, \pb$ & $(192.488 \pm 0.003) \, \pb$  \\ \hline
$\sigma_{{\rm dir}}^{{\rm NLO, only}}$  & $(97.903 \pm 0.058) \, \pb$  & $(97.75 \pm 0.14) \, \pb$     \\ \hline
$\sigma_{{\rm frag}}^{{\rm NLO}}$  & $(14.255 \pm 0.009) \, \pb$  & $(14.2550 \pm 0.0002) \, \pb$ \\ \hline
$\sigma_{{\rm frag}}^{{\rm NNLO, only}}$ & $(14.999 \pm 0.022) \, \pb$  & $(14.971 \pm 0.015) \, \pb$   \\ \hline
\end{tabular}
\caption{Comparison of direct and fragmentation contributions to the photon-plus-jet cross section obtained with \nnlojet and JETPHOX. The 'only' numbers indicate the 
increment from the coefficient at the respective order. }
\label{tab:com_NNLOJET_JP}
\end{table}

The comparison between the JETPHOX and \nnlojet results for the predictions of the different contributions to the photon-plus-jet cross sections are shown in Table\,\ref{tab:com_NNLOJET_JP}. The predictions for the direct and fragmentation contribution obtained with \nnlojet and JETPHOX agree within the respective integration uncertainties. However, we discovered a 1\% deviation between the NNLO fragmentation coefficient stemming from the numerically small 
gluon-to-photon fragmentation process alone, 
which is however within the technical uncertainties of the JETPHOX implementation. 
 
\section{Comparison with Experimental Measurements}
\label{sec:exp}

The cross section for photon production in hadronic collisions has been studied extensively in experimental measurements. Earlier measurements of isolated photon~\cite{Athens-Athens-Brookhaven-CERN:1979oxx, Anassontzis:1982gm, CMOR:1989qzc, UA1:1988zam, UA2:1991wce, D0:2005ofv, CDF:2017cuc}  and photon-plus-jet production~\cite{D0:2008chx, D0:2013lra} in hadronic collisions were followed by measurements at the LHC. These measurements for isolated photon production~\cite{ATLAS:2017nah, CMS:2018qao, ATLAS:2019buk, ALICE:2019rtd} as well as for the production of a photon in association with a jet~\cite{CMS:2018qao, ATLAS:2017xqp, CMS:2019jlq} are covering a large kinematic range and are reaching percent level accuracy demanding for theory predictions meeting this level of precision.

In the experimental analyses selection cuts on the photon are imposed. Typically these cuts limit the transverse momentum $p_T^{\gamma}$ and the pseudorapidity $\eta_{\gamma}$ of the photon to a specific kinematic range. Photon-plus-jet measurements use the anti-$k_T$ algorithm~\cite{Cacciari:2008gp} for jet reconstruction and apply
additional cuts on the transverse momentum of the (leading) jet $p_T^{\rm jet}$ as well as on its rapidity $y_{\rm jet}$ or pseudorapidity $\eta_{\rm jet}$. 

In the following we will compare our NNLO predictions for isolated photon production to 13\,\TeV ATLAS~\cite{ATLAS:2019buk} and 7\,\TeV ALICE data~\cite{ALICE:2019rtd} and our predictions for the photon-plus-jet cross section to 8\,\TeV CMS~\cite{CMS:2019jlq} and 13\,\TeV ATLAS data~\cite{ATLAS:2017xqp}. Throughout this comparison we set the value for the electromagnetic coupling constant to $\alpha = \alpha(0) = 1/137.036$, as is appropriate for the coupling of an external resolved photon.  
For the first time these NNLO predictions employ the same photon isolation as in the experimental measurements, thus avoiding the mismatch inherent in previous theory--data comparisons at this order and eliminating the associated source of systematic uncertainty.

\subsection{ATLAS 13 TeV Isolated Photon Measurement}
\label{subsec:ATLAS_iG}
The ATLAS 13\,\TeV isolated photon measurement~\cite{ATLAS:2019buk} provides data for the photon production cross section differential in the photon transverse momentum $p_T^{\gamma}$ for four different pseudorapidity bins 
\begin{equation}
|\eta_{\gamma}| < 0.6, \quad 0.6 < |\eta_{\gamma} | < 1.37, \quad 1.56 < |\eta_{\gamma} | < 1.81, \quad 1.81< |\eta_{\gamma}| < 2.37 \, .
\end{equation}
The photon is required to have a minimum transverse momentum $p_T^{\gamma} >125 \, \GeV$. Moreover, a fixed cone photon isolation with parameters
\begin{equation}
R=0.4, \quad \varepsilon = 0.0042, \quad E_T^{\rm thres} = 4.8 \, \GeV
\end{equation}
is imposed. For our theoretical predictions we use the same photon isolation and the scales are set according to \eqref{eq:default_scales}. 

\begin{figure}[!t]
\centering
\begin{subfigure}[b]{0.496\textwidth}
\centering
\includegraphics[width=\textwidth]{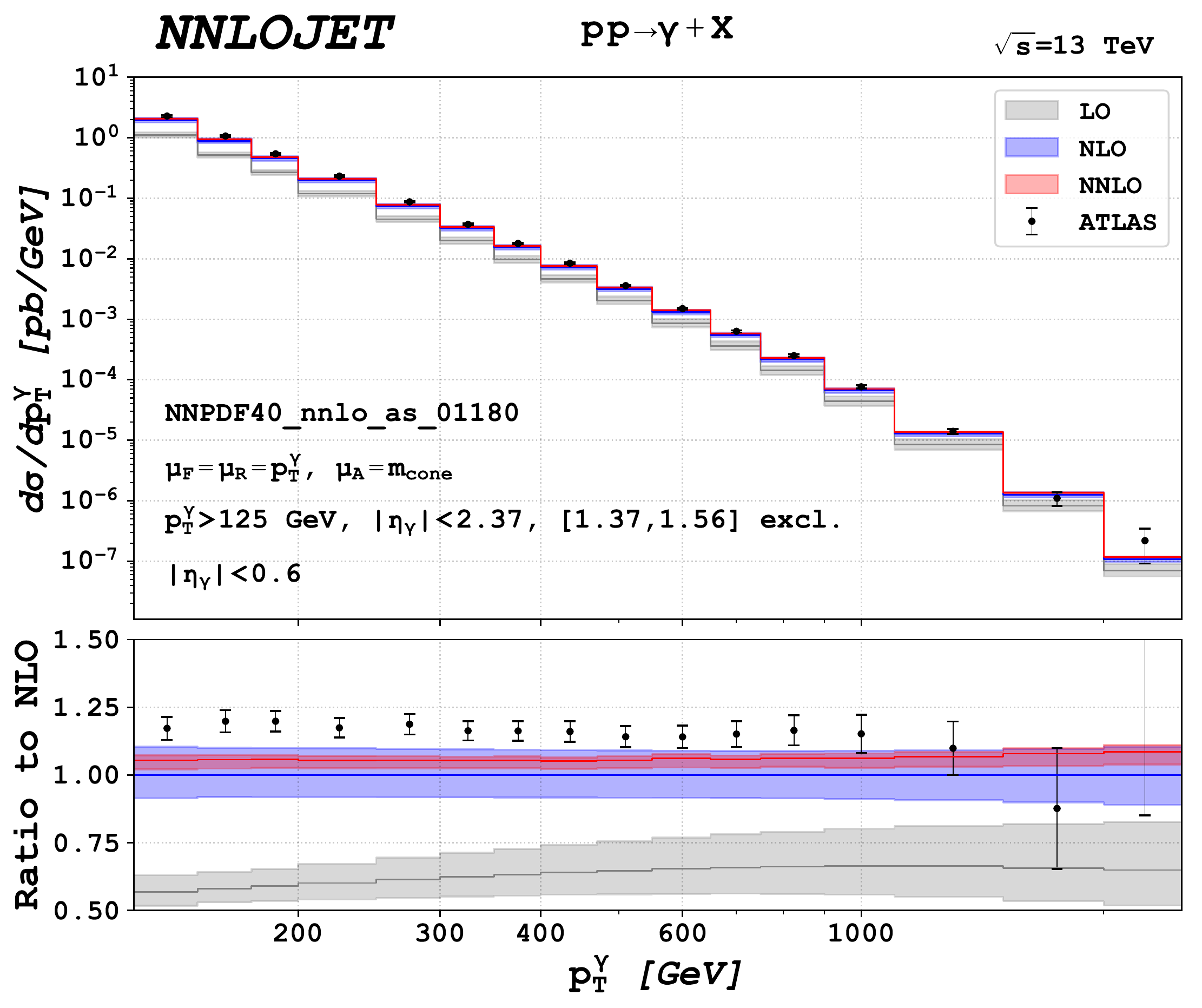}
\end{subfigure}
\hfill
\begin{subfigure}[b]{0.496\textwidth}
\centering
\includegraphics[width=\textwidth]{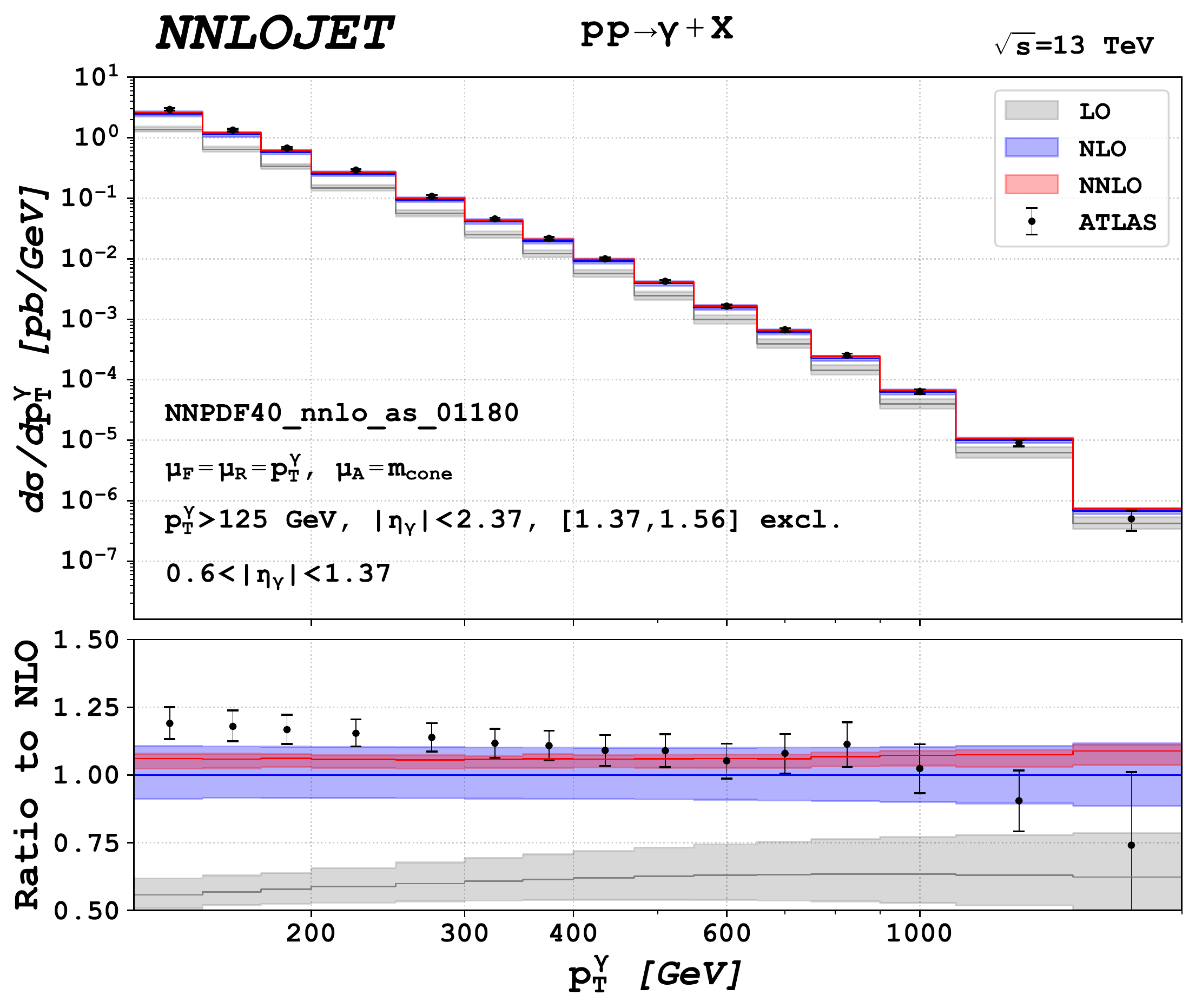}
\end{subfigure}
\vskip\baselineskip
\begin{subfigure}[b]{0.496\textwidth}   
\centering 
\includegraphics[width=\textwidth]{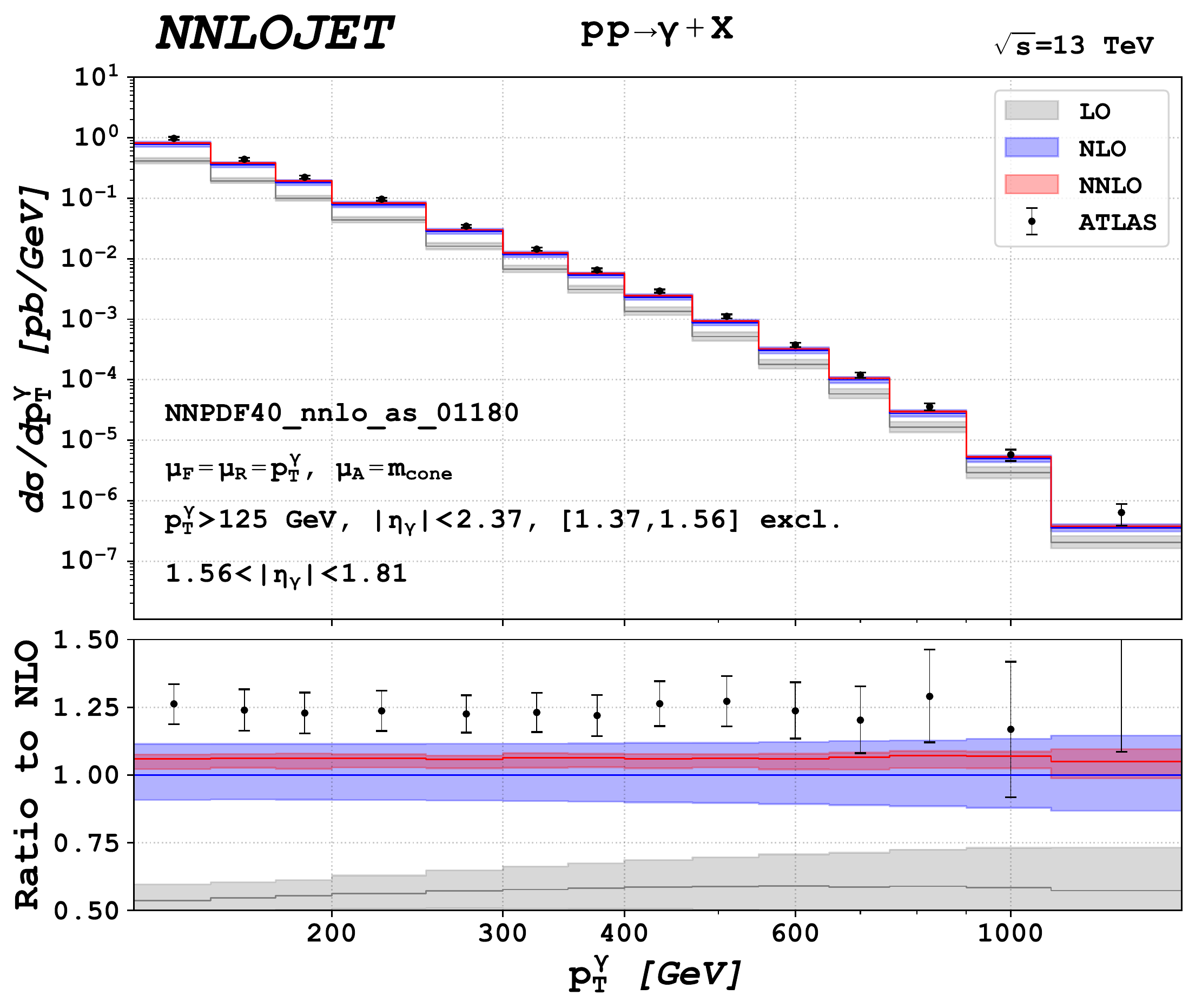}
\end{subfigure}
\hfill
\begin{subfigure}[b]{0.496\textwidth}   
\centering 
\includegraphics[width=\textwidth]{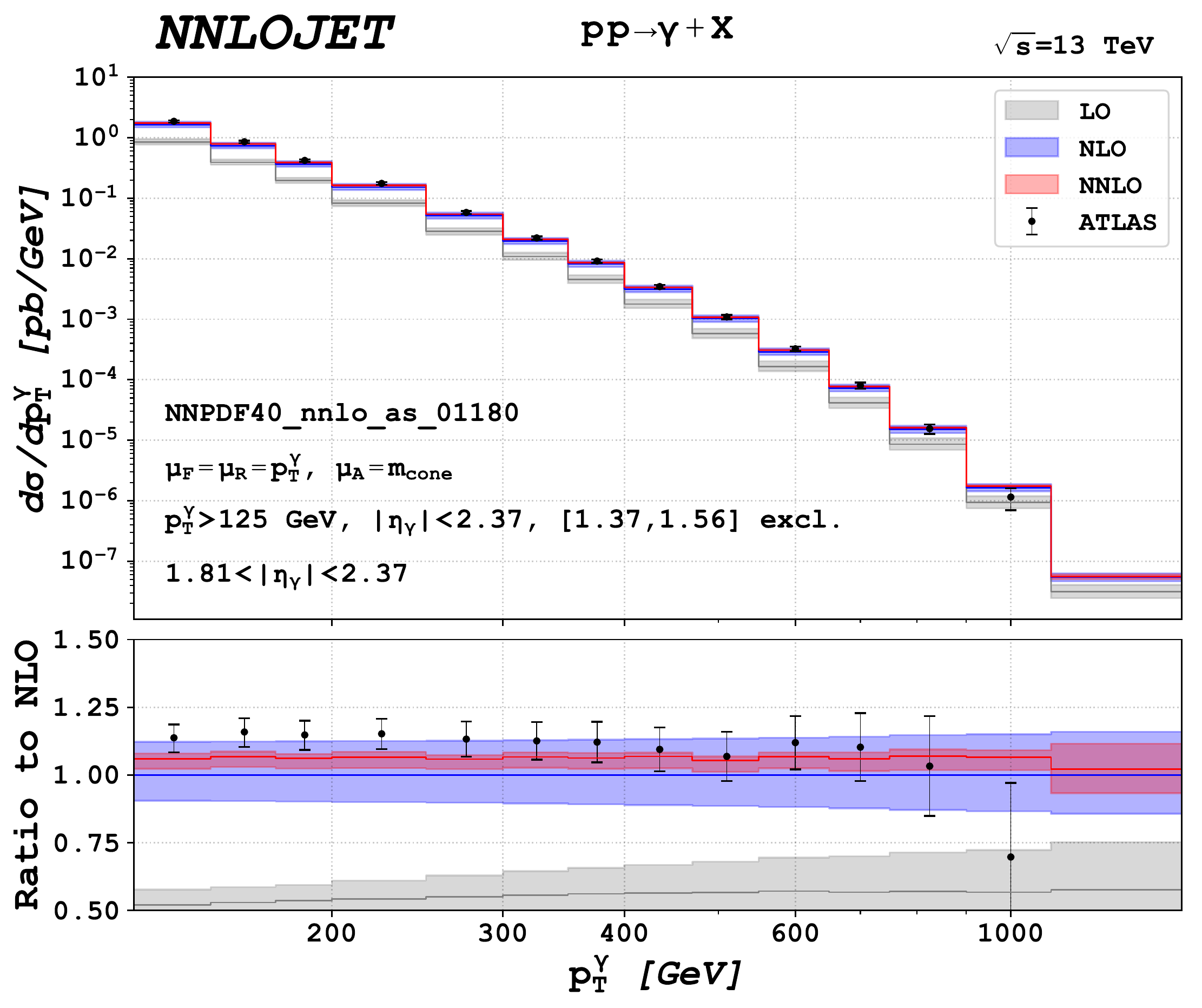}
\end{subfigure}
\caption{Photon transverse momentum distributions for isolated photon production at LO, NLO and NNLO for four different pseudorapidity bins. The prediction are compared to 13\,\TeV ATLAS data~\protect\cite{ATLAS:2019buk}.}
\label{fig:ATLAS_13_iG}
\end{figure}

Figure~\ref{fig:ATLAS_13_iG} shows the isolated photon cross section for the four photon pseudorapidity bins. For all bins the NNLO corrections are positive and increase the cross section by 5--6\% compared to the NLO cross section. These corrections are relatively constant over the whole transverse momentum range. However, we observe an increase in the NNLO-to-NLO K-factor for the first two pseudorapidity bins for very large transverse momenta where the corrections increase to roughly 9\% compared to the NLO predictions. In contrast, for the forward pseudorapidity bins the K-factor is decreasing for large transverse momenta where the difference between the NLO and NNLO cross section is only a few percent. 

The scale uncertainty at NLO is 18\% for the first and second pseudorapidity bin and increases to 24\% for the more forward pseudorapidity bins. In all bins the scale uncertainty is slightly increasing towards larger $p_T^{\gamma}$. Including NNLO corrections significantly reduces the scale uncertainty. Over most of the range in $p_T^{\gamma}$ and $\eta_{\gamma}$ the NNLO scale variation amounts to $(+1.5, -3.0)\%$ and it only increases moderately for large photon transverse momenta. 

Comparing our NNLO predictions with the ATLAS data, we observe that in the first, third and fourth pseudorapidity bin the slope of the measured distribution is well captured by the predictions. However, the predictions undershoot the data by 10\% (15\%) in the first (third) bin and are not compatible in normalisation with the data over the largest range of photon transverse momenta. In the second and fourth pseudorapidity bin, data and predictions agree in the mid $p_T^{\gamma}$ region. In the second bin, the slope of the measured distribution is slightly steeper than that of the theory predictions.

The fragmentation contribution to the NNLO cross section is very small and amounts to only 2\% of the full NNLO cross section in the low $p_T^{\gamma}$ region and is negligible for mid and large transverse momenta. Therefore, the NNLO predictions presented here are very similar to the NNLO predictions obtained with a hybrid isolation shown in~\cite{ATLAS:2019buk},  taking however into consideration a different overall normalisation
due to the different value for the electromagnetic coupling constant that was used there.

\subsection{ALICE 7 TeV Isolated Photon Measurement}
\label{subsec:ALICE_iG}
The ALICE 7\,\TeV isolated photon measurement~\cite{ALICE:2019rtd}  provides data for the photon production cross section differential in the photon transverse momentum in the range $10 \, \GeV < p_T^{\gamma} < 60 \, \GeV$. The photon is required to have a pseudorapidity $|\eta_{\gamma}| < 0.27$ and a fixed cone photon isolation with the parameters
\begin{equation}
R=0.4, \quad \varepsilon = 0, \quad E_T^{{\rm thres}} = 2 \, \GeV 
\label{eq:iso_para_ALICE}
\end{equation}
is imposed. 
\begin{figure}[!t]
\centering
\includegraphics[scale=0.40]{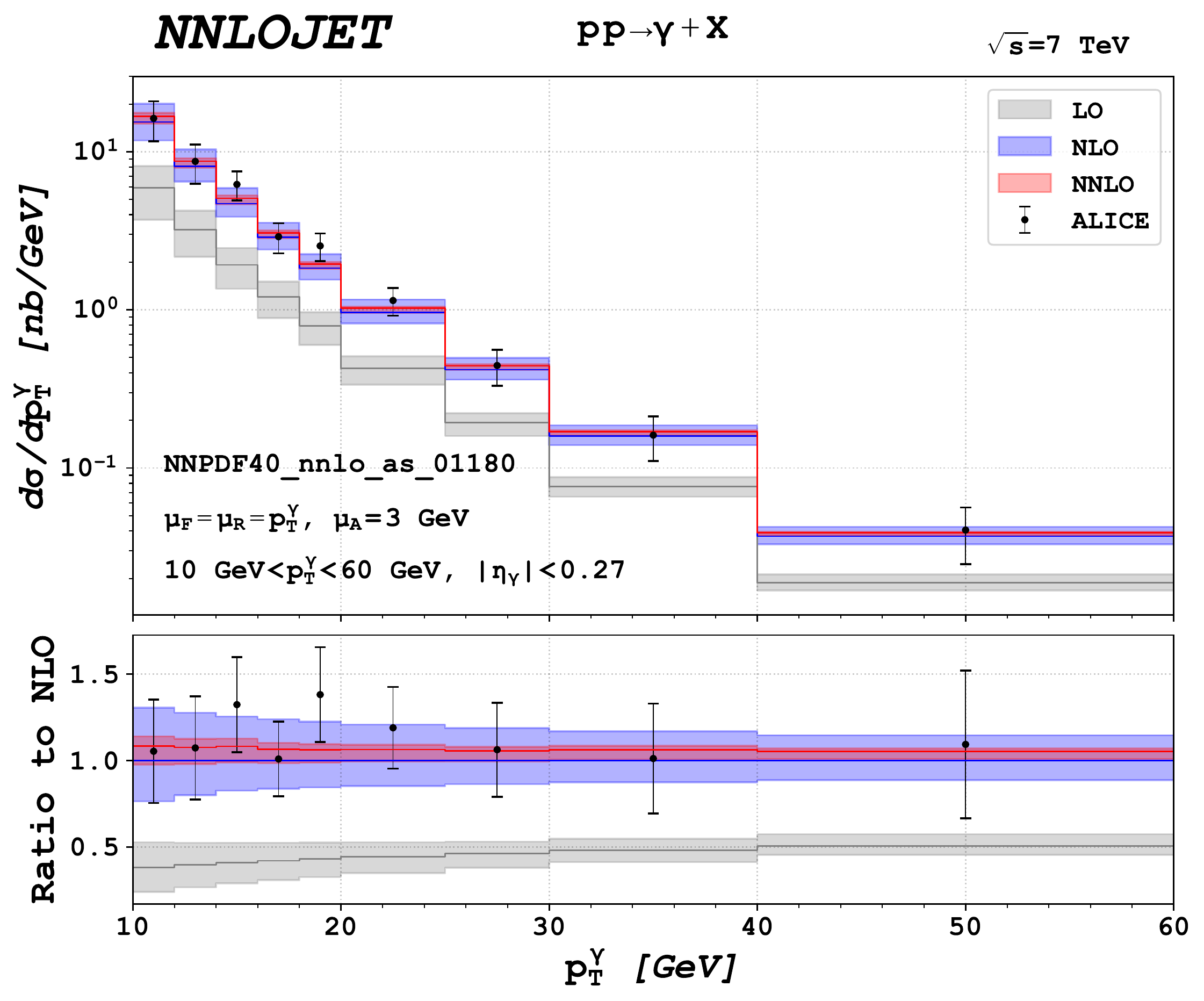}
\caption{Predictions for the photon transverse momentum distribution at LO, NLO, and NNLO. The predictions are compared to data from the ALICE measurement~\protect\cite{ALICE:2019rtd}.}
\label{fig:ALICE_7_iG}
\end{figure}

The prediction for the photon transverse momentum distribution and the comparison to the ALICE data is shown in Figure~\ref{fig:ALICE_7_iG}. In our predictions for the ALICE measurement we used a fixed fragmentation scale of $\mu_A = 3 \, \GeV$ since our default choice $\mu_A = m_{\rm cone}$ in Eq.~\eqref{eq:m_cone} yields
 numerical values for the fragmentation scale that are too small and outside of the supported range of the BFGII fragmentation functions. Lastly, we multiply our predictions by the correction factors accounting for hadronisation effects and underlying event contributions that are quoted in~\cite{ALICE:2019rtd}. We use a fixed cone photon isolation with the parameters of \eqref{eq:iso_para_ALICE}.

The NNLO corrections enhance the NLO predictions by approximately 8\% in the region $10 \, \GeV < p_T^{\gamma} < 16 \, \GeV$ and 6\% in the region $ 16 \, \GeV < p_T^{\gamma} < 60 \, \GeV$. A significant reduction of scale uncertainty is observed when going from NLO to NNLO. The NLO scale variation ranges from 
 54\% in the lowest bin to  26\% in the highest bin. In contrast, the scale uncertainty at NNLO is only $(+ 5, -10)\%$ in the low $p_T^{\gamma}$ regime and further decreases to $(+1.5, -4.5)\%$ in the last transverse momentum bin.

All data points are in agreement with our NNLO prediction within the experimental and theoretical uncertainties, noting that experimental errors exceed the scale uncertainty of the NNLO calculation across the entire kinematic range.

The contribution from fragmentation processes to the NNLO cross section is decreasing with increasing $p_T^{\gamma}$. In the lowest bin the fragmentation contribution amounts to 10\% of the cross section while in the highest bin it only contributes 2\% to the cross section.

\subsection{CMS 8 TeV Photon-Plus-Jet Measurement}
\label{subsec:CMS_GJ}
The CMS 8\,\TeV photon-plus-jet study~\cite{CMS:2019jlq} provides data for the photon-plus-jet cross section as a function of the transverse momentum of the photon, the pseudoradidity of the photon, and the pseudorapidity of the jet. The photon is required to have $p_T^{\gamma} > 40 \, \GeV$ and $|\eta_{\gamma}|<1.44$ or $1.57 < |\eta_{\gamma}|<2.5$. 
Jets are identified using the anti-$k_T$ algorithm with a cone radius $R_{{\rm jet}}=0.5$. The leading jet is required to have $p_T^{{\rm jet}} > 25 \, \GeV$ and $|\eta_{{\rm jet}}| < 2.5$. Moreover, it must be separated from the photon candidate by $\Delta R_{\gamma{\rm jet}} > 0.5$. 

While the ATLAS and ALICE measurements apply their photon isolation procedures at the fiducial cross section level, photon isolation in the CMS 
measurement~\cite{CMS:2019jlq} 
is performed only at the level of object reconstruction. The photon isolation parameters used in the CMS measurement can thus not be taken over to the 
theory predictions (which are always at the level of fiducial cross sections) in a direct manner. The CMS 8\,\TeV study recommends~\cite{CMS:2019jlq}  
that theory predictions should be computed with the fixed-cone isolation parameters
\begin{equation}
R = 0.2, \quad \varepsilon=0 , \quad E_T^{{\rm thres}} = 4 \, \GeV
\label{eq:iso_para_CMS}
\end{equation}
to mimic the effect of photon isolation in their measurement. 

Data for the photon transverse momentum distribution are given for different bins in the pseudorapidity of the photon
\begin{equation}
|\eta_{\gamma}| < 0.8, \quad  0.8 <|\eta_{\gamma}| < 1.44, \quad 1.57 < |\eta_{\gamma} | < 2.1, \quad 2.1 < |\eta_{\gamma}| < 2.5 \, ,
\end{equation}
and the jet
\begin{equation}
|\eta_{{\rm jet}}| < 0.8, \quad  0.8 <|\eta_{{\rm jet}}| < 1.5, \quad 1.5 < |\eta_{{\rm jet}} | < 2.1, \quad 2.1 < |\eta_{{\rm jet}}| < 2.5 \, .
\end{equation}

\begin{figure}[!t]
\centering
\begin{subfigure}[b]{0.496\textwidth}
\centering
\includegraphics[width=\textwidth]{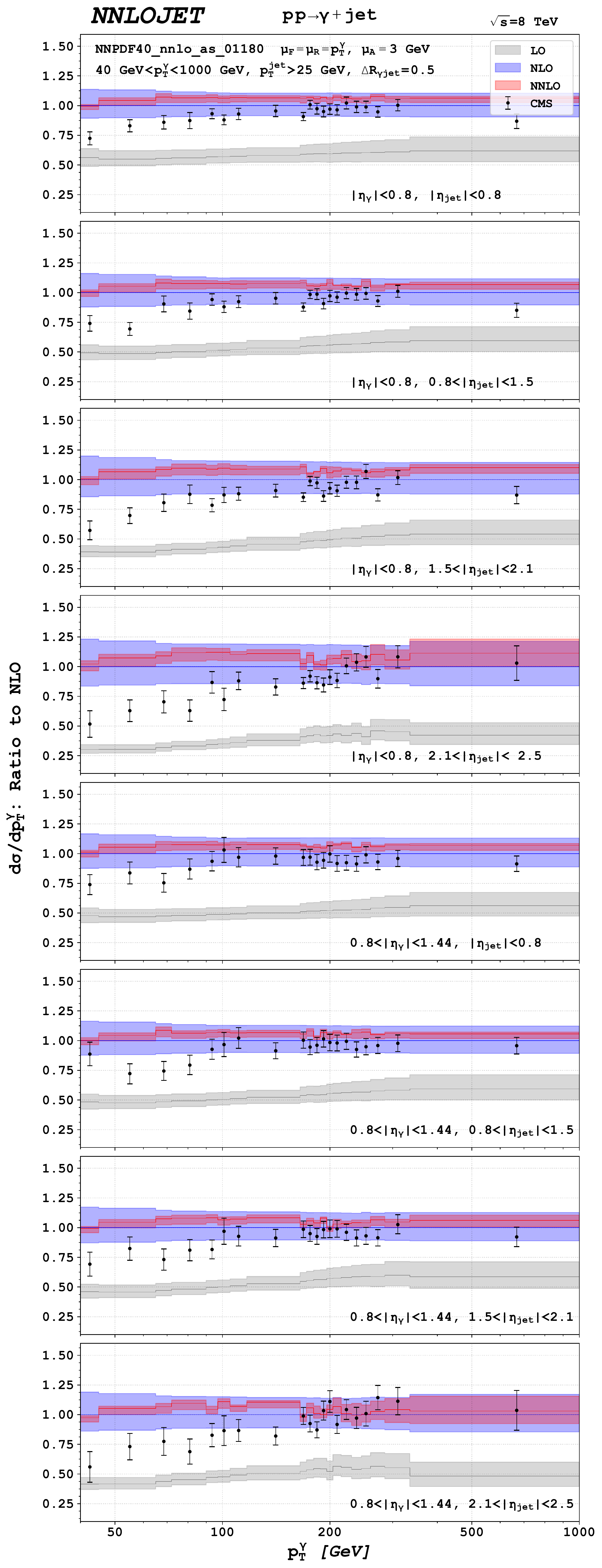}
\end{subfigure}
\hfill
\begin{subfigure}[b]{0.496\textwidth}
\centering
\includegraphics[width=\textwidth]{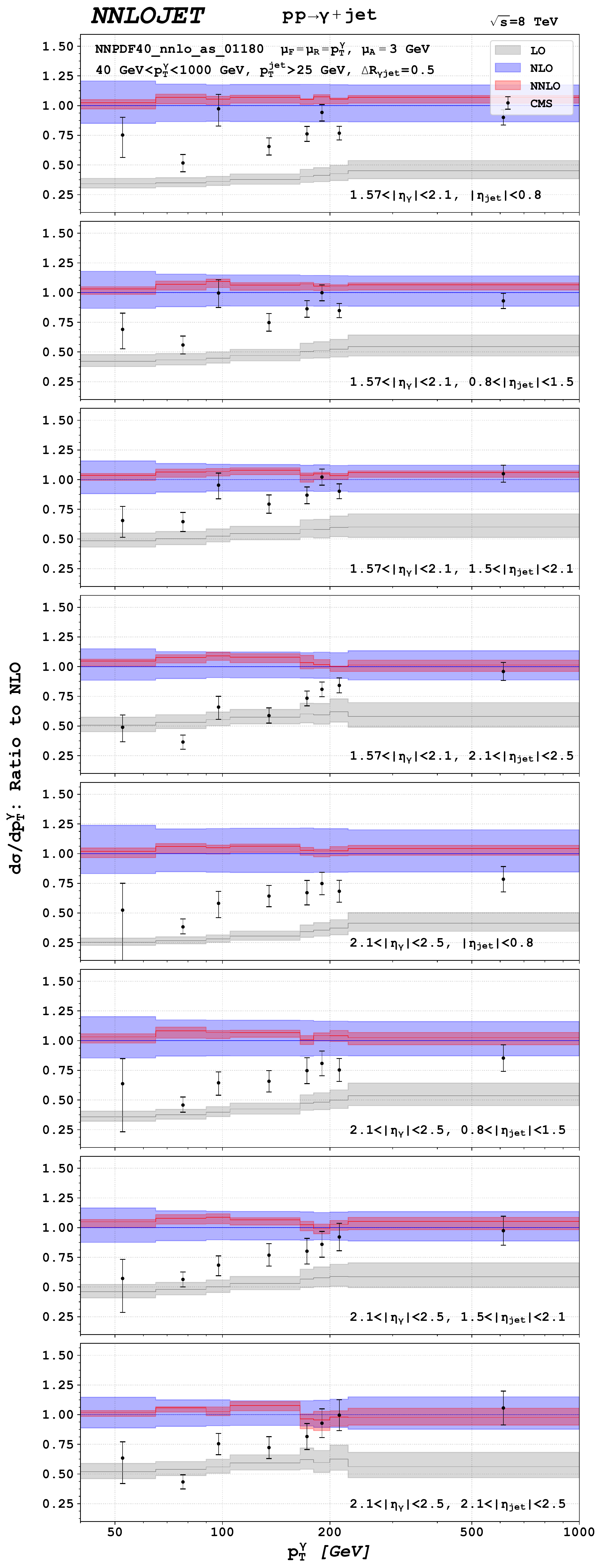}
\end{subfigure}
\caption{Photon transverse momentum distributions for different bins in the pseudorapidity of the photon and the jet. The predictions are compared to CMS 8\,\TeV data~\protect\cite{CMS:2019jlq}.}
\label{fig:CMS_8_GJ}
\end{figure}

For the theory predictions we use a fixed cone isolation with the parameters of \eqref{eq:iso_para_CMS}. Since the numerical values of our default choice of the fragmentation scale $\mu_A = m_{\rm cone}$ are again too small and not supported by the BFGII set for these isolation parameters, we use a fixed fragmentation scale of $\mu_A = 3 \, \GeV$. Our NNLO predictions are compared to CMS data in the 16 different pseudorapidity ranges in Figure~\ref{fig:CMS_8_GJ}. For most of the range of photon and jet pseudorapidities the NNLO corrections are positive and increase the cross section by $7-10\%$. Compared to NLO the shape of the NNLO distribution is slightly modified. In the low $p_T^{\gamma}$ regime the corrections are small (for some pseudorapidity bins even negative) and increase towards larger transverse momenta. For some of the forward pseudorapidity bins a decline of the NNLO-to-NLO K-factor for high photon transverse momenta is observed. The scale uncertainties at NLO amount to approximately 25\% for similar photon and jet pseudorapidities and increase up to 40\% for unbalanced pseudorapidity configurations. At NNLO the scale uncertainty is reduced to $(+1, -3)\%$ for comparable pseudorapidities of the photon and the jet. For less balanced configurations we also observe a substantial reduction of scale uncertainty compared to NLO for small and mid $p_T^{\gamma}$. However, in these configurations the NNLO scale uncertainty increases in the high $p_T^{\gamma}$ regime.

Comparing the NNLO predictions to the CMS data, we observe a disagreement, which is most prominent in the low $p_T^{\gamma}$ regime where the predictions overshoot the data by up to 50\%. This tension between data and theory predictions is reduced for mid and large photon transverse momenta. However, also in this $p_T^{\gamma}$ range we observe an inconsistency between theory and data in some of the forward pseudorapidity bins of the photon and the jet. 
We note that the reduction by more than a factor of two of the scale uncertainty bands  going from NLO to NNLO is essential to fully expose this tension.

Due to the tight isolation criterion \eqref{eq:iso_para_CMS} the fragmentation contribution to the NNLO cross section is small. Its contribution to the cross section is 3--5\% in the lowest photon transverse momentum bin and declines steeply for increasing $p_T^{\gamma}$. However, given that these fiducial-level photon isolation criteria are only an approximation to the object-level isolation criteria used in the CMS measurement~\cite{CMS:2019jlq}, this quantification of the fragmentation contributions is quite uncertain. 

\subsection{ATLAS 13 TeV Photon-Plus-Jet Measurement}
\label{subsec:ATLAS_GJ}
\begin{figure}[!t]
\centering
\includegraphics[scale=0.40]{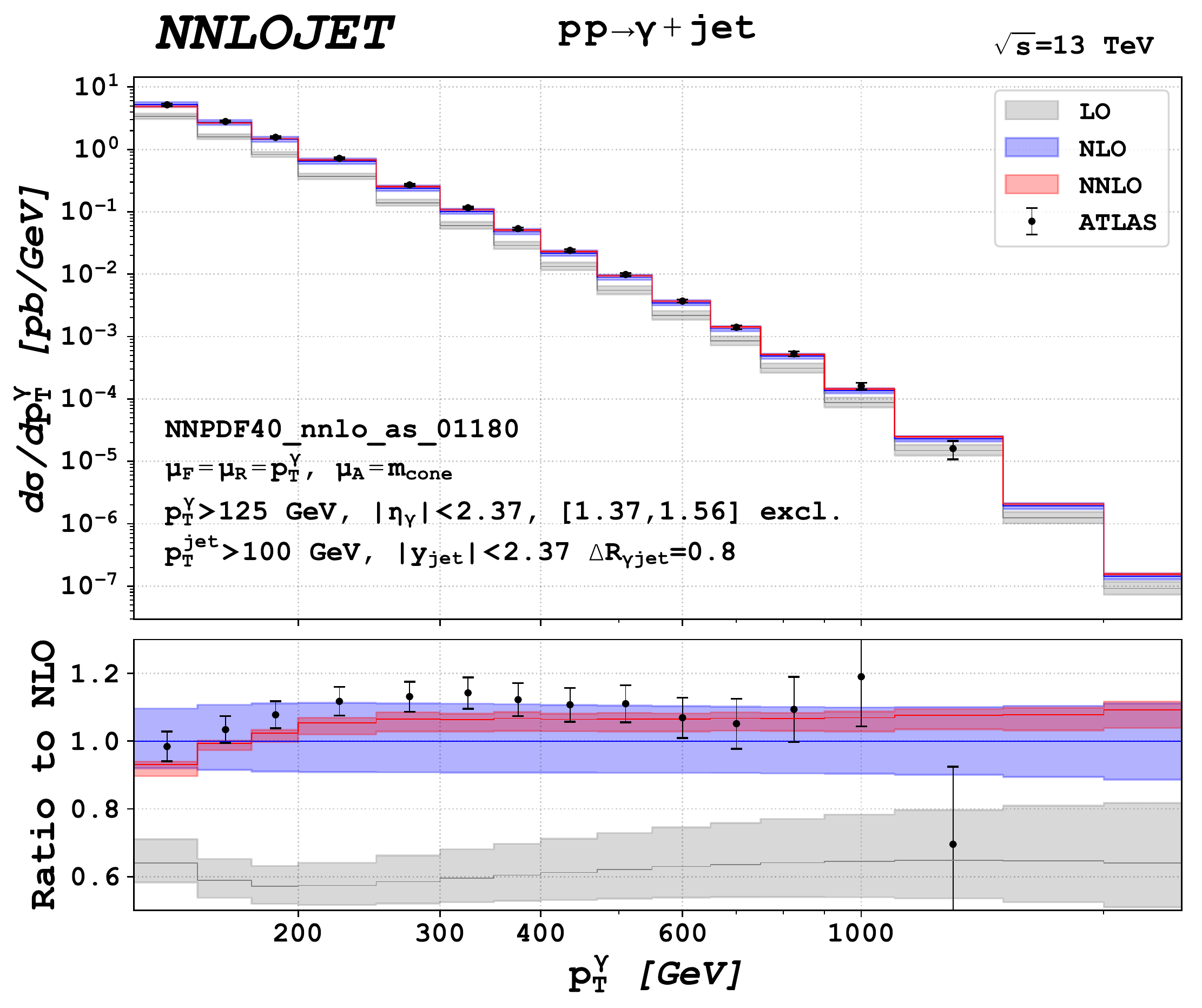}
\caption{Predictions for the photon transverse momentum distributions at LO, NLO, and NNLO. The predictions are compared to data from the ATLAS measurement~\protect\cite{ATLAS:2017xqp}.}
\label{fig:ATLAS_13_GJ_ptgam}
\end{figure}

\begin{figure}[t]
\centering
\includegraphics[scale=0.40]{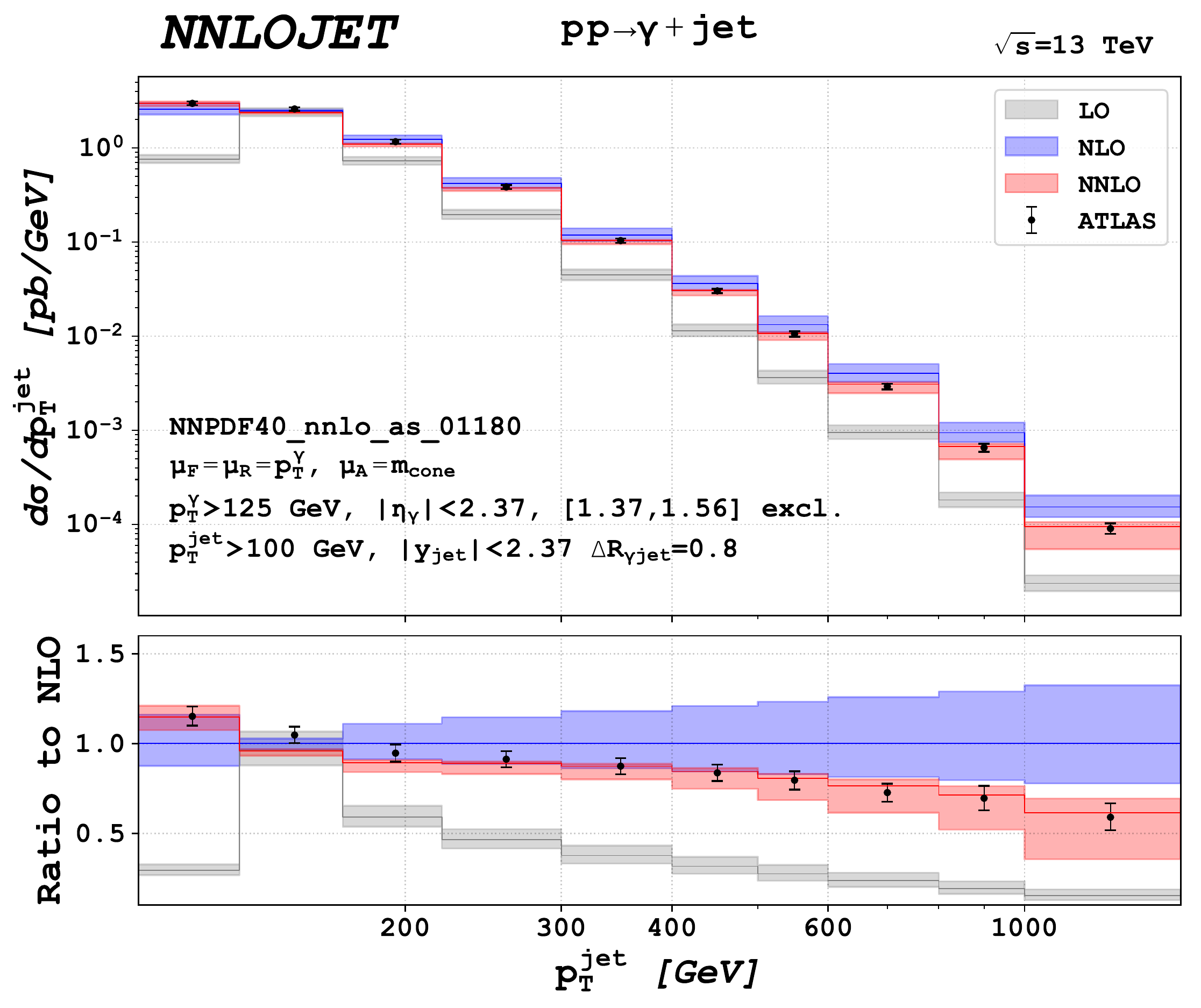}
\caption{Predictions for the leading jet transverse momentum distributions at LO, NLO, and NNLO. The predictions are compared to data from the ATLAS measurement~\protect\cite{ATLAS:2017xqp}.}
\label{fig:ATLAS_13_GJ_ptjet}
\end{figure}

\begin{figure}[!t]
\centering
\includegraphics[scale=0.40]{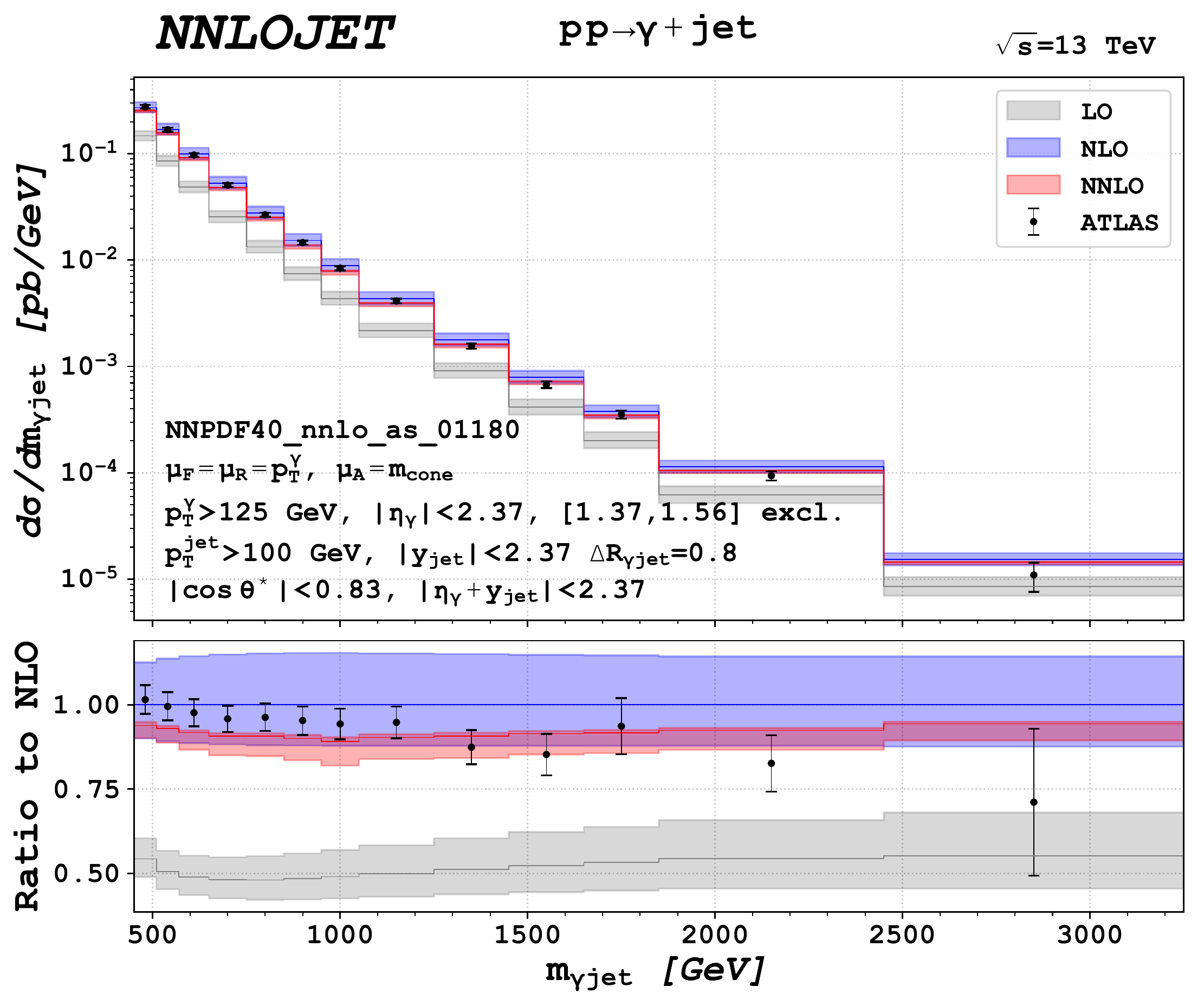}
\caption{Predictions for the photon-jet invariant mass distributions at LO, NLO, and NNLO. The predictions are compared to data from the ATLAS measurement~\protect\cite{ATLAS:2017xqp}.}
\label{fig:ATLAS_13_GJ_mgamjet}
\end{figure}

\begin{figure}[!t]
\centering
\includegraphics[scale=0.40]{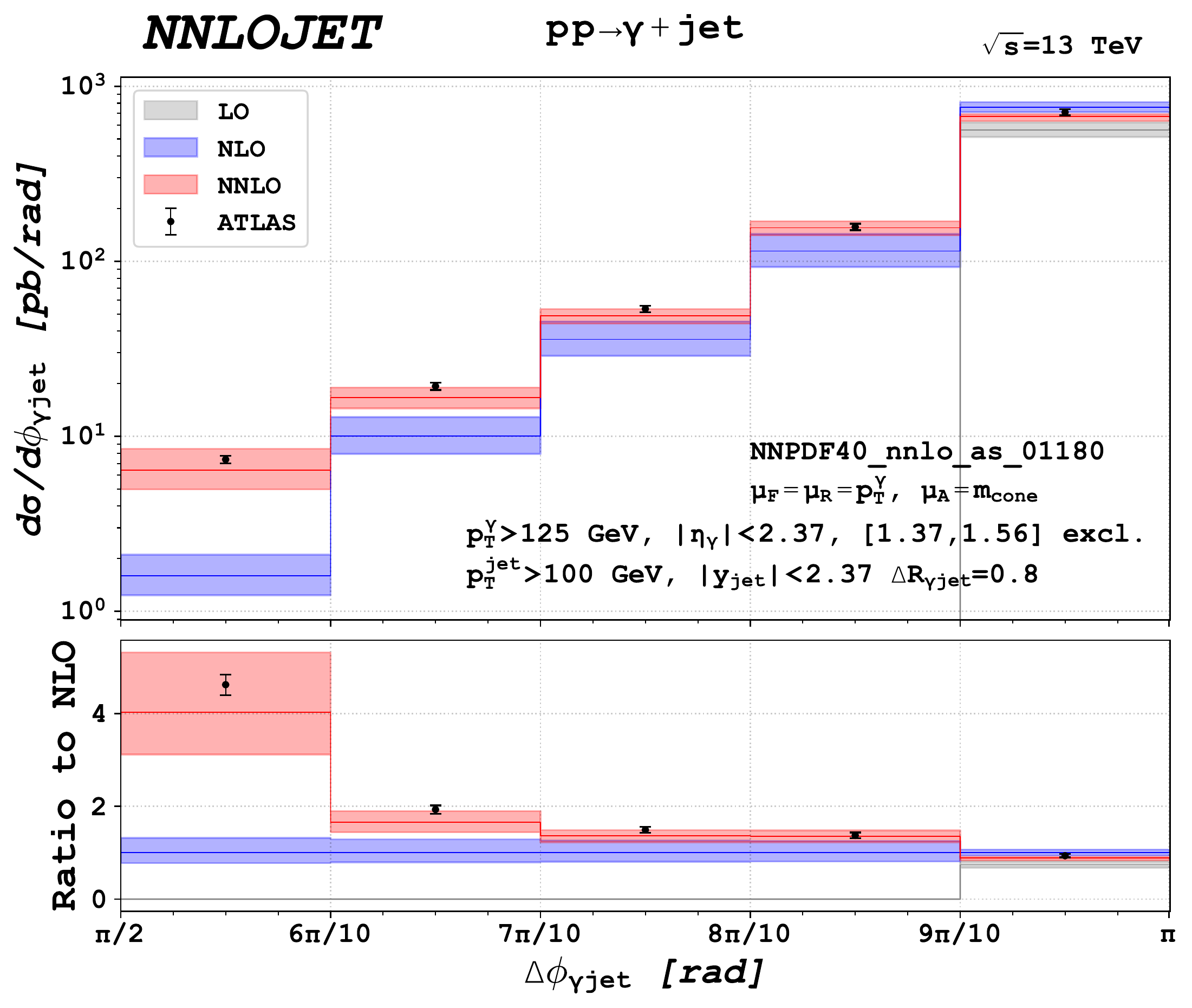}
\caption{Predictions for the photon-plus-jet cross section as a function of the azimuthal angular separation between the photon and the jet at LO, NLO, and NNLO. The predictions are compared to data from the ATLAS measurement~\protect\cite{ATLAS:2017xqp}.}
\label{fig:ATLAS_13_GJ_dphi}
\end{figure}

\begin{figure}[!t]
\centering
\includegraphics[scale=0.40]{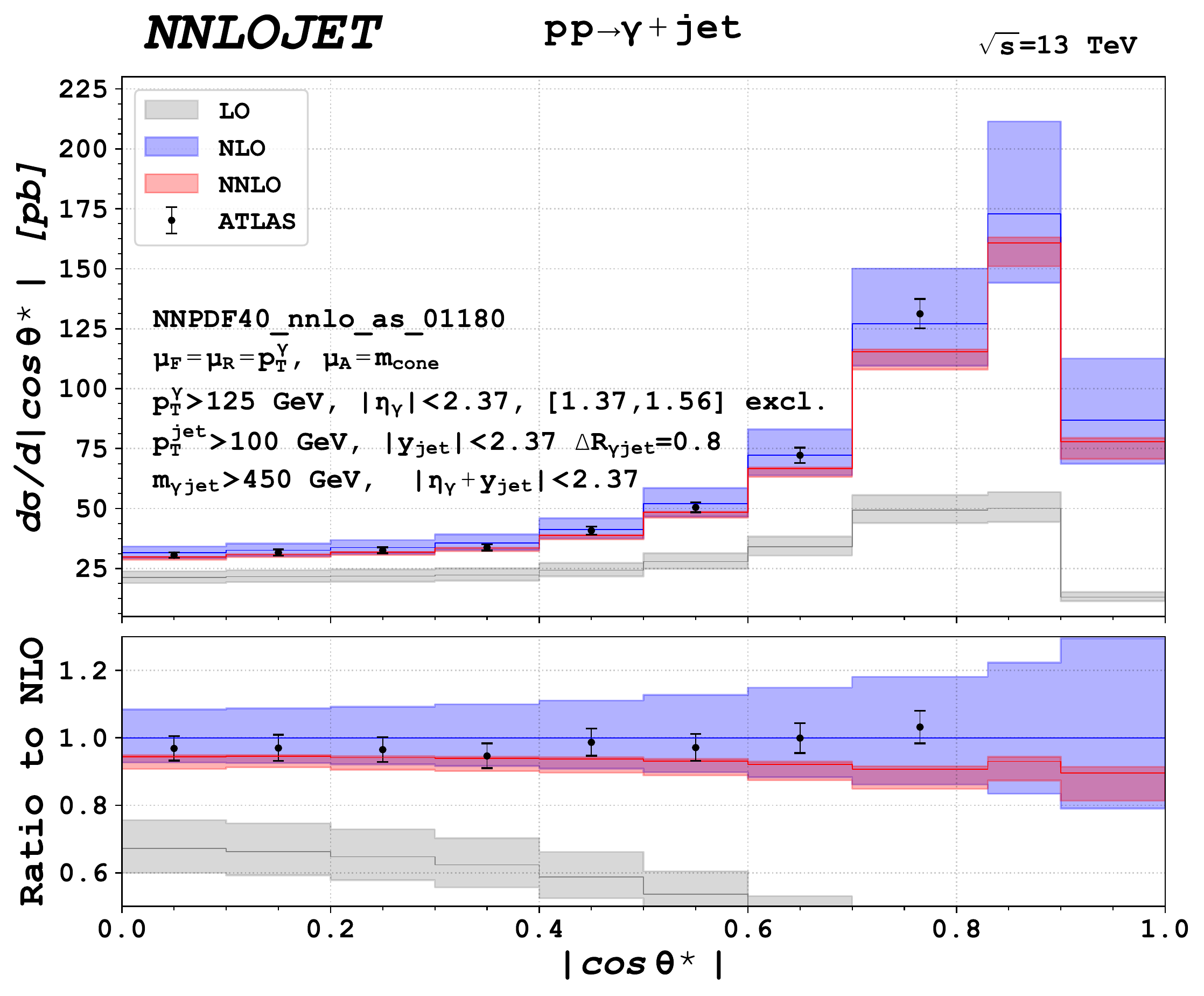}
\caption{Predictions for the photon-plus-jet cross section differential in $|\cos \theta^*|$ at LO, NLO, and NNLO. The predictions are compared to data from the ATLAS measurement~\protect\cite{ATLAS:2017xqp}.}
\label{fig:ATLAS_13_GJ_costheta}
\end{figure}

The ATLAS 13\,\TeV photon-plus-jet measurement~\cite{ATLAS:2017xqp} provides data for the cross section differential in the photon transverse momentum, the leading jet transverse momentum, the azimuthal angular separation between the photon and the jet, the photon--jet invariant mass, and the cosine of the scattering angle in the photon--jet centre-of-mass system $\cos \theta^*$. Jets are identified using the anti-$k_T$ algorithm with a cone radius $R_{{\rm jet}}=0.4$. In the measurement the leading jet fulfilling $|y_{{\rm jet}}| < 2.37$ and $\Delta R_{\gamma{\rm jet}} > 0.8$ is selected and it has to have a transverse momentum $p_T^{{\rm jet}} > 100 \, \GeV$. Photons are required to have a minimum transverse momentum of $p_T^{\gamma} > 125 \, \GeV$ and they have to satisfy $|\eta_{\gamma}| < 2.37$. Moreover, photons in the transition region between the barrel and the endcap calorimeters ($1.37< |\eta_{\gamma}| < 1.56$) are excluded. A photon isolation with parameters
\begin{equation}
R=0.4, \quad \varepsilon = 0.0042,\quad E_T^{\rm thres} = 10 \, \GeV 
\end{equation} 
is imposed. For the invariant mass and the $|\cos \theta^*|$ distribution the additional cuts 
\begin{equation}
| \eta_{\gamma} + y_{\rm jet}| < 2.37, \quad |\cos\theta^*| < 0.83, \quad m_{\gamma {\rm jet}} >450 \, \GeV
\label{eq:ATLAS_hm_cuts}
\end{equation}
are applied. Our NNLO predictions and the comparison to the ATLAS data for the different distributions are shown in Figure~\ref{fig:ATLAS_13_GJ_ptgam}--\ref{fig:ATLAS_13_GJ_costheta}. 

The $p_T^{\gamma}$ distribution is shown in Figure~\ref{fig:ATLAS_13_GJ_ptgam}. Including NNLO corrections strongly impacts the shape of the distribution compared to NLO: In the low $p_T^{\gamma}$ region the cross section is reduced by $-7\%$ while in the medium to large photon transverse momentum regime the NNLO corrections are positive and amount to  approximately $+7\%$. The NLO scale uncertainty of $(17-22)\%$ is reduced to $(+1,-4)\%$ for low and $(+2, -5)\%$ for high photon transverse momenta at NNLO.


Comparing our NNLO predictions to data, we see excellent agreement for the shape of the distributions. In the low and mid $p_T^{\gamma}$ region the data falls slightly above the NNLO prediction. However, theory and data remain consistent within the respective uncertainties in this regime. In the highest bin for which data is available, the NNLO prediction overestimates the data. This tension could indicate the onset of electroweak Sudakov logarithms~\cite{Becher:2012xr,Becher:2013zua}, which become sizeable in this region of phase space. The contribution from fragmentation processes to the full NNLO cross section is 5\% in the low $p_T^{\gamma}$ regime. For larger values of the photon transverse momentum this contribution declines strongly. A more detailed discussion of the different contributions to the NNLO cross section will be provided in Section~\ref{subsec:iso_comp_breakdown} below. 

The leading jet transverse momentum distribution is shown in Figure~\ref{fig:ATLAS_13_GJ_ptjet}. At LO the jet recoils against the photon alone and therefore the transverse momenta of the jet and the photon coincide. Beyond LO this one-to-one correspondence does no longer hold and the phase space region $100 \, \GeV < p_T^{\rm jet} < 125 \, \GeV$, which is not accessible at LO, is populated. This region of phase space is contained in the first bin ($[100,130] \, \GeV$) of the distribution. In this bin the NNLO corrections enhance the NLO cross section by 15\% and reduce the scale uncertainty from 28\% to 12\%. Since the NNLO corrections are effectively only of NLO-type in the bulk of phase space that contributes to this bin, the scale uncertainty remains rather large. For larger $p_T^{\rm jet}$ the NNLO corrections are negative and reduce the cross section by
between  $- 4\%$ in the second bin and $-40\%$ in the very high $p_T^{\rm jet}$ region. Especially for small and intermediate jet transverse momenta the scale uncertainty is strongly reduced by including NNLO corrections. In the second bin the NNLO scale uncertainty is only $(+ 1, -3) \%$ and it increases moderately to $(+2 , -11) \%$ in the region around $p_T^{\rm jet} \approx 500 \, \GeV$. For even larger transverse momenta the scale uncertainty increases significantly. This kinematic regime is dominated by events with two hard recoiling jets and a relatively soft photon. These configurations are effectively described at NLO accuracy resulting in increasing scale uncertainties. 

While at NLO the theory prediction fails to describe the shape of the data, including NNLO corrections a very good agreement with the data in shape and normalisation is observed. It is only in the second bin that the data within its uncertainties does not overlap with the NNLO scale uncertainty band. However, this deviation remains below two standard deviations.

As for the $p_T^{\gamma}$ distribution, the contribution from fragmentation processes to the cross section is small. It amounts to $5 \%$ in the first bin and further decreases for larger transverse momenta. 

Figure~\ref{fig:ATLAS_13_GJ_mgamjet} shows the invariant mass distribution of the photon-jet system. In the measurement of this distribution additional constraints on $|\cos  \theta^* |$ and the sum of the photon pseudorapidity and jet rapidity are imposed. The NNLO correction reduces the cross section by between $-7\%$ and $-11\%$ compared to NLO and shifts the cross section to the lower edge of the NLO uncertainty band over the entire invariant mass spectrum. The impact of the NNLO corrections is most prominent in the region around $m_{\gamma {\rm jet}} \approx 1000 \, \GeV$. As far as scale uncertainties are concerned, a strong improvement going from NLO to NNLO is observed. While the scale uncertainty at NLO is 23\% in the low mass regime and increases up to 27\% in the tail of the distribution, the maximal scale uncertainty at NNLO is $(+1,-8) \%$ for $m_{\gamma {\rm jet}} \approx 1000 \, \GeV$ and further decreases for both smaller and larger invariant masses.

As in the photon transverse momentum spectrum, the NNLO distribution underestimates the data in the low mass region by 6--8\%. In this regime data and theory predictions are not fully consistent. For larger masses we see agreement between theory and data within the increasing experimental errors. We also observe that the shape of the measured distribution is better described at NNLO compared to NLO. The observation that the measured distribution in the low mass region as well as in the low to mid $p_T^{\gamma}$ regime falls systematically above the NNLO predictions could point to an underestimation of the NNPDF4.0 PDF set in this region of phase space.

The angular separation distribution of the leading jet and the photon is shown in Figure~\ref{fig:ATLAS_13_GJ_dphi}. At LO the photon is recoiling against a single parton, forming the jet. In this back-to-back configuration $\Delta \phi = \pi$ holds. Smaller angular separations are accessible only from NLO onwards. The scale uncertainty at NLO in this region increases with smaller $\Delta \phi$ from 44\% up to 55\%. In the back-to-back bin the NLO scale uncertainty is 13\%. Including NNLO corrections reduces the scale uncertainty in this bin  to $(+3,-6) \%$. In the region $ 3/5 \pi < \Delta \phi < 9/10 \pi$ the scale uncertainty at NNLO amounts to $(19 - 28)\%$. The lowest $\Delta \phi$ bin is effectively populated only at NNLO, resulting in a large NNLO-to-NLO K-factor of almost four and a large scale uncertainty of $(+ 33 , -22 ) \%$. NNLO corrections are necessary to describe the data accurately beyond the back-to-back configuration. Over the complete $\Delta \phi$ spectrum data and predictions are consistent within their respective uncertainties.

The contribution from fragmentation processes to the NNLO cross section is largest in the back-to-back bin where it yields 5 \% of the total cross section. Its relative contribution decreases for smaller azimuthal separations.

The ATLAS study also provides data for the cross section as a function of $\cos  \theta^*  \equiv \tanh(\Delta y/2)$, where $\Delta y$ is the rapidity difference of the jet and the photon. $\theta^*$ corresponds to the scattering angle in the center of mass frame of the underlying $2 \to 2$ scattering event. The distribution in $|\cos  \theta^* |$ is shown in Figure~\ref{fig:ATLAS_13_GJ_costheta}. For the measurement of this distribution the additional cuts (\ref{eq:ATLAS_hm_cuts}) on the invariant mass of the photon-jet system and the sum of jet rapidity and photon pseudorapidity are imposed. As for the invariant mass distribution, the NNLO corrections are negative and reduce the NLO cross section by $-6 \%$ for small values of $|\cos \theta^*|$ and by roughly $-10 \%$ for larger values. The scale uncertainty at NLO ranges from 15\% to more than 50\% for $|\cos \theta^*|$ values close to unity. The inclusion of NNLO corrections yields a strong reduction of scale uncertainty. For small values of $|\cos \theta^*|$ the NNLO scale band is $(+1,-4) \%$ and it moderately increases to $(+2,-9) \%$ at large $|\cos \theta^*|$. 

Comparing the NNLO prediction to data, we observe good agreement in most bins, except for the last two for which data are provided. In these two bins the NNLO prediction falls below the data by approximately 10\%. In~\cite{ATLAS:2017xqp} the $|\cos \theta^*|$ distribution was discussed as a candidate for providing sensitivity on the photon fragmentation contribution in particular at high values of $|\cos \theta^*|$. As discussed in more detail in Section~\ref{subsec:iso_comp_breakdown} below, we indeed observe an increase of the fragmentation contribution to the full NNLO cross section from 1.5\% in the first bin to almost 10\% in the region $|\cos \theta^*| \approx 1$. Therefore, the disagreement between theory and data in the last two bins could point to an underestimation of the fragmentation contribution by the BFGII set at large photon momentum fractions.

\section{Comparison of Different Isolation Prescriptions}
\label{sec:iso_comp}

In this section we investigate the impact of different isolation prescriptions on the distributions of the photon-plus-jet cross section. For this comparison we use the set-up of the 13\,\TeV ATLAS study~\cite{ATLAS:2017xqp} and only modify the photon isolation. 

Our comparison includes two calculations with a fixed cone isolation. As the default isolation the ATLAS isolation with parameters 
\begin{equation}
R=0.4, \quad \varepsilon = 0.0042, \quad E_T^{\rm thres} = 10 \, \GeV \qquad \mbox{(default cone)}
\end{equation} 
 is used. To investigate the change in magnitude and in composition (direct vs.\ fragmentation contribution) of the cross section with increasing maximal hadronic energy inside the cone $E_T^{\rm max}$, we consider a loose fixed cone isolation with parameters 
 \begin{equation}
 R=0.4, \quad \varepsilon = 0.0042, \quad E_T^{\rm thres} = 50 \, \GeV \qquad \mbox{(loose cone)}\,.
 \end{equation}

Previous NNLO calculations for isolated photon~\cite{Campbell_2017,Chen_2020} and photon-plus-jet production~\cite{Chen_2020,Campbell:2017dqk} used idealised photon isolations, i.e.\ smooth cone or hybrid cone prescriptions. In the following comparison we confront a hybrid isolation with parameters 
\begin{equation}
R=0.4, \; \varepsilon = 0.0042, \; E_T^{\rm thres} = 10 \, \GeV, \; R_d = 0.1, \; \varepsilon_d = 0.1, \; n=2  \quad \mbox{(hybrid)}
\label{eq:hybrid_comp}
\end{equation}
with the default fixed cone isolation. The parameters of the inner cone are tuned to mimic the ATLAS fixed cone isolation~\cite{Chen_2020}. Therefore, one objective of the following comparison is to reveal to which extent the hybrid isolation faithfully reproduce the experimental isolation prescription at NNLO accuracy.

In the democratic clustering isolation approach~\cite{Glover:1993xc,Hall:2018jub} photons and partons are clustered together. This democratic treatment of photons and partons can be particularly useful for calculations of electroweak corrections were additional photons are emitted. To test the impact of democratic clustering on QCD corrections we apply an anti-$k_T$ algorithm with cone radius 
$R_{{\rm jet}}=0.4$ and a $z_{\rm cut} = 0.926$. The criterion $z_{\rm em} > z_{\rm cut}$ with $z_\mathrm{em}$ denoting the electromagnetic energy fraction~\eqref{eq:z_em} can be reformulated as a criterion on the maximal hadronic energy inside the cone of the democratic jet. 
The maximal integrated hadronic energy inside this cone may not exceed
\begin{equation}
E_T^{\rm max} = \frac{1-z_{\rm cut}}{z_{\rm cut}} \, p_T^{\gamma} \, .
\label{eq:ETmax_demo}
\end{equation}
The value $z_{\rm cut}=0.926$ is chosen such that at $p_T^{\gamma} = 125 \, \GeV$ the maximal hadronic energy is $10 \, \GeV$, which coincides with the value of $E_T^{\rm thres}$ of the ATLAS isolation criterion.

In addition to these isolated photon-plus-jet calculations, we also present results for the photon-plus-jet cross section without imposing any photon isolation. Even if this inclusive set-up may not be feasible in an experimental measurement, it can provide valuable insights on the overall impact of applying an isolation criterion to the photon.

The comparison of the isolation prescriptions is performed in two steps. In Section~\ref{subsec:iso_comp_NNLO} we examine the impact of the isolations on the magnitude of the cross sections. A detailed analysis of the composition of the cross sections in terms of direct and fragmentation components follows in Section~\ref{subsec:iso_comp_breakdown}.

\subsection{Comparison of the NNLO Cross Sections}
\label{subsec:iso_comp_NNLO}

\begin{figure}[!t]
\centering
\begin{subfigure}[b]{0.496\textwidth}
\centering
\includegraphics[width=\textwidth]{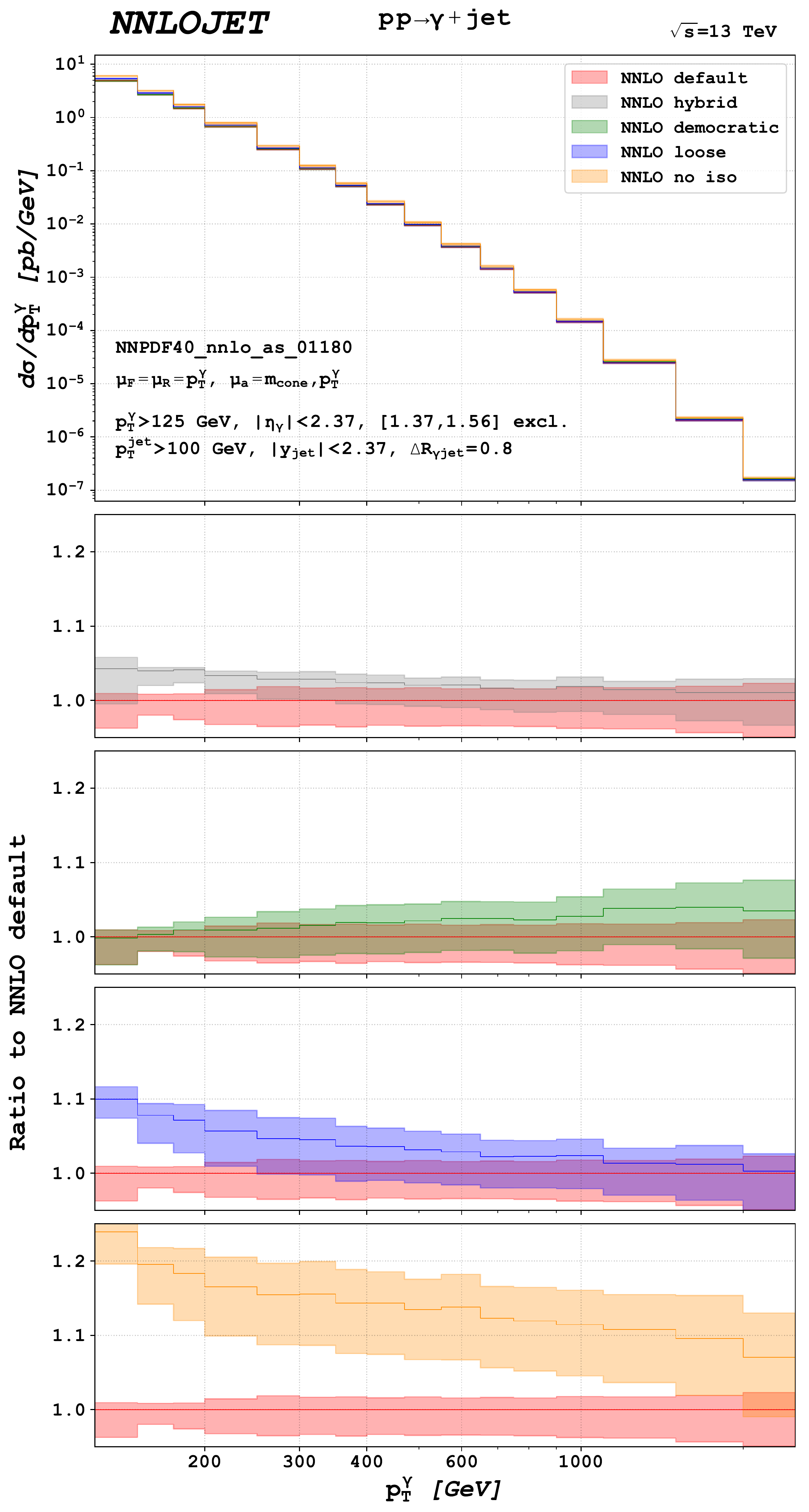}
\end{subfigure}
\hfill
\begin{subfigure}[b]{0.496\textwidth}
\centering
\includegraphics[width=\textwidth]{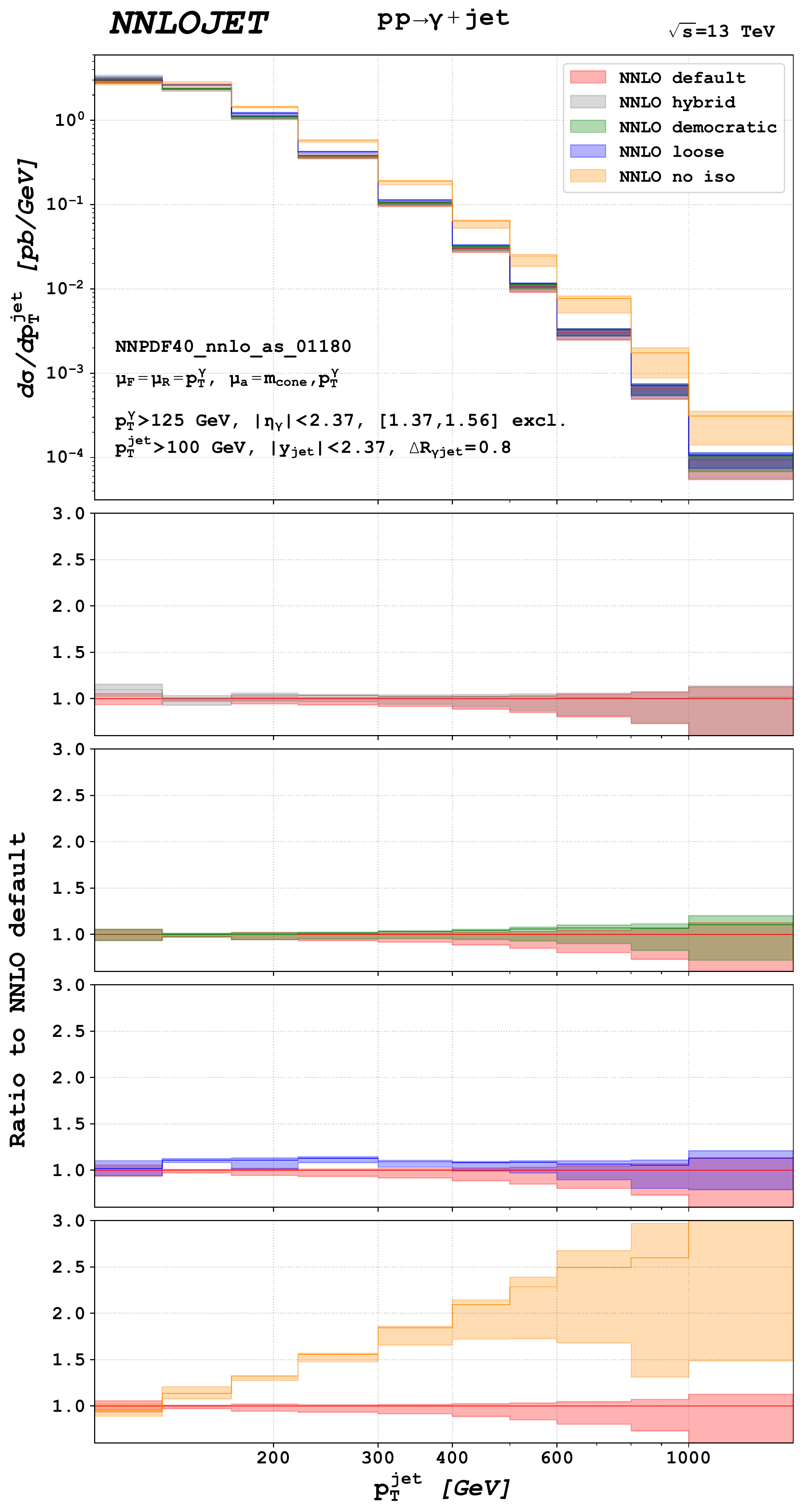}
\end{subfigure}
\caption{Predictions for the photon transverse momentum (left) and jet transverse momentum distributions (right) with different isolation prescriptions.}
\label{fig:comp_ptgam_ptjet}
\end{figure}

For the comparison of the isolation prescriptions we use our default scale choices $\mu_R = \mu_F = p_T^{\gamma}$ and $\mu_A = m_{\rm cone}$. The predictions with hybrid isolation do not contain a contribution from fragmentation processes and therefore they do not depend on a fragmentation scale. In this case the scale uncertainty is estimated by a seven-point scale variation of the renormalisation and factorisation scales. It is noted that for the hybrid isolation an additional uncertainty arises from the choice of the isolation parameters $(R_d, \, \varepsilon_d, \, n)$ in~\eqref{eq:hybrid_comp}. We do not quantify this uncertainty here but stress that it is not captured by the scale uncertainty. For the inclusive predictions without photon isolation the standard choice for the fragmentation scale cannot be applied as there is no isolation cone. For this inclusive set-up we use $\mu_A = p_T^{\gamma}$, since the fragmentation process is in this case inclusive with respect to all extra radiation up to the characteristic scale $p_T^{\gamma}$ of the final state.

The different predictions for the $p_T^{\gamma}$ and $p_T^{\rm jet}$ distributions of the photon-plus-jet cross section are compared in Figure~\ref{fig:comp_ptgam_ptjet}. 
In the lower panels we show the ratio of the individual predictions with respect to the default ATLAS isolation. 

Comparing the photon transverse momentum distributions, we observe a difference between the hybrid and default isolation of roughly 4\% for small $p_T^{\gamma}$. For larger values this difference is reduced and at $p_T^{\gamma} \approx 1000 \, \GeV$ the two predictions largely overlap. A similar behaviour is found for the loose isolation and the inclusive photon-plus-jet cross section. At small transverse momenta a 10\% (24\%) difference compared to the default isolation is observed for the prediction with a loose (without) isolation. With increasing $p_T^{\gamma}$ the distributions approach each other. The loose isolation cross section fully coincides with the default cross section for $p_T^{\gamma} = 2 \, \TeV$, while the inclusive remains 6\% larger. This behaviour indicates that photons at large transverse momenta are likely to be accompanied by only little hadronic energy so that the impact of the photon isolation is only moderate. 

The slope of the democratic clustering distribution is not as steep as the slope of the default distribution. This behaviour can be understood from the functional form of the effective maximal hadronic energy inside the democratic jet given in \eqref{eq:ETmax_demo}. At $p_T^{\gamma} = 125 \, \GeV$ the maximal hadronic energy for democratic clustering is $E_T^{\rm max} =10\,\GeV$ and therefore similar to the corresponding value for the default isolation $E_T^{\rm max} = 10.5\,\GeV$. However, in the democratic clustering approach it increases more strongly with increasing $p_T^{\gamma}$ compared to the default isolation. At $p_T^{\gamma} = 1000 \, \GeV$ the maximal hadronic energy is $E_T^{\rm max} = 80 \, \GeV$ for democratic clustering, while for the default criterion it is $E_T^{\rm max} = 14.2 \, \GeV$. Therefore, in the high $p_T^{\gamma}$ regime the democratic clustering cross section does even exceed the cross section obtained with a loose isolation. An important observation is that the region of largest discrepancy between the isolation prescription coincides with the region where the contribution from fragmentation processes is strongest. This will be further analysed in Section~\ref{subsec:iso_comp_breakdown}, where the composition of the  cross sections with different isolation prescriptions is discussed.

The difference between the hybrid isolation and the default isolation in the  $p_T^{\rm jet}$ distribution is also most prominent at low transverse momenta. In the first bin the hybrid isolation overshoots the default isolation by 10\% while in the second bin it falls below the default isolation by 5\%. For $p_T^{\rm jet} > 170 \, \GeV$ the shapes of the two distributions coincide and the hybrid isolation exceeds the default isolation by only a few percent. The distribution obtained with a loose isolation falls above the default isolation over the entire range of jet transverse momenta. The largest difference between the two isolations is observed in the regime $130 \, \GeV < p_T^{\rm jet} < 300 \, \GeV$ where the corresponding ratio reaches almost 1.13.

In the first bin of the $p_T^{\rm jet}$ distribution the inclusive (no isolation) photon-plus-jet cross section is smaller than the cross section obtained with the default isolation. This peculiar behaviour stems from the Sudakov shoulder~\cite{Catani:1997xc} at $p_T^{\rm jet}=125 \, \GeV$ causing a poor convergence of the perturbative expansion in this regime. Towards larger jet transverse momenta a strong increase in the ratio is observed. As discussed in Section~\ref{subsec:ATLAS_GJ} this regime is dominated by events with two hard recoiling jets accompanied by a relatively soft photon (still fulfilling the $p_T^{\gamma}$ cut). A large contribution to this configuration stems from fragmentation processes, where a high-$p_T$ jet fragments into a photon and passes a moderate momentum fraction of its original momentum to the photon. Such fragmentation processes in this kinematic configuration are almost fully vetoed by all isolation criteria, since the photon is accompanied by a large amount of hadronic energy. The strongly increasing scale uncertainty in the tail of the distribution is caused by the particular scale choice $\mu_R = \mu_F = \mu_A = p_T^{\gamma}$, yielding large logarithms of the ratio $p_T^{\rm jet}/p_T^{\gamma}$ in this regime.

An increasing ratio between the democratic clustering isolation and the default isolation can also be seen in the $p_T^{\rm jet}$ distribution. In the lowest bin the cross section obtained with democratic clustering is 7\% smaller than the default cross section while at high $p_T^{\rm jet}$ the democratic clustering cross section exceeds the default cross section by 10\%.

\begin{figure}[!t]
\centering
\begin{subfigure}[b]{0.496\textwidth}
\centering
\includegraphics[width=\textwidth]{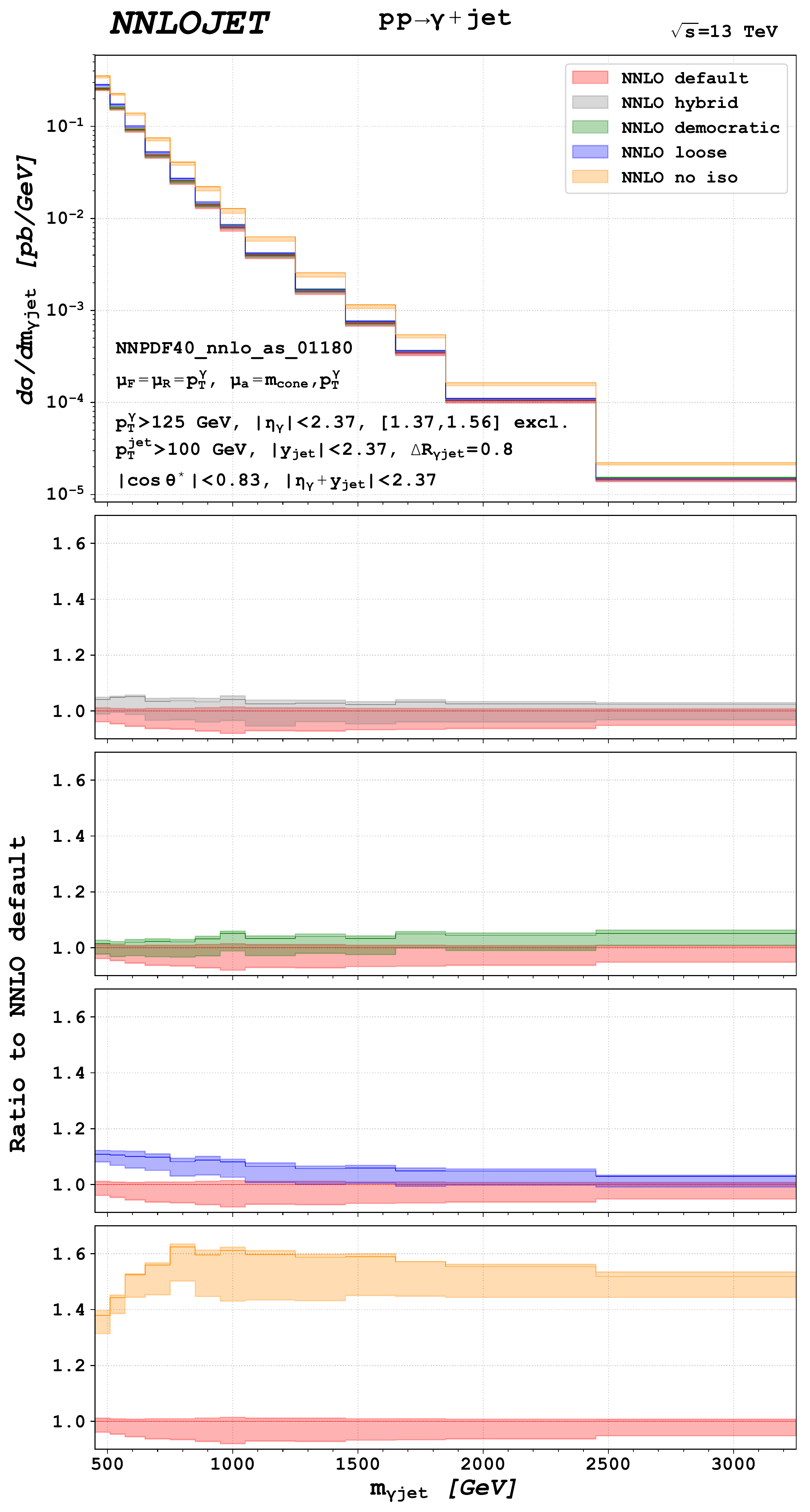}
\end{subfigure}
\hfill
\begin{subfigure}[b]{0.496\textwidth}
\centering
\includegraphics[width=\textwidth]{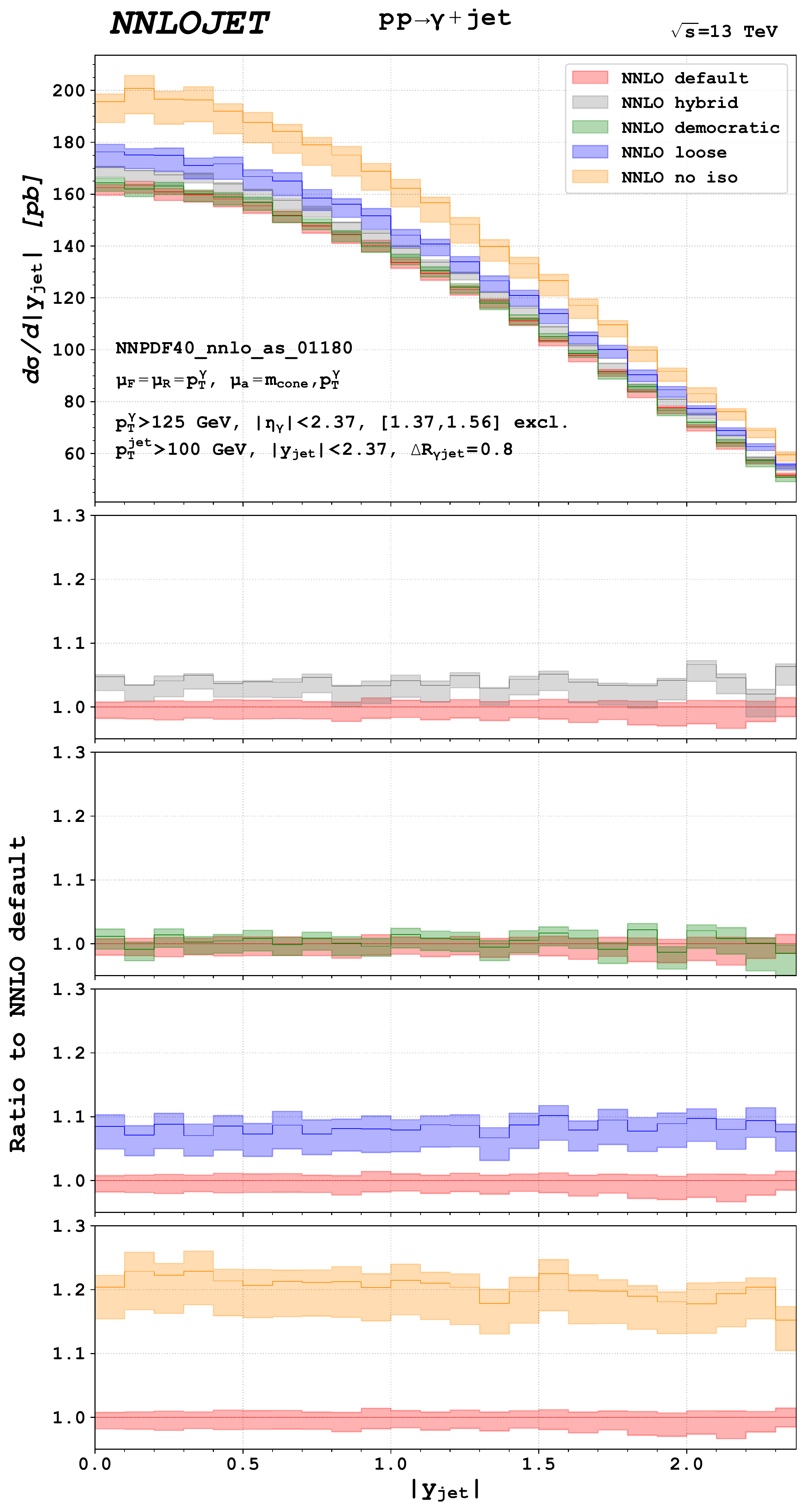}
\end{subfigure}
\caption{Predictions for the photon-jet invariant mass (left) and the leading jet rapidity distribution (right) using different isolation prescriptions.}
\label{fig:comp_mgamj1_yjet}
\end{figure}

Figure~\ref{fig:comp_mgamj1_yjet} shows the invariant mass distribution of the photon--jet system as well as the rapidity distribution of the leading jet. For the invariant mass distribution the additional cuts \eqref{eq:ATLAS_hm_cuts} are imposed. Comparing the invariant mass distribution of the default and the hybrid isolation, we observe the largest difference in the low mass region. At $m_{\gamma {\rm jet}} = 500 \, \GeV$ the hybrid isolation overestimates the ATLAS isolation by 4\%. The difference between the two isolation prescriptions vanishes at very high invariant masses. For $m_{\gamma {\rm jet}} = 3000 \, \GeV$ we observe only a 1\% difference and the scale uncertainty bands of both predictions largely overlap.

\begin{figure}[!t]
\centering
\begin{subfigure}[b]{0.496\textwidth}
\centering
\includegraphics[width=\textwidth]{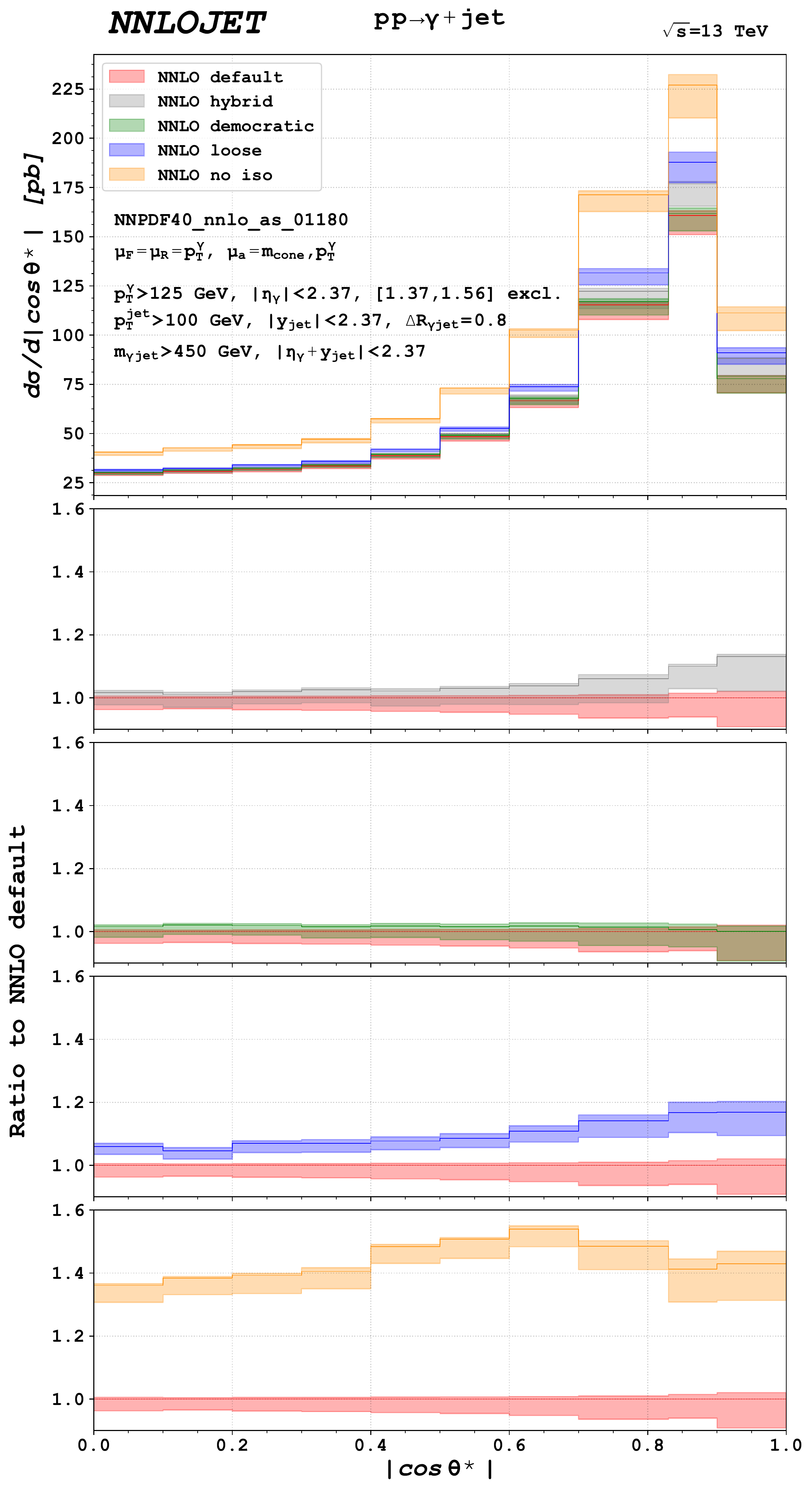}
\end{subfigure}
\hfill
\begin{subfigure}[b]{0.496\textwidth}
\centering
\includegraphics[width=\textwidth]{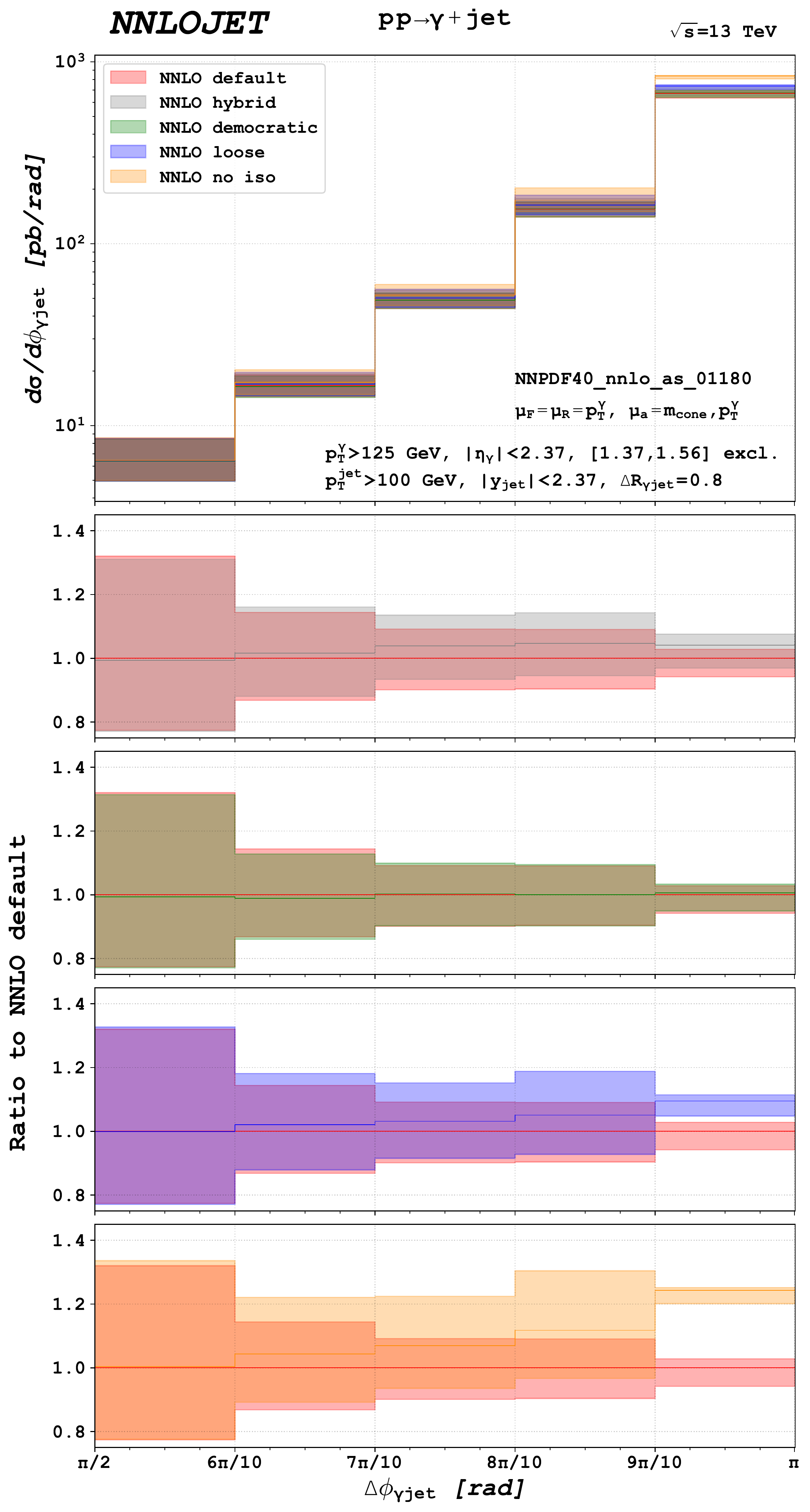}
\end{subfigure}
\caption{Predictions for the photon-plus-jet cross section differential in $|\cos \theta^*|$ (left) and in the azimuthal angular separation between the photon and the jet (right) using different isolation prescriptions.}
\label{fig:comp_costheta_deltaphi}
\end{figure}

Comparing the loose and the default isolation, the largest difference is found in the low mass regime where the loose isolation prediction exceeds the default isolation by 11\%. With increasing masses the corresponding ratio decreases and it amounts to only 1.02 in the highest mass bin.

In the mass regime around $m_{\gamma {\rm jet}} \approx 600 \, \GeV$ a strong increase of the inclusive photon-plus-jet distribution compared to the default distribution is seen. The relative difference of the two distributions reaches its maximum around 1000\,\GeV. For larger masses the corresponding ratio moderately decreases to a value of 1.5.

The distribution obtained with a democratic clustering isolation shows a slightly different slope compared to the default distribution and the corresponding ratio increases from 1.01 to roughly 1.05 in the very high mass regime. This indicates that the large $m_{\gamma {\rm jet}}$ regime is dominated by high-$p_T$ photons for  which the two isolation criteria deviate. 

The shapes of the rapidity distributions of the leading jet coincide for the different isolation prescriptions, yielding constant ratios over the whole rapidity range. This is in line with the expectation that the entire rapidity range is dominated by low-$p_T$ jets events. The corresponding ratio can be inferred from the ratios obtained for the first few bins of the jet transverse momentum distribution. The hybrid isolation systematically falls above the default isolation by about $4\%$, the loose isolation by about $9 \%$ and the inclusive photon-plus-jet cross section by about $20 \%$. The democratic isolation prescription matches the default isolation not only in shape but also in magnitude, as is expected given that the parameters were tuned to match at low $p_T^\gamma$ where the cross section is largest. Both isolations yield very similar distributions. The fragmentation contribution to the different distributions is constant over the considered rapidity range. 

The $|\cos \theta^*|$ distribution and the $\Delta \phi$ distribution are shown in Figure~\ref{fig:comp_costheta_deltaphi}. $\theta^*$ corresponds to the polar scattering angle of the underlying $2 \to 2$ scattering event. Due to a different functional dependence on $\cos \theta^*$ of the underlying matrix elements, the contribution from fragmentation is most prominent in the high $|\cos \theta^*|$ regime (see also Section~\ref{subsec:iso_comp_breakdown}). The largest difference between the hybrid and the default isolation is observed in this regime, where it amounts to almost 13\%. The loose isolation distribution also deviates strongest from the default isolation for large $|\cos \theta^*|$. The corresponding ratio increases from 1.06 at small $|\cos \theta^*|$ to 1.17 at $|\cos \theta^*|=1$.  In contrast to the loose and hybrid isolation, the democratic clustering distribution agrees with the default isolation distribution over the full range of $|\cos \theta^*|$ within the respective scale uncertainty bands. This indicates that the impact of events with very high-$p_T$ photons to the $|\cos \theta^*|$ distribution is small. The inclusive photon-plus-jet cross section exceeds all isolated photon-plus-jet cross sections. The largest difference between the inclusive distribution and the default distribution is observed at $|\cos \theta^*| = 0.65$ where it amounts to roughly 55\%. The corresponding ratio decreases for larger and smaller values of $|\cos \theta^*|$.

The different $\Delta \phi$ distributions almost fully coincide for $\pi/2 < \Delta \phi < 3 \pi/5$. In this regime the photon is recoiling against at least two additional partons. Therefore, it is likely to be well separated from additional hadronic energy and the isolation prescription does not strongly impact the cross section. This is in line with the observation that fragmentation processes yield only a negligible contribution to the cross section in this kinematic range. The difference between the different isolations increases with increasing $\Delta \phi$. The largest discrepancy is observed in the back-to-back bin. In this bin the inclusive distribution is 25\% larger than the default distribution and the loose isolation distribution overshoots the default isolation by 9\%. The distribution obtained with democratic clustering matches the default distribution over the whole range of $\Delta \phi$, which shows that the impact of photons with large transverse momenta to this distribution is very small. The hybrid isolation overestimates the default isolation for $7/10 \pi < \Delta \phi < \pi$ by approximately 4\%. Since the fragmentation contribution to the cross section increases with $\Delta \phi$, this confirms observations in other distributions that the difference between the hybrid and default isolation is largest in regions where also the fragmentation contribution is largest.

A particularly remarkable feature of all distributions is the close agreement between democratic isolation and default cone-based isolation, which is observed everywhere except 
for the high-$p_T^{\gamma}$ tail of the photon transverse momentum distribution. This agreement is most likely a consequence of the fact that the anti-$k_T$ 
algorithm~\cite{Cacciari:2008gp} 
that is used for the clustering in the democratic isolation approach results in an almost perfectly cone-shaped capture area of radius $R_{\rm jet}$~\cite{Cacciari:2008gn} (which is chosen identical to the isolation cone radius 
$R$ here), and that the $z_{{\rm cut}}$ applied subsequently on the photon jet is determined to reproduce $E_T^{\rm max}$ of the default isolation 
 at the lower end of the  $p_T^{\gamma}$ distribution, where the cross section is largest. With parameters chosen appropriately, the 
  effect of the democratic isolation procedure based on the anti-$k_T$ algorithm  is thus very similar to a fixed cone isolation.

\subsection{Decomposition of the Cross Sections}
\label{subsec:iso_comp_breakdown}

\begin{figure}[!t]
\centering
\begin{subfigure}[b]{0.496\textwidth}
\centering
\includegraphics[width=\textwidth]{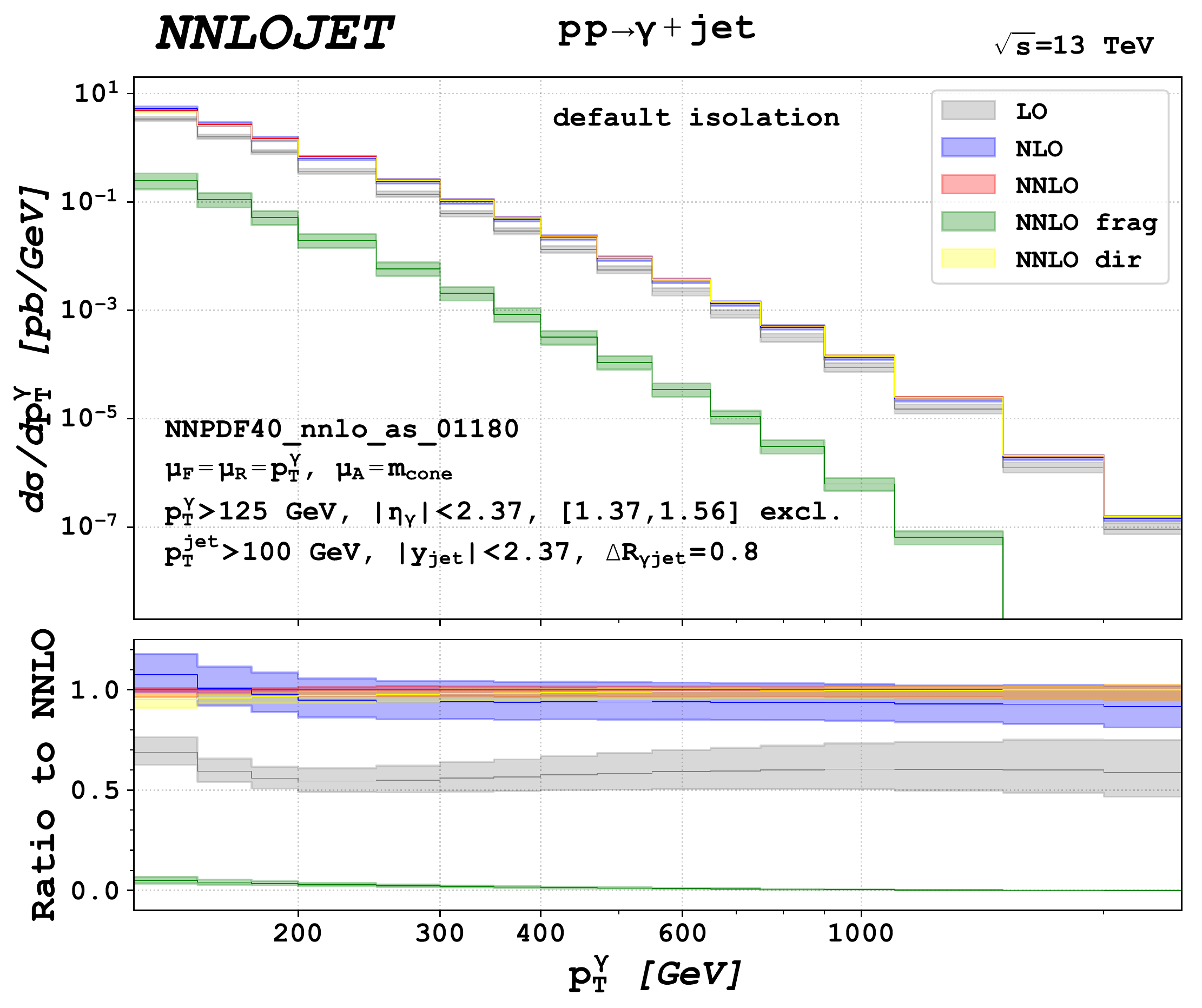}
\end{subfigure}
\hfill
\begin{subfigure}[b]{0.496\textwidth}
\centering
\includegraphics[width=\textwidth]{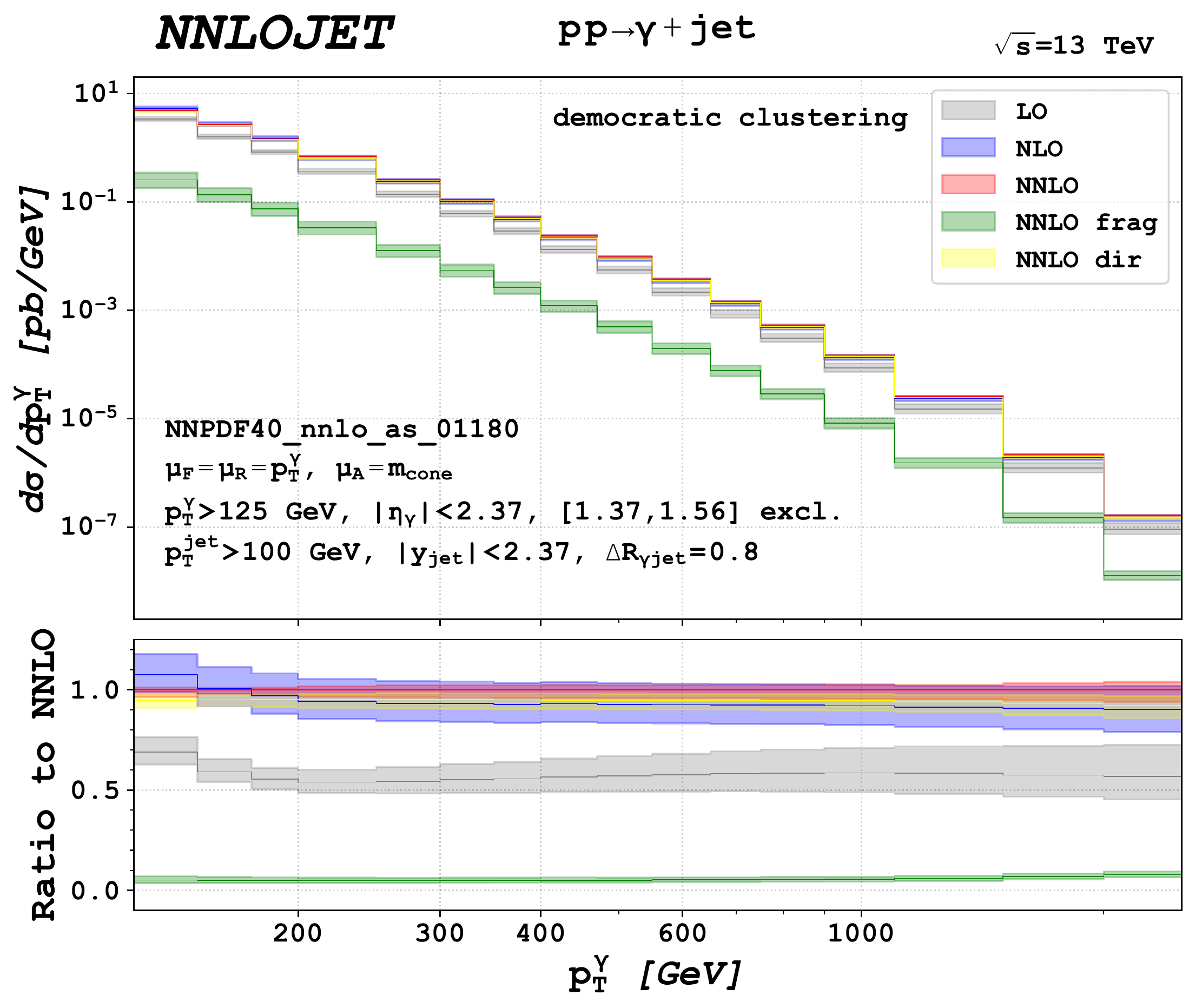}
\end{subfigure}
\vskip\baselineskip
\begin{subfigure}[b]{0.496\textwidth}   
\centering 
\includegraphics[width=\textwidth]{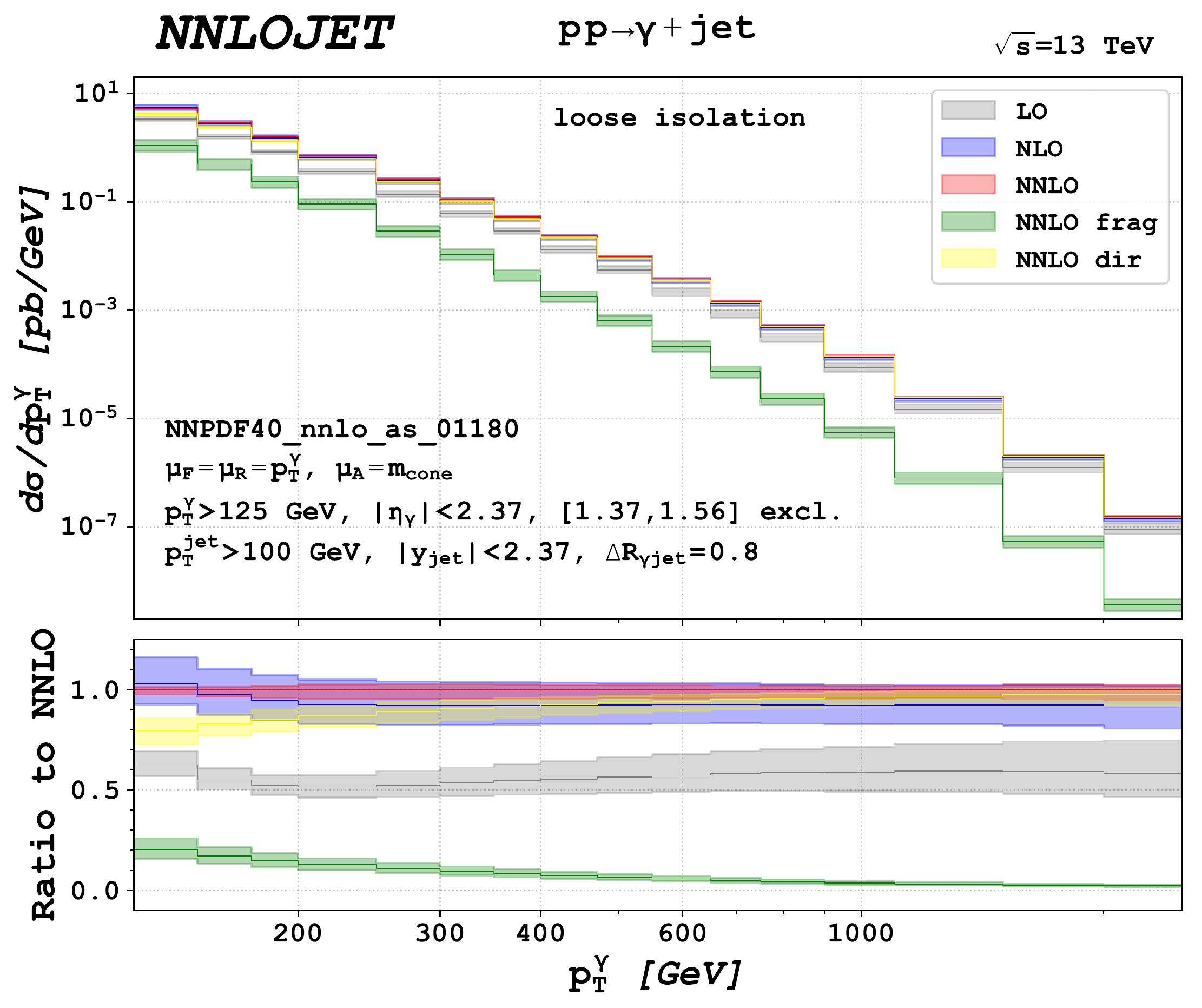}
\end{subfigure}
\hfill
\begin{subfigure}[b]{0.496\textwidth}   
\centering 
\includegraphics[width=\textwidth]{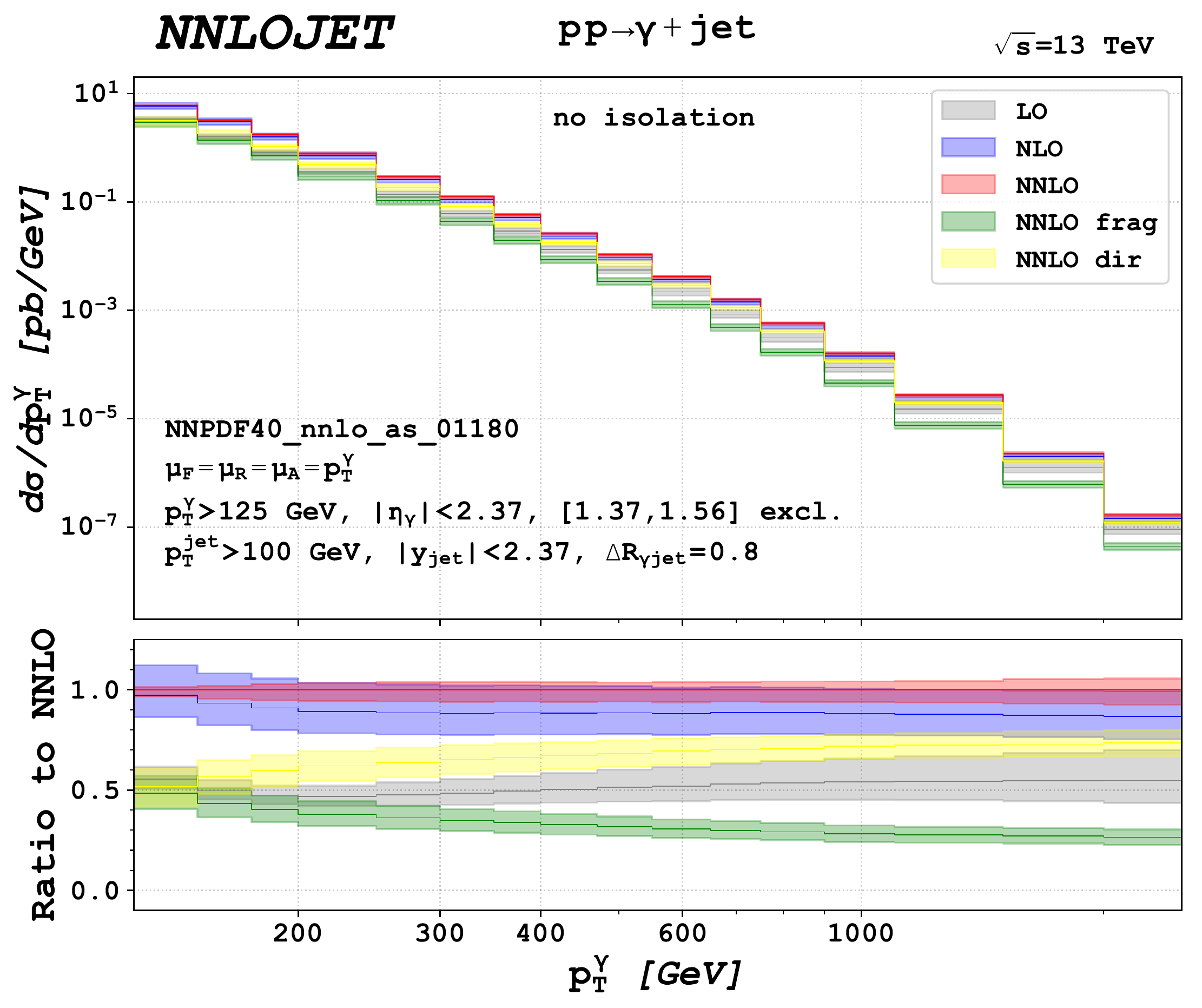}
\end{subfigure}
\caption{Decomposition of the cross section differential in the photon transverse momentum. Results obtained with the default isolation, a loose isolation, democratic clustering and without isolation are shown.}
\label{fig:bd_pt_gam}
\end{figure}

In this section we investigate the impact of the isolation prescription on the composition of the photon-plus-jet cross section. As the hybrid isolation only contains a contribution from direct photon production processes, this isolation is not considered in the following. 

In Figure~\ref{fig:bd_pt_gam} the photon transverse momentum distributions for the default isolation, the loose isolation, democratic clustering isolation, and the inclusive set-up are shown. Besides the decomposition of the full NNLO cross section (red) into the direct (yellow) and fragmentation contribution (green)
according to (\ref{eq:sigma}), the LO and NLO predictions are displayed. In the LO approximation the photon is in a back-to-back configuration with one additional parton. Therefore, it is well separated from any hadronic energy and the isolation prescription does not affect the cross section at this order. This does no longer hold from NLO onward, where additional real radiation processes as well as fragmentation processes start to contribute.

\begin{figure}[!t]
\centering
\begin{subfigure}[b]{0.496\textwidth}
\centering
\includegraphics[width=\textwidth]{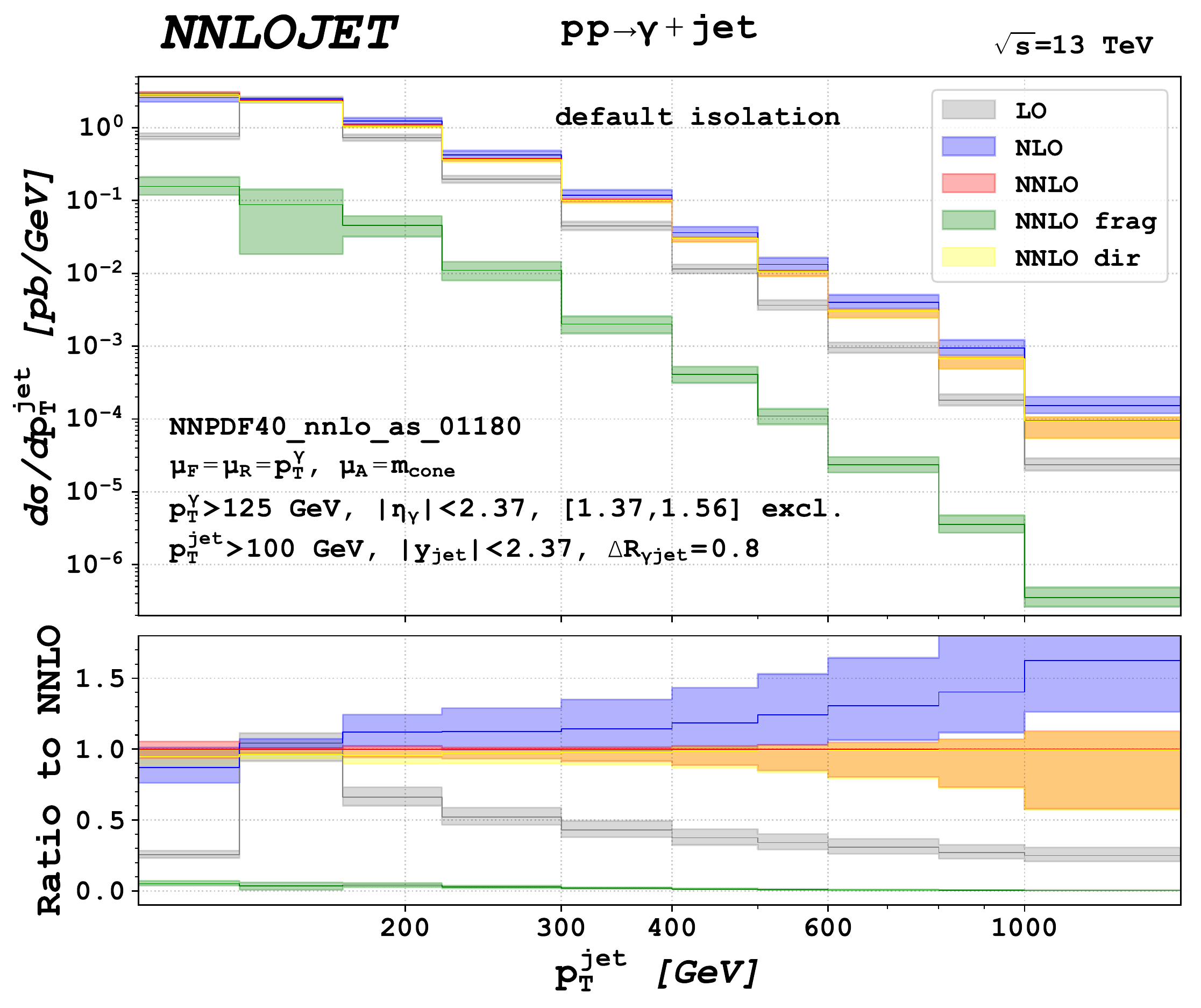}
\end{subfigure}
\hfill
\begin{subfigure}[b]{0.496\textwidth}
\centering
\includegraphics[width=\textwidth]{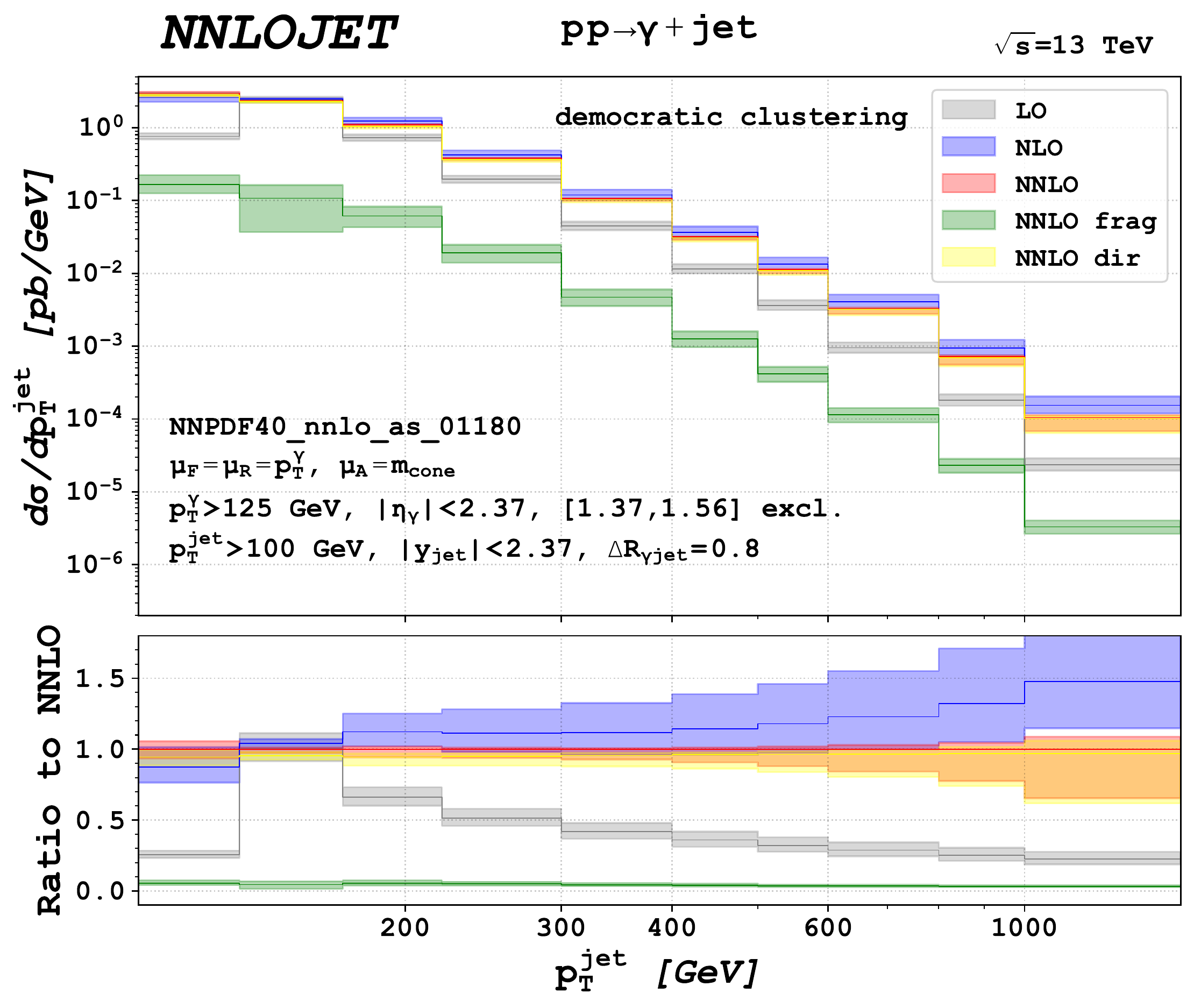}
\end{subfigure}
\vskip\baselineskip
\begin{subfigure}[b]{0.496\textwidth}   
\centering 
\includegraphics[width=\textwidth]{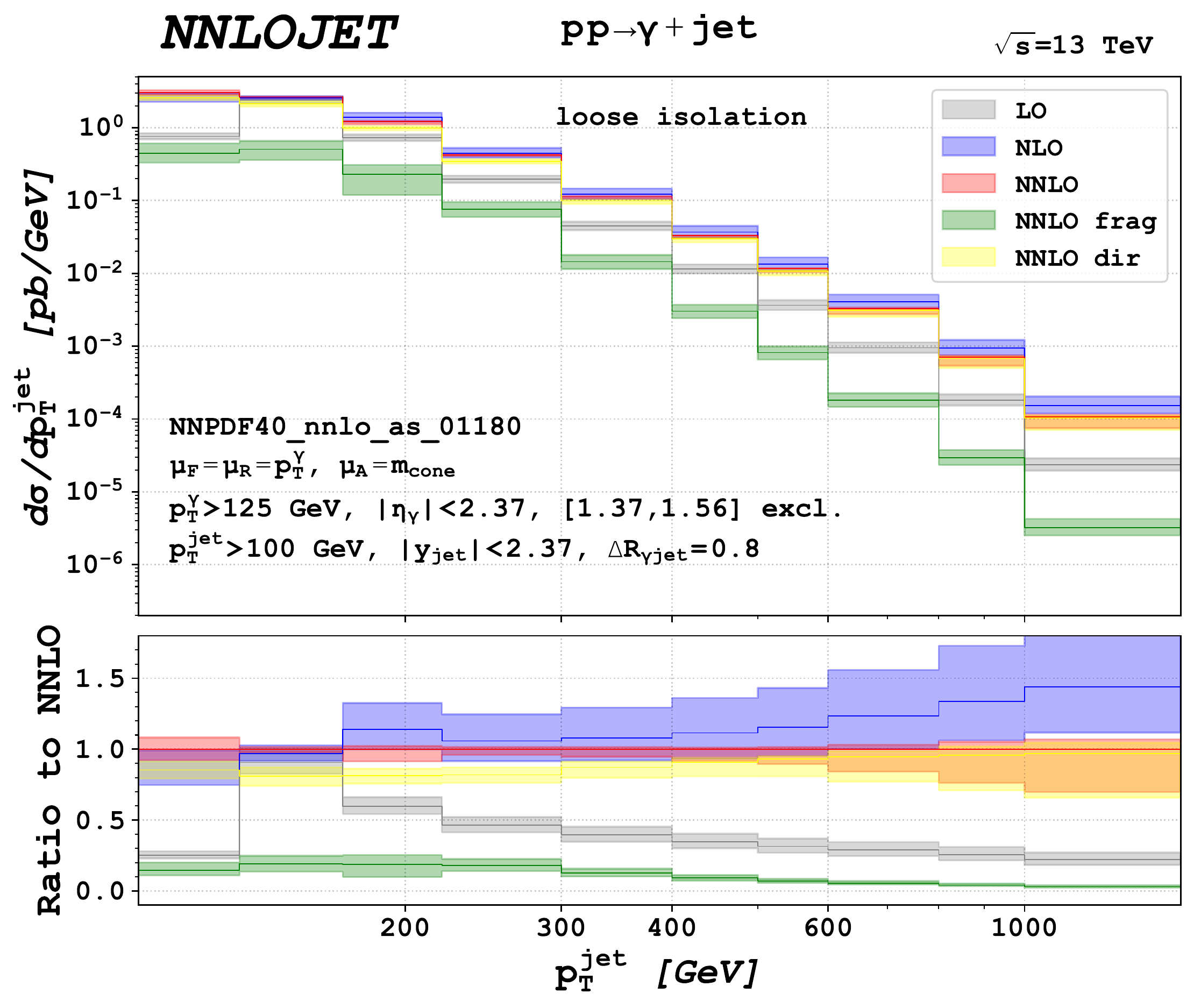}
\end{subfigure}
\hfill
\begin{subfigure}[b]{0.496\textwidth}   
\centering 
\includegraphics[width=\textwidth]{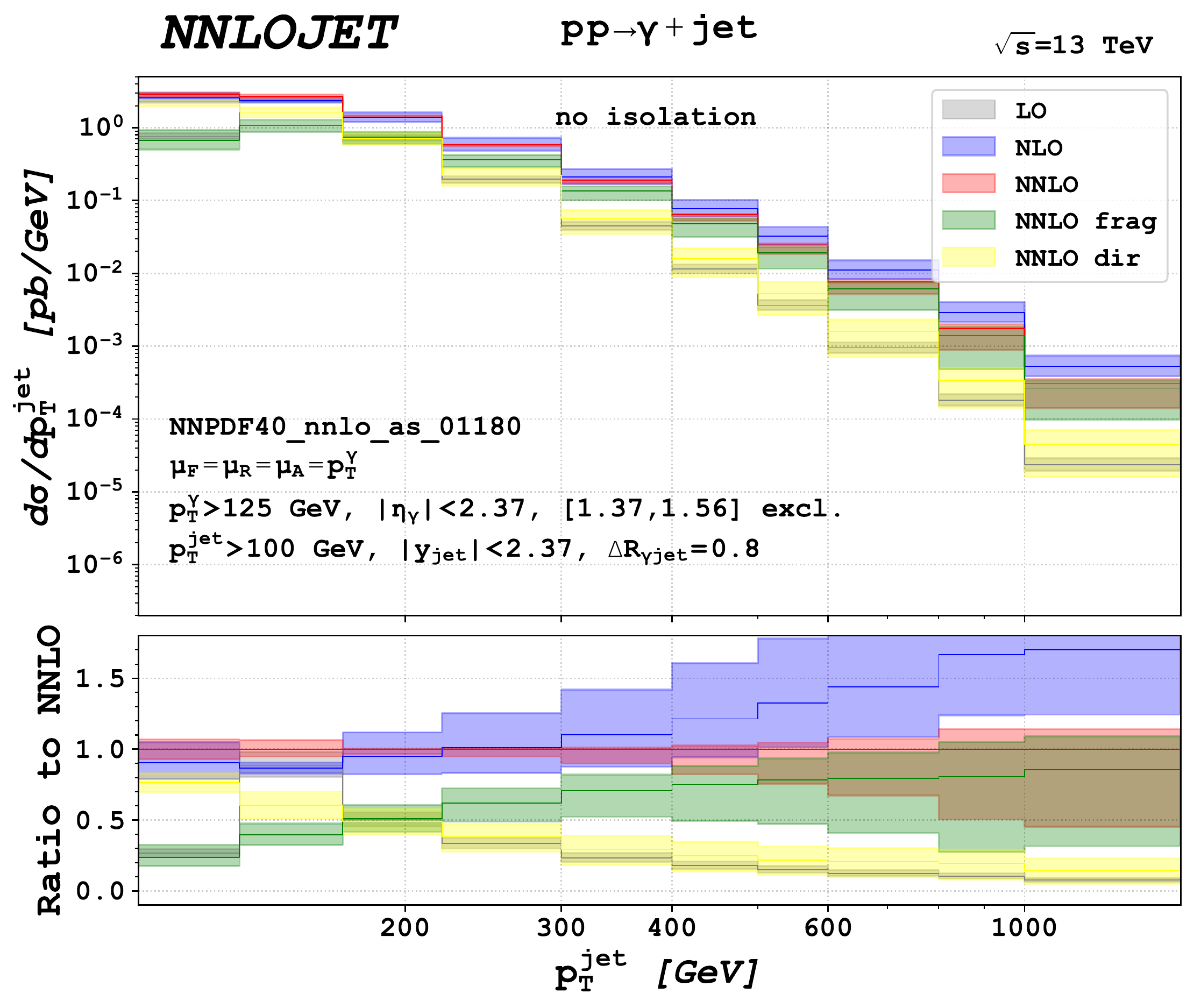}
\end{subfigure}
\caption{Decomposition of the cross section differential in the jet  transverse momentum. Results obtained with the default isolation, a loose isolation, democratic clustering and without isolation are shown.}
\label{fig:bd_pt_j1}
\end{figure}

For the default isolation, the loose isolation and the inclusive set-up the contribution from fragmentation processes to the full NNLO cross section declines with increasing $p_T^{\gamma}$. Moreover, the looser the isolation parameters are the larger is the fragmentation contribution to the photon transverse momentum distribution. For the default isolation the impact of the fragmentation contribution to the full NNLO cross section is 5\% in the lowest bin and beyond $p_T^{\gamma} = 600 \, \GeV$ it decreases to less than 1\%. 

With a loose photon isolation a higher sensitivity on the fragmentation processes is obtained. At small photon transverse momenta 20\% of the full NNLO cross section stems from these processes. Without photon isolation, direct photons and fragmentation each yield roughly 50\% of the cross section at small $p_T^{\gamma}$ and in the high $p_T^{\gamma}$ regime the fragmentation contribution still amounts to 25\% of the cross section. For the democratic clustering isolation the fragmentation contribution moderately increases with increasing $p_T^{\gamma}$ from 5\% to 8\%. This is in line with the observation in Figure~\ref{fig:comp_ptgam_ptjet} where the strongest discrepancy to the default isolation was found in the high $p_T^{\gamma}$ regime.

\begin{figure}[!t]
\centering
\begin{subfigure}[b]{0.496\textwidth}
\centering
\includegraphics[width=\textwidth]{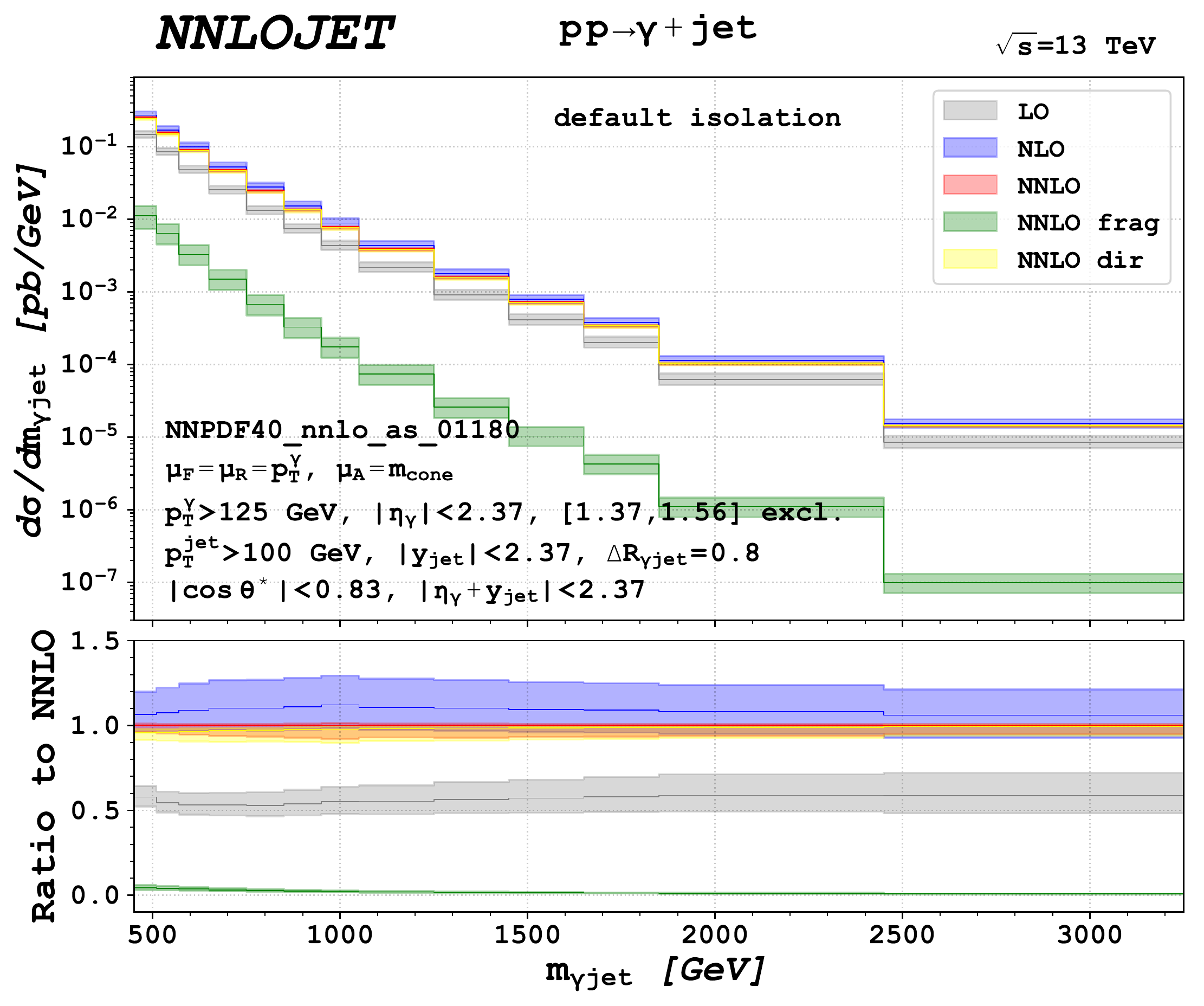}
\end{subfigure}
\hfill
\begin{subfigure}[b]{0.496\textwidth}
\centering
\includegraphics[width=\textwidth]{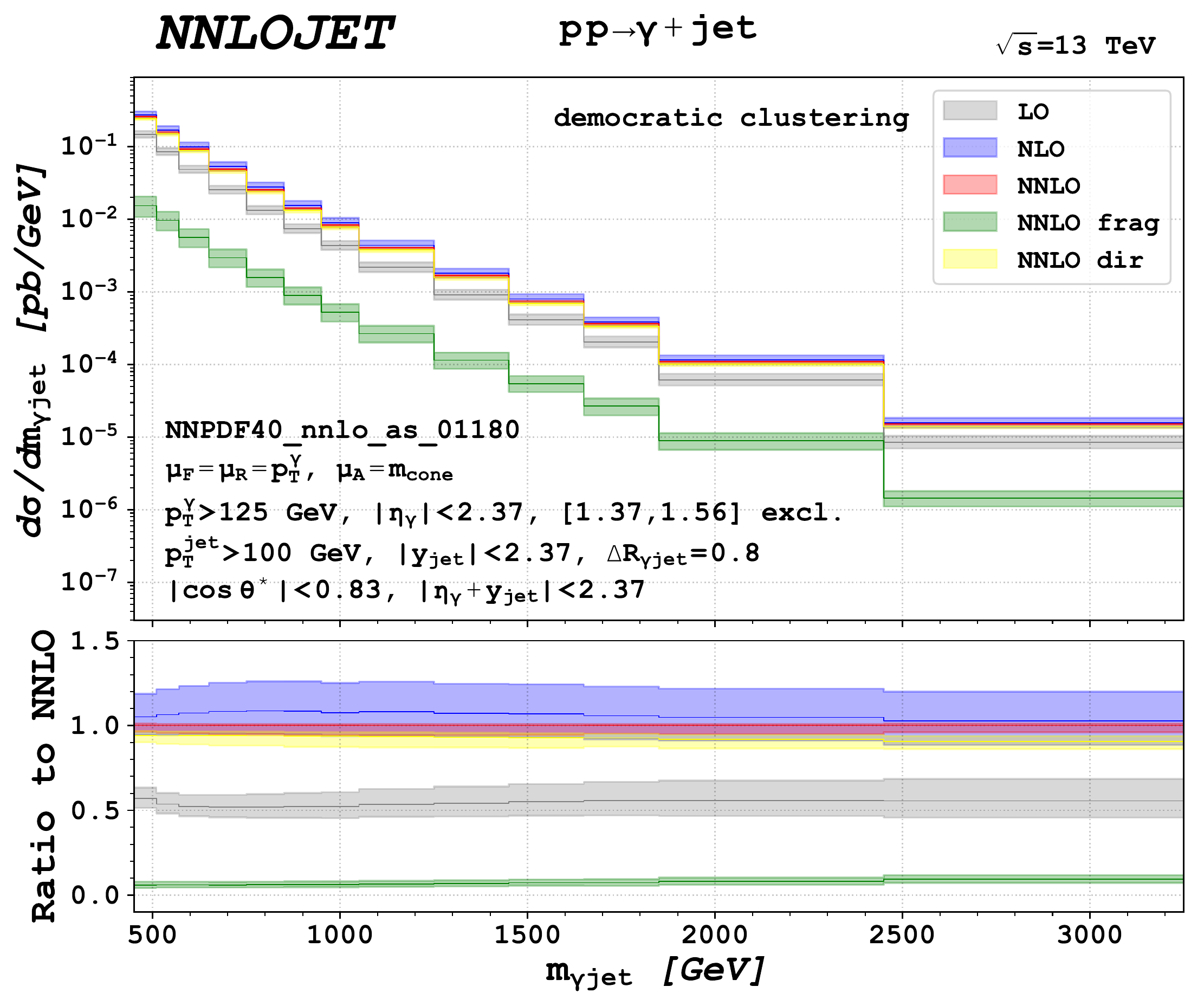}
\end{subfigure}
\vskip\baselineskip
\begin{subfigure}[b]{0.496\textwidth}   
\centering 
\includegraphics[width=\textwidth]{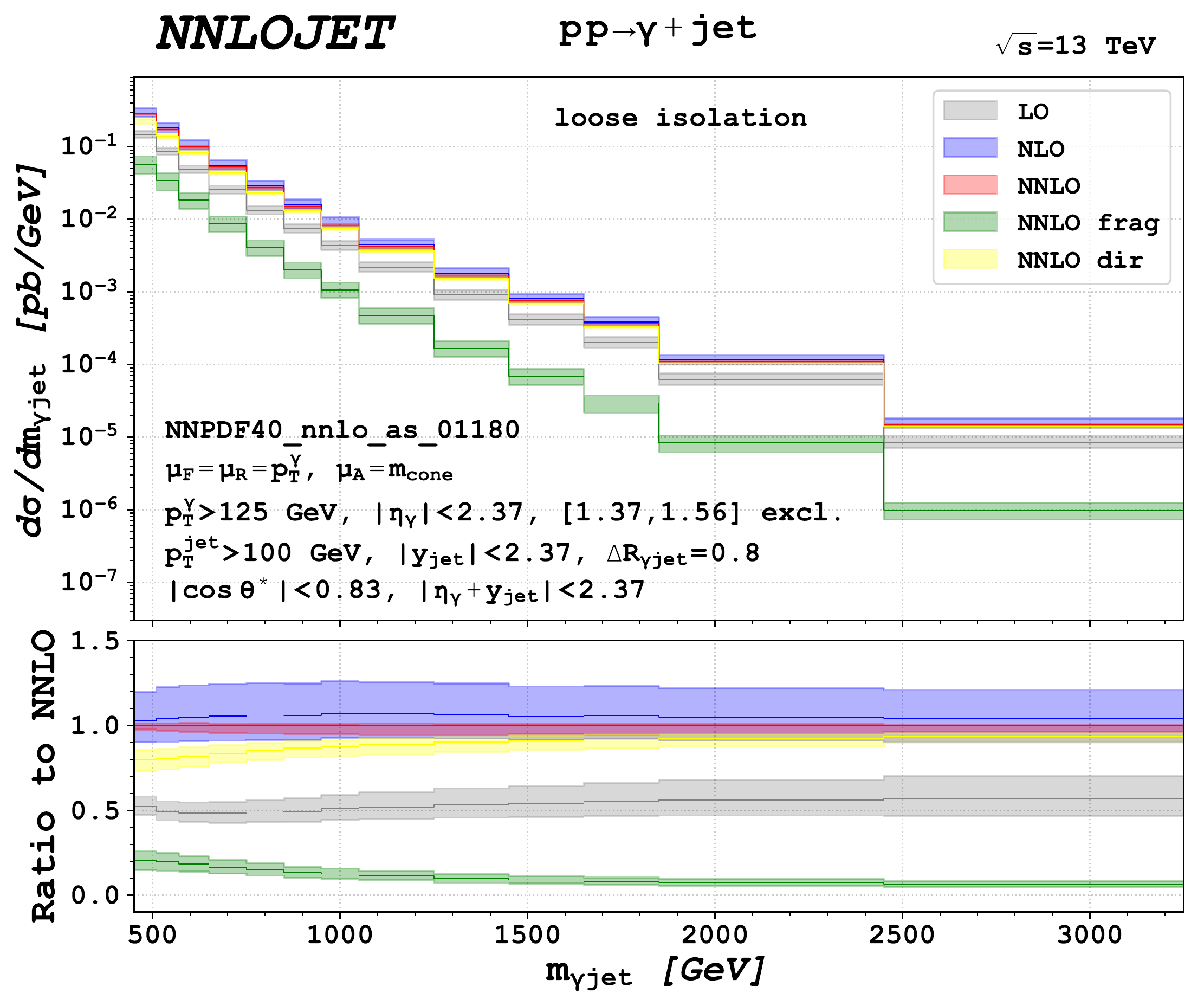}
\end{subfigure}
\hfill
\begin{subfigure}[b]{0.496\textwidth}   
\centering 
\includegraphics[width=\textwidth]{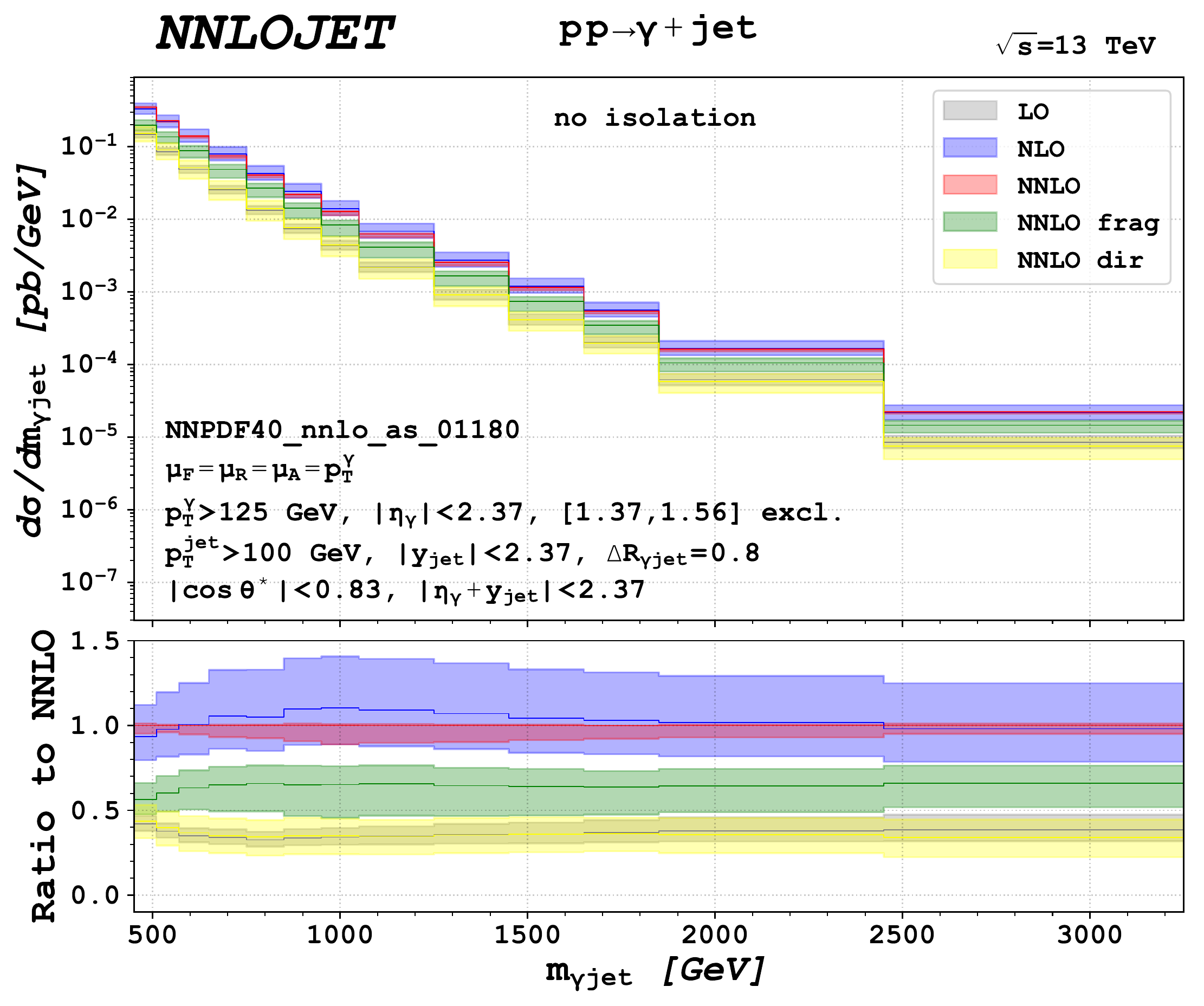}
\end{subfigure}
\caption{Decomposition of the cross section differential in the photon-jet invariant mass. Results obtained with the default isolation, a loose isolation, democratic clustering and without isolation are shown.}
\label{fig:bd_m_gam_j1}
\end{figure}

Figure~\ref{fig:bd_pt_j1} shows the decomposition of the leading jet transverse momentum distribution for the different isolations. Comparing the default and loose isolation, an increase in the sensitivity on fragmentation processes with looser isolation parameters is observed. The predictions obtained with the default isolation contain a relative fragmentation contribution of $5 \%$ for low $p_T^{\rm jet}$ whereas for the loose isolation the fragmentation contribution is $ 20\%$ in this regime. As for the $p_T^{\gamma}$ distribution, the impact from fragmentation processes decreases with increasing transverse momenta.

In the democratic clustering approach the relative contributions from direct photons and photon fragmentation remain almost constant over the complete kinematic range. For this isolation roughly 4\% of the photons originate from fragmentation processes.

For the inclusive set-up the high $p_T^{\rm jet}$ region is completely dominated by fragmentation processes. Here the direct contribution only amounts to 15\% of the NNLO cross section. This strong increase of fragmentation processes causes the large difference between the inclusive and isolated photon cross sections displayed in Figure~\ref{fig:comp_ptgam_ptjet}. At large jet transverse momenta, events with two hard recoiling jets accompanied by a relatively soft photon yield the largest contribution to the cross section. In this configuration the allowed range of momentum fractions which one of the jet passes to the photon during fragmentation increases with increasing leading jet transverse momentum. The accessible momentum fraction range for the fragmentation of a 1\,\TeV jet is $z \in [0.125,1]$, while for a jet with 150\,GeV transverse momentum it is only $z \in [0.83,1]$. The fragmentation functions are strongly increasing towards lower momentum fractions. Therefore, in the majority of events in the tail of the inclusive $p_T^{\rm jet}$ distribution the photon is accompanied by a large amount of hadronic energy.

The breakdown of the invariant mass distribution of the photon--jet system is shown in Figure~\ref{fig:bd_m_gam_j1}. The composition of the cross section as a function of the invariant mass is very similar to the composition of the $p_T^{\gamma}$ distributions, reflecting the functional dependence of $m_{\gamma {\rm jet}}$ on $p_T^{\gamma}$. For the fixed cone isolations, the contribution from fragmentation is largest for small masses where it amounts to 20\% for the loose isolation and 4\% for the default isolation. Towards larger masses the fragmentation contribution declines strongly. The invariant mass distribution obtained with democratic clustering contains an increasing contribution from fragmentation processes for larger masses. For $m_{\gamma {\rm jet}} \approx 500 \, \GeV$ the contribution from fragmentation to the NNLO cross section is 6\% and it increases to 9\% at $m_{\gamma {\rm jet}} \approx 3 \, \TeV$. 

The fragmentation contribution dominates the non-isolated NNLO cross section over the entire range of masses. Its relative size increases from 56\% for low masses to 66\% in the tail of the distribution. The invariant mass distribution also shows that the scale uncertainties of the fragmentation and the direct contribution largely compensate each other yielding only a very small scale band of the full NNLO cross section.

Comparing Figure~\ref{fig:bd_m_gam_j1} to the invariant mass distribution in Figure~\ref{fig:comp_mgamj1_yjet}, we observe a similar correlation between the sensitivity on fragmentation processes and the sensitivity on the isolation prescription, i.e.\ the regions in which the isolation prescriptions deviate are the regions with the largest fragmentation contribution.

\begin{figure}[!t]
\centering
\begin{subfigure}[b]{0.496\textwidth}
\centering
\includegraphics[width=\textwidth]{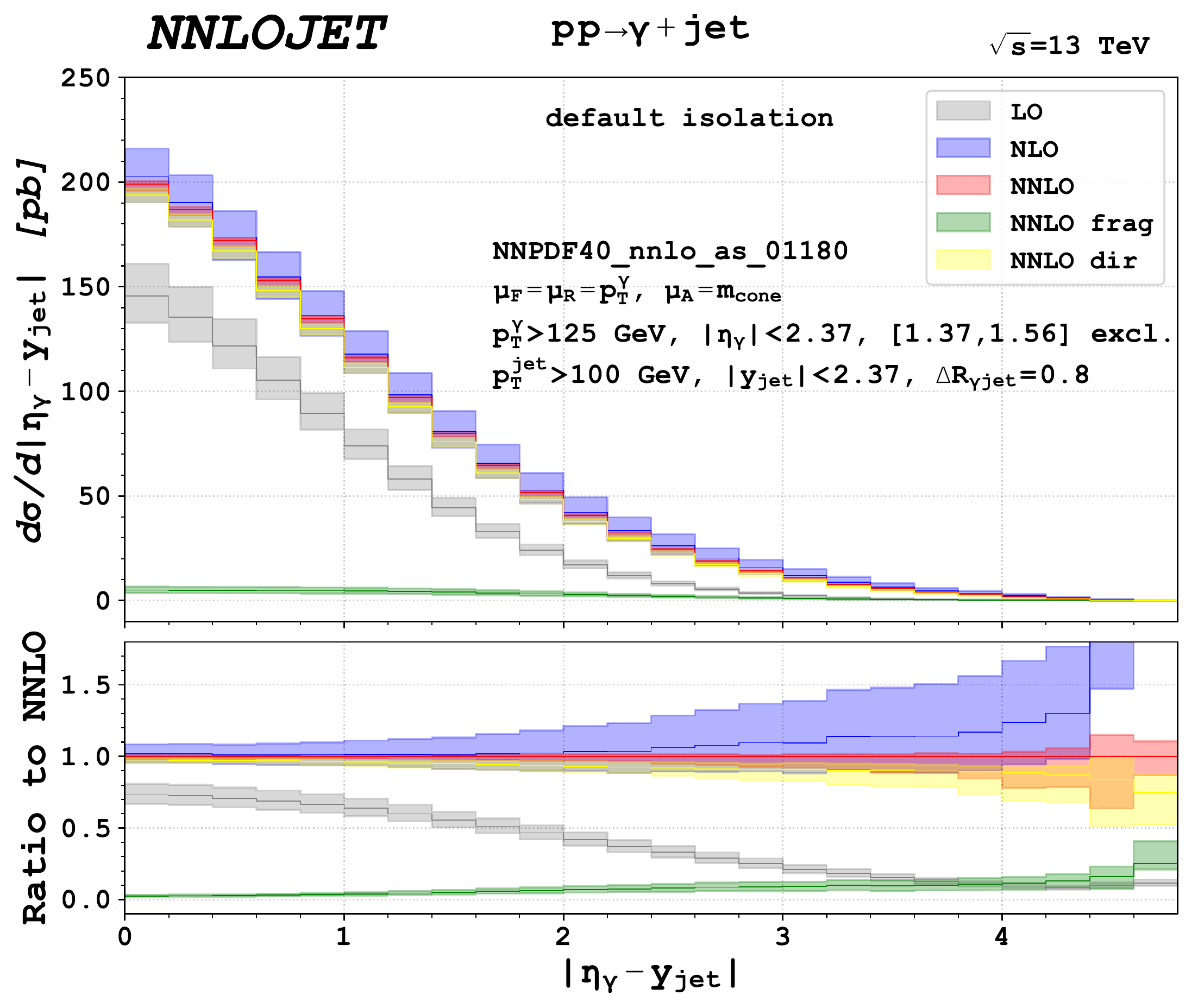}
\end{subfigure}
\hfill
\begin{subfigure}[b]{0.496\textwidth}
\centering
\includegraphics[width=\textwidth]{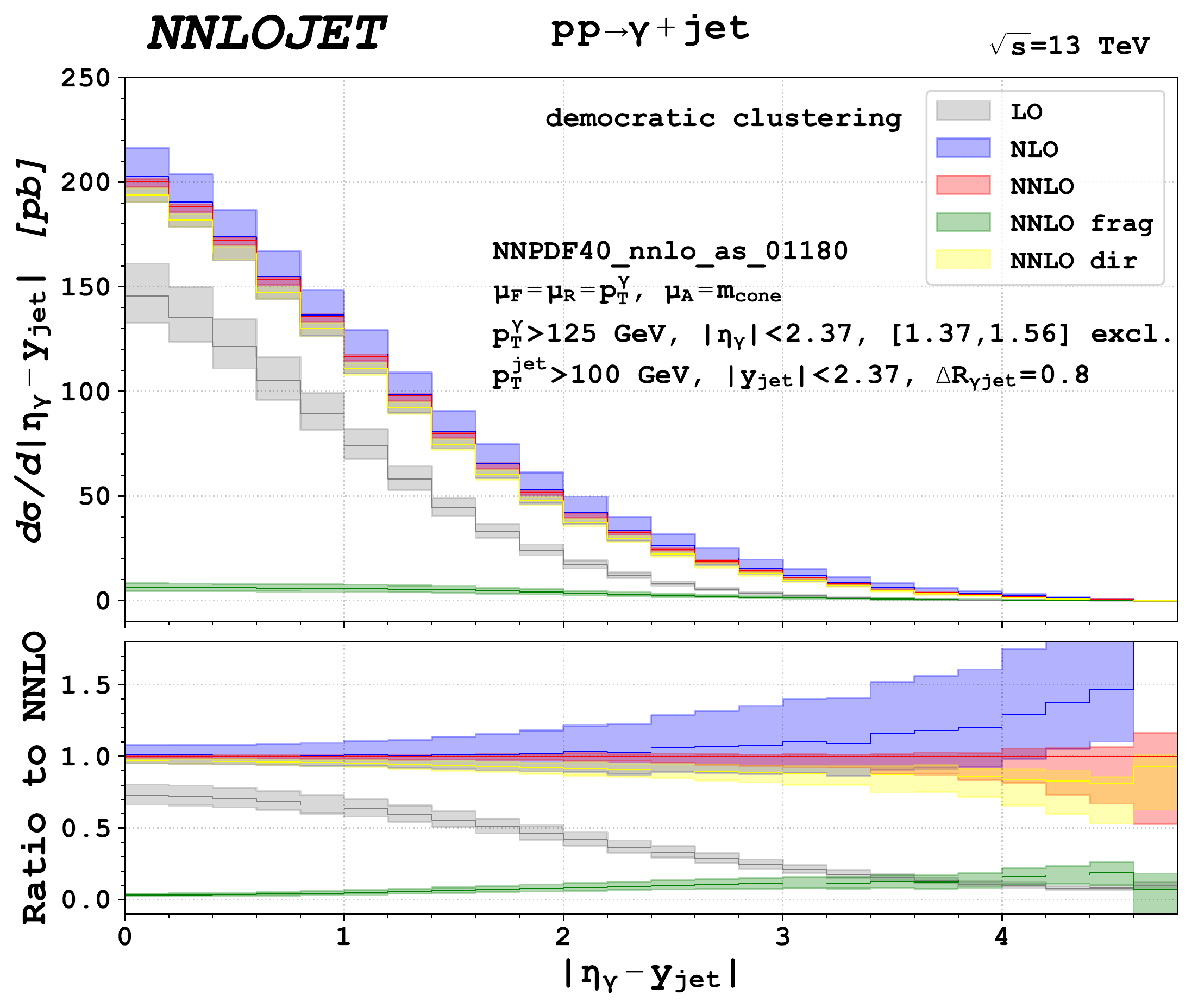}
\end{subfigure}
\vskip\baselineskip
\begin{subfigure}[b]{0.496\textwidth}   
\centering 
\includegraphics[width=\textwidth]{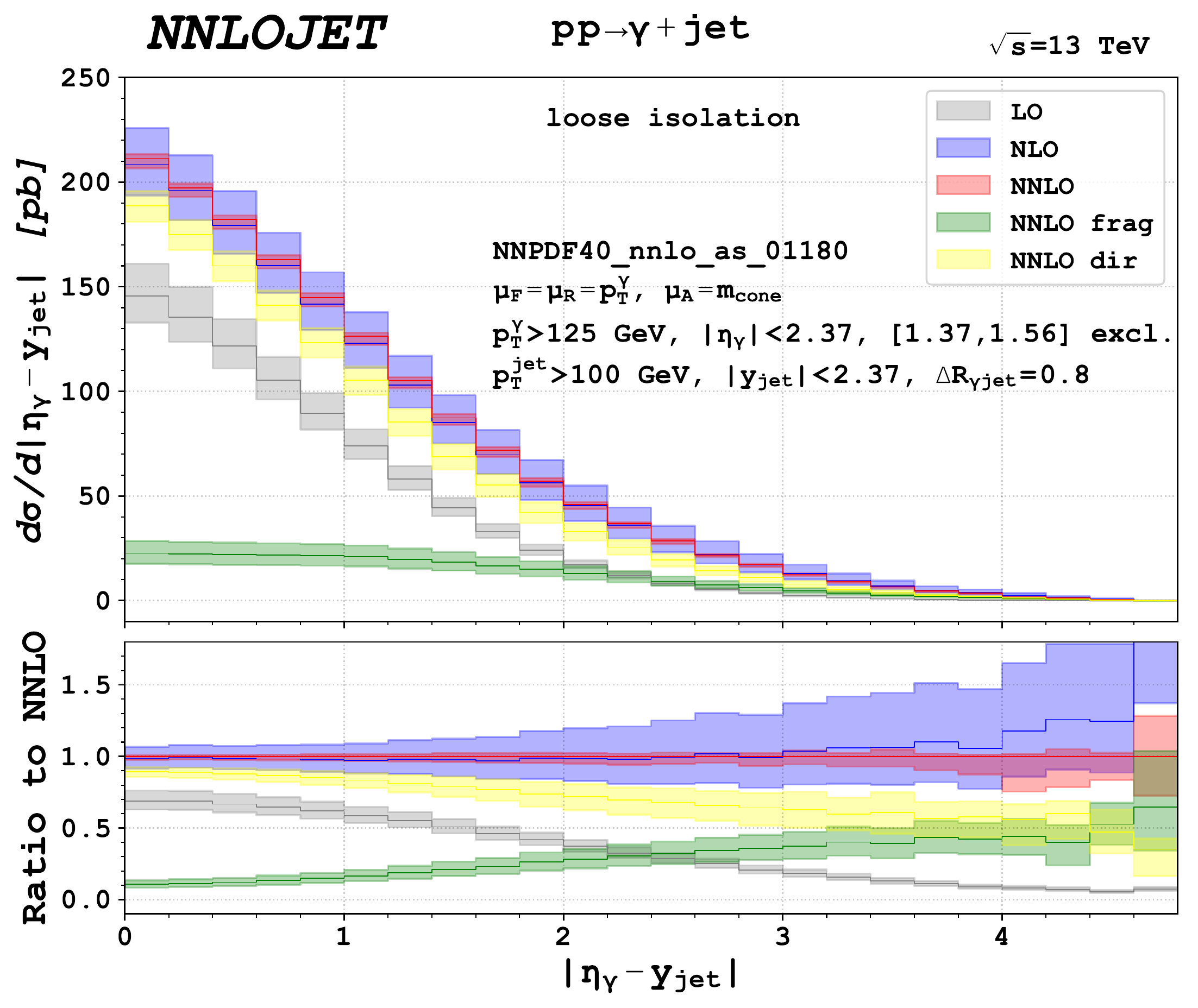}
\end{subfigure}
\hfill
\begin{subfigure}[b]{0.496\textwidth}   
\centering 
\includegraphics[width=\textwidth]{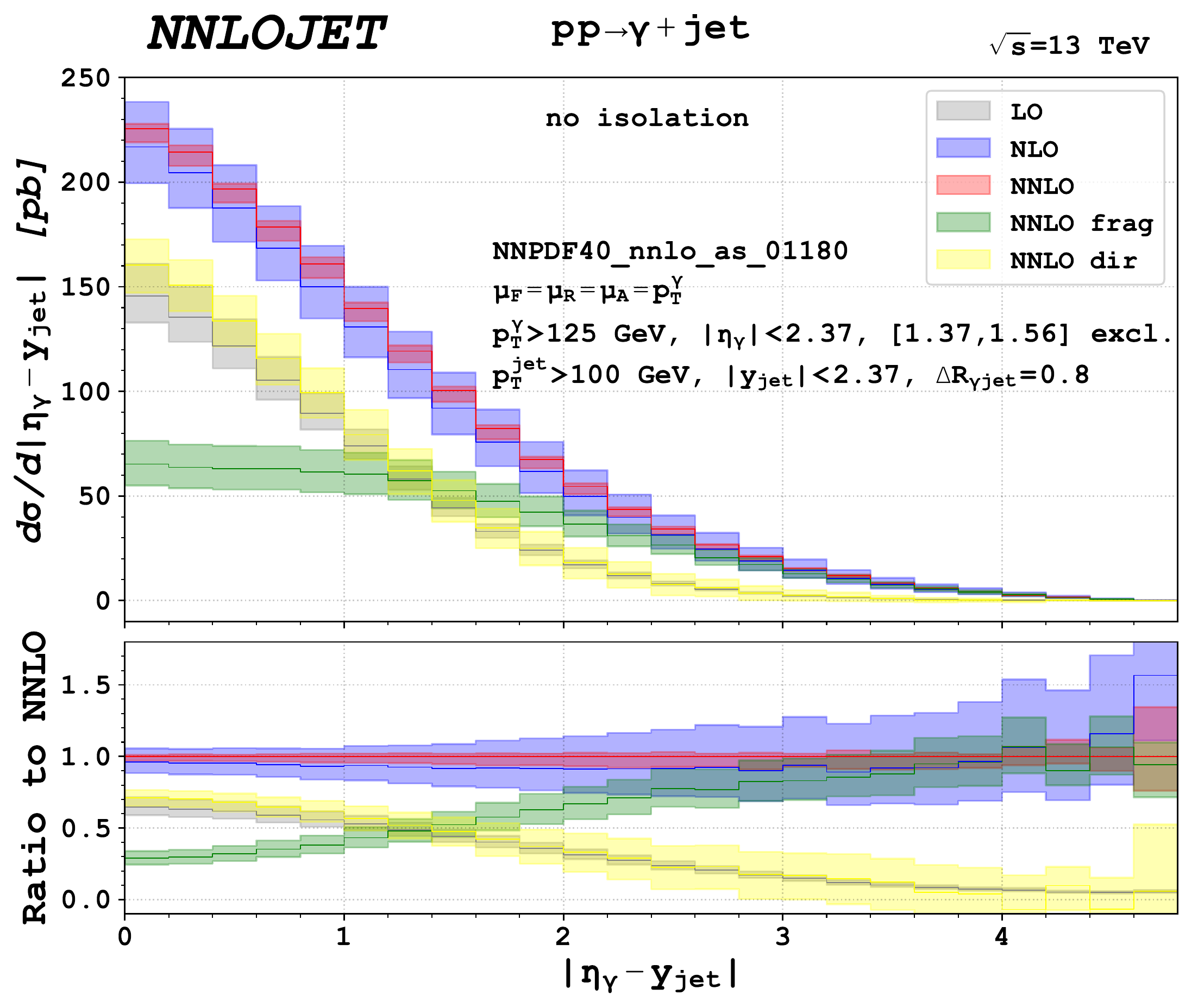}
\end{subfigure}
\caption{Decomposition of the cross section differential in the rapidity difference between the photon and the jet. Results obtained with the default isolation, a loose isolation, democratic clustering and without isolation are shown.}
\label{fig:bd_y_gam_minus_y_j1}
\end{figure}

The difference between photon pseudorapidity and jet rapidity is related to the scattering angle in the parton-parton centre-of-momentum frame, with large rapidity difference corresponding to small scattering angles. 
Predictions for the cross section as a function of the rapidity difference are shown in Figure~\ref{fig:bd_y_gam_minus_y_j1}. For all isolations the cross section is falling with increasing $|\eta_{\gamma} - y_{\rm jet}|$. Moreover, for very large rapidity differences the NNLO predictions fall below the scale uncertainty band of the NLO predictions indicating a poor perturbative convergence of the cross section in this kinematic regime.

The relative size of the fragmentation contribution increases towards larger rapidity differences for all considered isolations. At $|\eta_{\gamma} - y_{\rm jet}|=4$, fragmentation processes amount to 11\% of the full NNLO cross section for the default isolation. For even larger rapidity differences a strong decline of the direct contribution is observed, yielding even larger relative contributions from fragmentation processes. In the democratic clustering approach we observe a very similar behaviour of the cross section. Using this isolation the relative size of the fragmentation contribution also increases towards very unbalanced configurations and at $|\eta_{\gamma} - y_{\rm jet}|=4$ fragmentation processes amount to 16\% of the NNLO cross section.

No qualitative change is found for the loose isolation prescription. The impact of the fragmentation contribution grows from 10\% for small rapidity differences to 44\% at $|\eta_{\gamma} - y_{\rm jet}|=4$. We note that it is only in these very unbalanced configurations that the fragmentation contribution is roughly as large as the direct contribution for the loose isolation. 

For the inclusive set-up the direct contribution dominates for $|\eta_{\gamma} - y_{\rm jet}|<1.4$. For larger rapidity differences fragmentation processes dominate the NNLO cross section. For rapidity differences larger than four the contribution from direct photons is negligible. 

\begin{figure}[!t]
\centering
\begin{subfigure}[b]{0.496\textwidth}
\centering
\includegraphics[width=\textwidth]{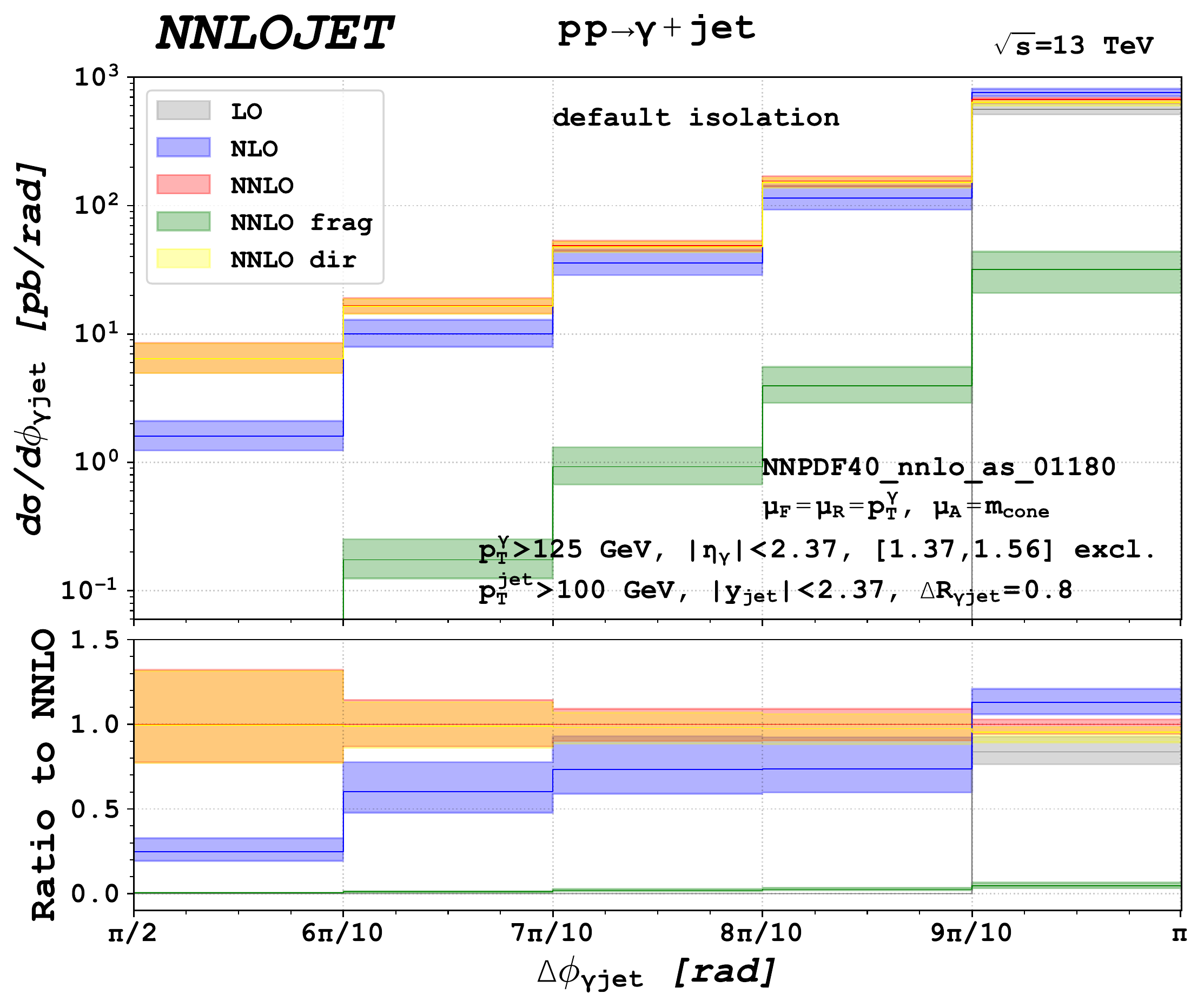}
\end{subfigure}
\hfill
\begin{subfigure}[b]{0.496\textwidth}
\centering
\includegraphics[width=\textwidth]{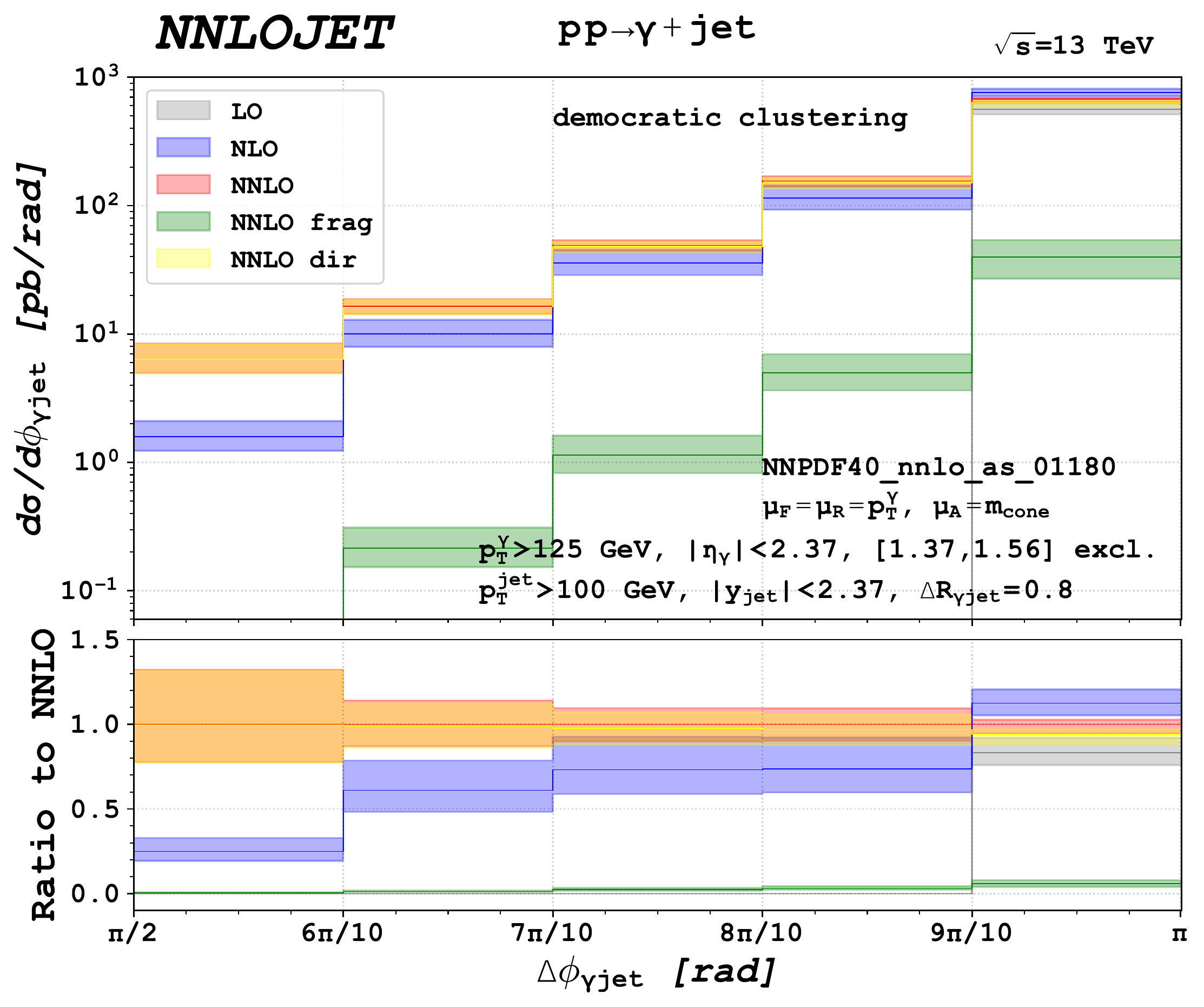}
\end{subfigure}
\vskip\baselineskip
\begin{subfigure}[b]{0.496\textwidth}   
\centering 
\includegraphics[width=\textwidth]{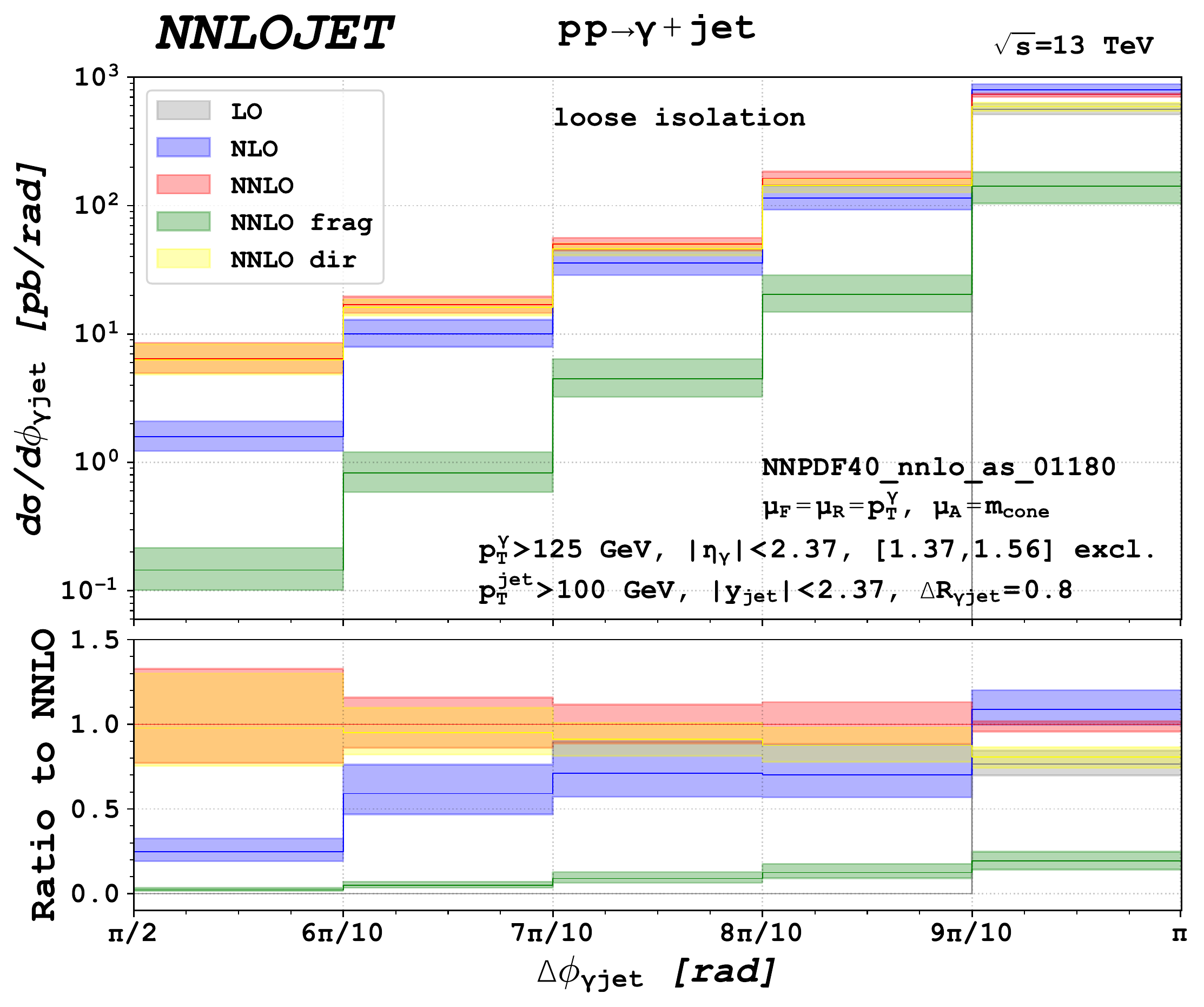}
\end{subfigure}
\hfill
\begin{subfigure}[b]{0.496\textwidth}   
\centering 
\includegraphics[width=\textwidth]{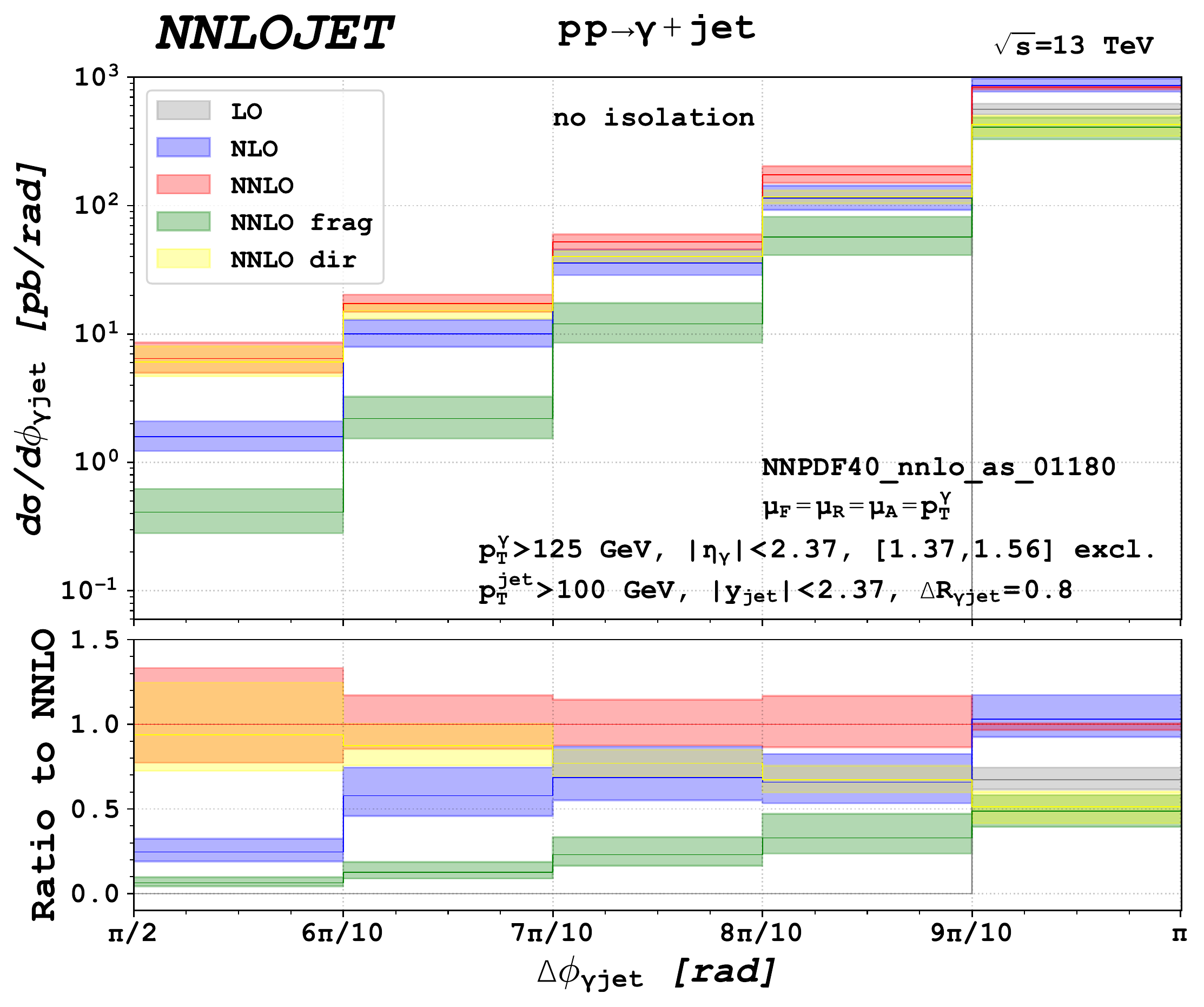}
\end{subfigure}
\caption{Decomposition of the cross section differential in the azimuthal separation between the photon and the jet. Results obtained with the default isolation, a loose isolation, democratic clustering and without isolation are shown.}
\label{fig:bd_dphi_gam_j1}
\end{figure}

The cross sections as a function of the angular separation of the photon and the jet are shown in Figure~\ref{fig:bd_dphi_gam_j1}. For all isolation set-ups the relative size of the fragmentation contribution grows with increasing $\Delta \phi$ and it is largest in the back-to-back bin. Figure~\ref{fig:comp_costheta_deltaphi} shows that the default isolation and the democratic clustering yield very similar $\Delta \phi$ distributions. This observation also holds for the composition of the cross section. For both isolations the relative contribution from fragmentation increases from 0.5 \% at $\pi/2 < \Delta \phi < 3 \pi/5$ to roughly 5\% in the back-to-back bin. 

Direct photon production also dominates the $\Delta \phi$ distribution obtained with a loose isolation. For small $\Delta \phi$ direct photons contribute almost exclusively, while in back-to-back configurations 20\% of the cross section is given by fragmentation processes.

\begin{figure}[!t]
\centering
\begin{subfigure}[b]{0.496\textwidth}
\centering
\includegraphics[width=\textwidth]{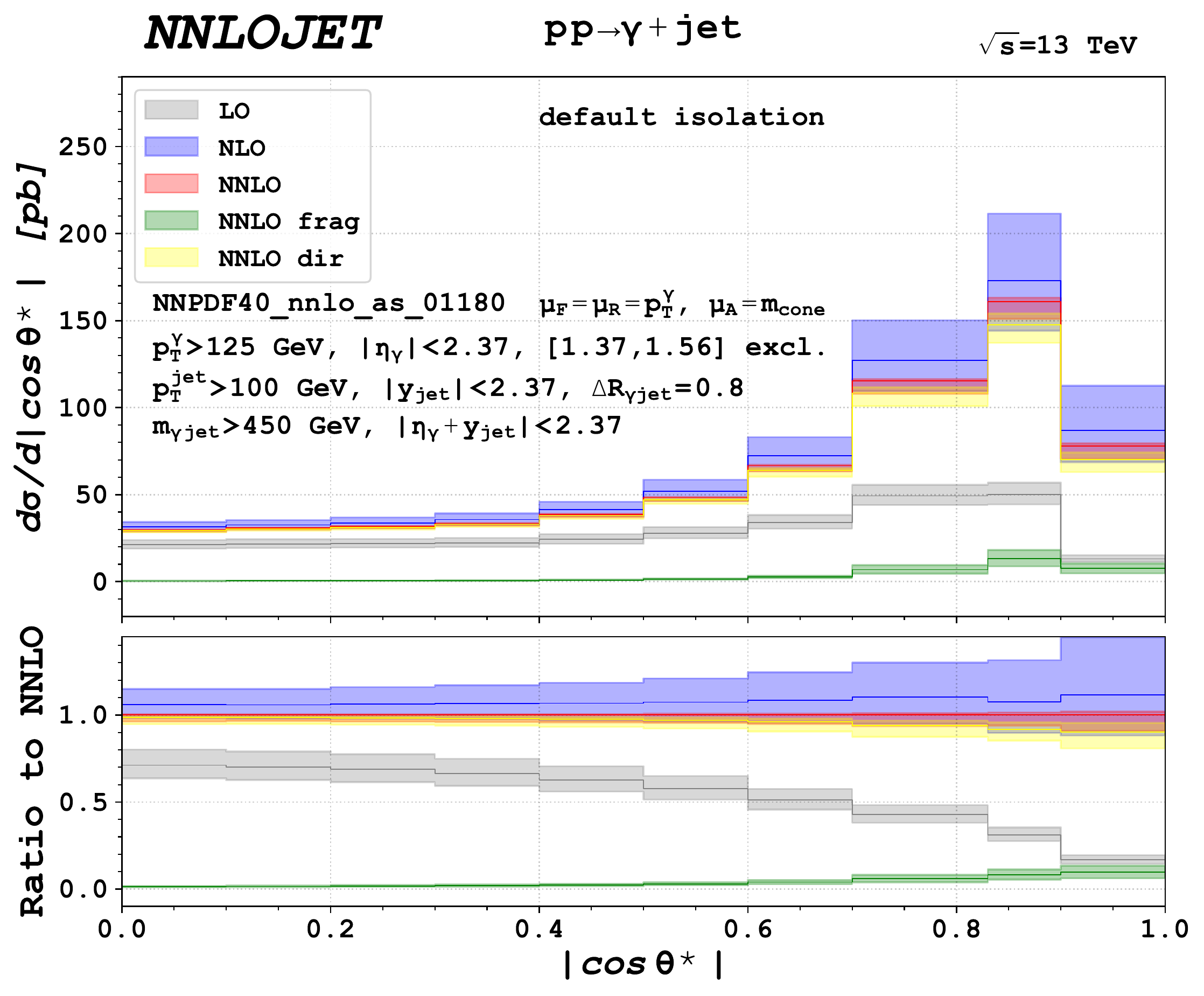}
\end{subfigure}
\hfill
\begin{subfigure}[b]{0.496\textwidth}
\centering
\includegraphics[width=\textwidth]{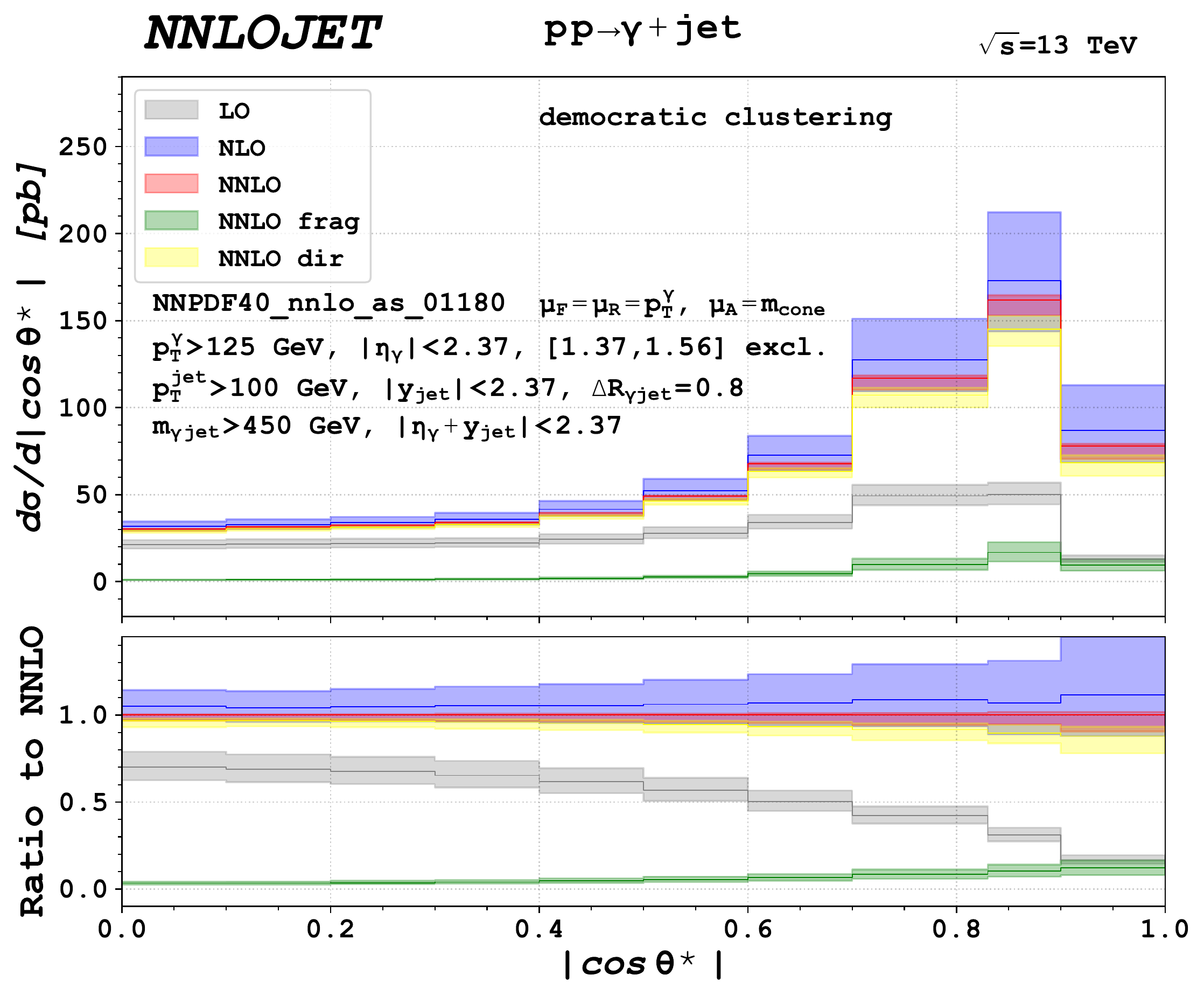}
\end{subfigure}
\vskip\baselineskip
\begin{subfigure}[b]{0.496\textwidth}   
\centering 
\includegraphics[width=\textwidth]{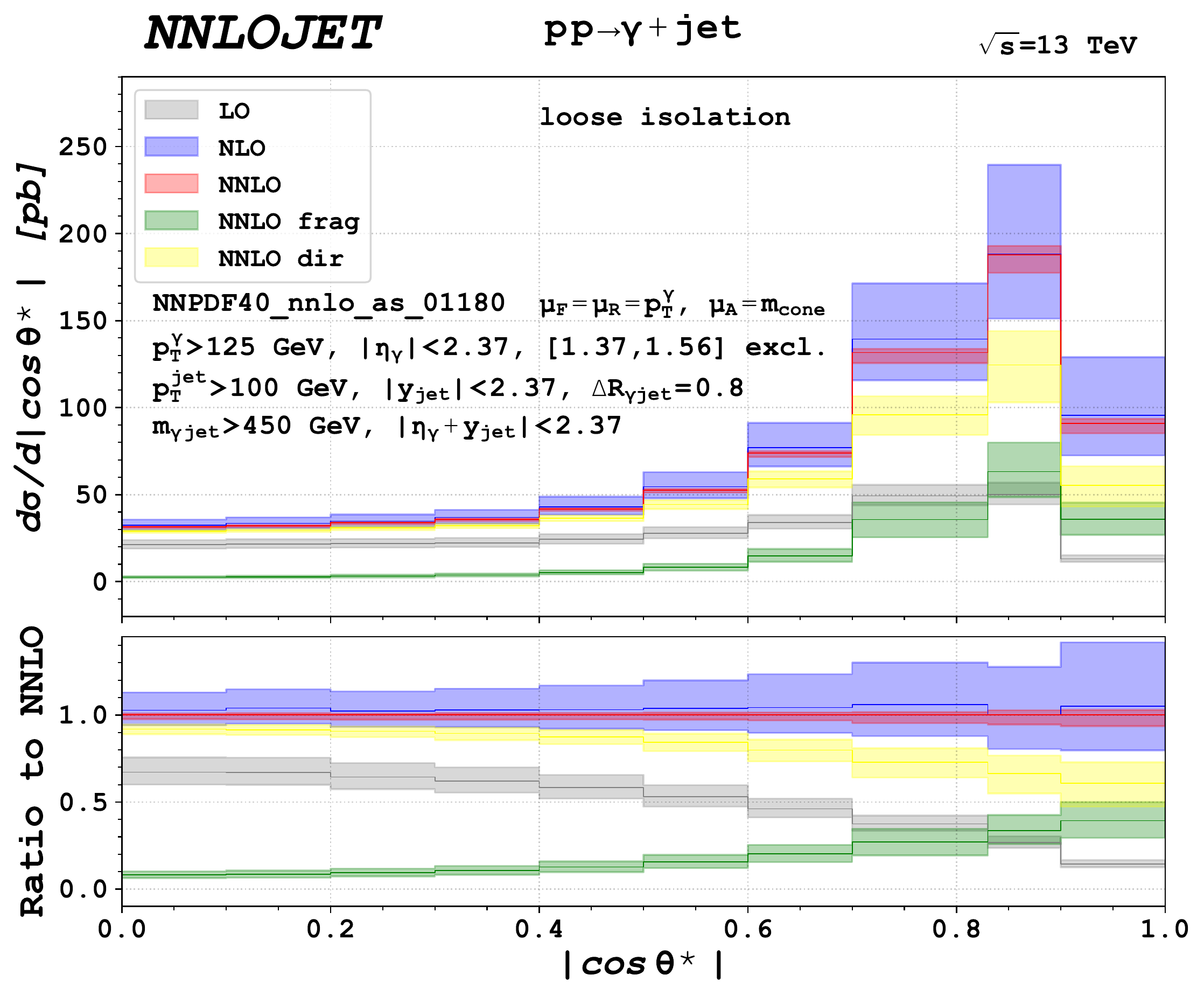}
\end{subfigure}
\hfill
\begin{subfigure}[b]{0.496\textwidth}   
\centering 
\includegraphics[width=\textwidth]{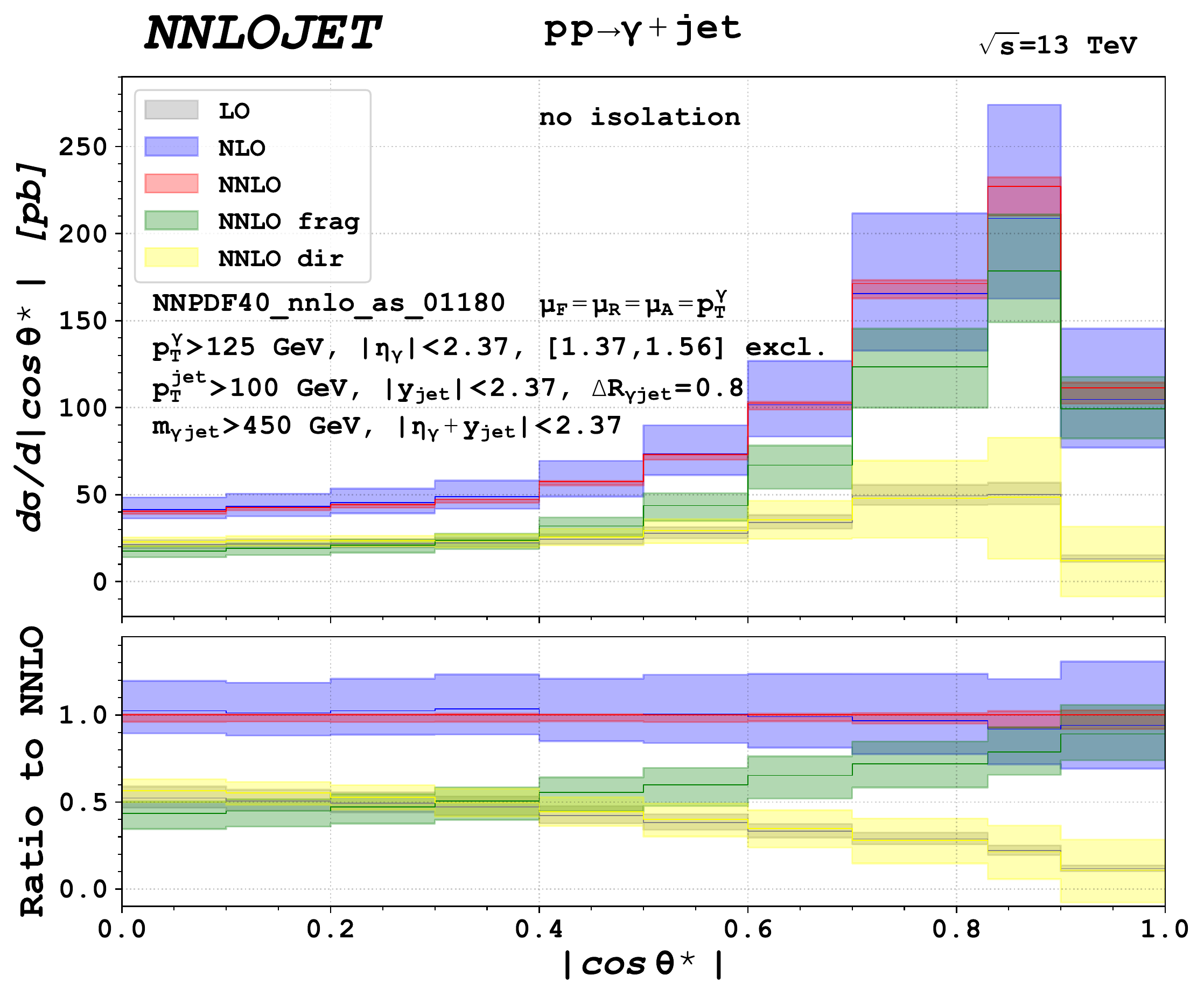}
\end{subfigure}
\caption{Decomposition of the cross section differential in $|\cos \theta^*|$. Results obtained with the default isolation, a loose isolation, democratic clustering and without isolation are shown.}
\label{fig:bd_abs_costheta}
\end{figure}

Dropping the isolation requirement, the fragmentation contribution increases strongly. However, also with this set-up the lowest $\Delta \phi$ bin is largely dominated by direct photons. The fragmentation contribution in this bin is only 6\%. In the back-to-back bin the fragmentation contribution is roughly 50\%. The large suppression of the fragmentation contribution in the small $\Delta \phi$-regime is due to an insufficient number of real radiation partons in fragmentation processes to effectively populate this kinematic regime. The large NNLO-to-NLO K-factor for the full cross section in the first bin shows that at least three additional partons besides the photon are needed to effectively populate this bin. In our predictions for the fragmentation contribution only processes with up to two additional partons are included. Therefore, it is expected that including higher order fragmentation processes would yield a strong increase in the fragmentation contribution for small $\Delta \phi$.

The observable $|\cos \theta^*|$ was discussed by ATLAS~\cite{ATLAS:2017xqp}  as a potential candidate for providing sensitivity on fragmentation processes in particular at large values of $|\cos \theta^*|$. The different isolated cross sections as well as the non-isolated cross section differential in $|\cos \theta^*|$ are shown in Figure~\ref{fig:bd_abs_costheta}. As expected from the different dependence of the underlying Born matrix elements for direct and fragmentation 
processes on $\cos \theta^*$, the relative fragmentation contribution increases for all set-ups towards larger $|\cos \theta^*|$.

For small $|\cos \theta^*|$ direct photons dominate the cross section for all considered isolations. The fragmentation contribution for $|\cos \theta^*|<0.3$ amounts to only 1.5\% of the full NNLO cross section with the default isolation. For democratic clustering 3\% of the cross section originate from fragmentation processes and for the loose isolation 8\%. In the bin $0.7<|\cos \theta^*|<0.83$, which is the highest bin for which the ATLAS measurement provides data, the relative contribution from fragmentation is 6\% for the default isolation, 8\% for democratic clustering and 27\% for the loose fixed cone isolation. For even larger values of $|\cos \theta^*|$, which have not been accessed by experimental measurements yet, the relative fragmentation contribution increases even further.

For the non-isolated cross section direct photon and fragmentation contribute almost equally for $|\cos \theta^*| < 0.4$. Towards larger values of $|\cos \theta^*|$ a strong decline of the direct contribution is observed and at $|\cos \theta^*| \approx 1$ fragmentation contributes 90\% to the full NNLO cross section.

\section{Imbalance between Photon and Jet Transverse Momentum: \texorpdfstring{$z_{\rm rec}$}{z rec}}
\label{sec:z_rec}

At large longitudinal momentum transfers, the photon fragmentation functions are only weakly constrained through 
experimental measurements at LEP from the  ALEPH~\cite{Buskulic:1995au} and OPAL~\cite{Ackerstaff:1997nha} experiments. However, for (tightly) isolated photon production the fragmentation contribution almost exclusively stems from events in which a large fraction of transverse momentum is passed to the photon. 
To constrain the photon fragmentation functions in this kinematical region,  several observables in electron-positron 
annihilation~\cite{Glover:1993xc}, deep inelastic scattering~\cite{Gehrmann-DeRidder:2006zbx,Gehrmann-DeRidder:2006lpc} and at hadron 
colliders~\cite{Belghobsi:2009hx,Kaufmann:2016nux} have been proposed. These proposed observables derive their sensitivity on the photon 
fragmentation functions at large momentum transfer from either investigating the substructure of a jet containing a highly energetic photon or from kinematic 
photon-jet correlations.  
Having very precise experimental data and theory predictions which incorporate the fragmentation processes at NNLO, it can now be envisaged to determine the photon fragmentation functions 
in this regime using  isolated photon and photon-plus-jet data from the LHC. 

In this section we study the observable $z_{\rm rec}$, which provides differential sensitivity on the photon fragmentation functions and therefore could help to constrain the photon fragmentation functions in future studies. 
The variable $z_{\rm rec}$ is a measure for the imbalance of the photon and leading jet transverse momenta, i.e.\
\begin{equation}
z_{\rm rec} \equiv \frac{p_T^{\gamma}}{p_T^{\rm jet}} \, .
\end{equation}
A similar variable  $z_{\gamma}$, including a projection of the photon momentum onto the leading jet axis, has been proposed in~\cite{Belghobsi:2009hx}.
For the LO cross section of direct photon production, the two objects are exactly balanced in the transverse plane and thus $z_{\rm rec} = 1$ holds. From NLO onwards additional parton radiation can generate configurations for which $z_{\rm rec}$ is not equal to unity. In this case we can distinguish two different scenarios. If the additional parton is radiated in the hemisphere of the photon we have $z_{\rm rec} < 1$ while if the additional parton is radiated in the opposite hemisphere and it is not clustered with the other parton by the jet algorithm we have $z_{\rm rec} > 1$.

For photon fragmentation at Born level, two back-to-back jets (partons) are radiated from which one subsequently fragments into a photon and passes a fraction $z$ of its momentum to the photon. This process is described by the parton-to-photon fragmentation function $D_{p \to \gamma}(z)$. In this case the imbalance between the transverse momentum of the photon and the leading jet is given by
\begin{equation}
z_{\rm rec} = \frac{p_T^{\gamma}}{p_T^{\rm jet}} = \frac{z \, p_T^{\rm jet}}{p_T^{\rm jet}} = z \, .
\end{equation}
Consequently, at this order there is a one-to-one correspondence between $z_{\rm rec}$ and the argument of the fragmentation function $z$. The range of $z_{\rm rec}$ values which is populated by lowest order fragmentation processes can be inferred from the corresponding isolation parameters. For a fixed cone isolation with parameter $(R, \, \varepsilon, \, E_T^{\rm thres})$ and assuming that $\varepsilon \ll 1$, we have
\begin{equation}
1 > z_{\rm rec} > z_{\rm rec}^{\rm min} = \frac{p_{T_{\rm min}}^{\gamma}}{p_{T_{\rm min}}^{\gamma} + E_T^{\rm thres}} \, ,
\label{eq:z_rec_min}
\end{equation}
where $p_{T_{\rm min}}^{\gamma}$ is the minimal photon transverse momentum. Beyond lowest order, the one-to-one correspondence between $z_{\rm rec}$ and the argument of the fragmentation function does no longer hold. Moreover, the radiation of additional partons extends the range of populated $z_{\rm rec}$ values to $z_{\rm rec} > 1$ and $z_{\rm rec} < z_{\rm rec}^{\rm min}$.
\begin{figure}[!t]
\centering
\begin{subfigure}[b]{0.496\textwidth}
\centering
\includegraphics[width=\textwidth]{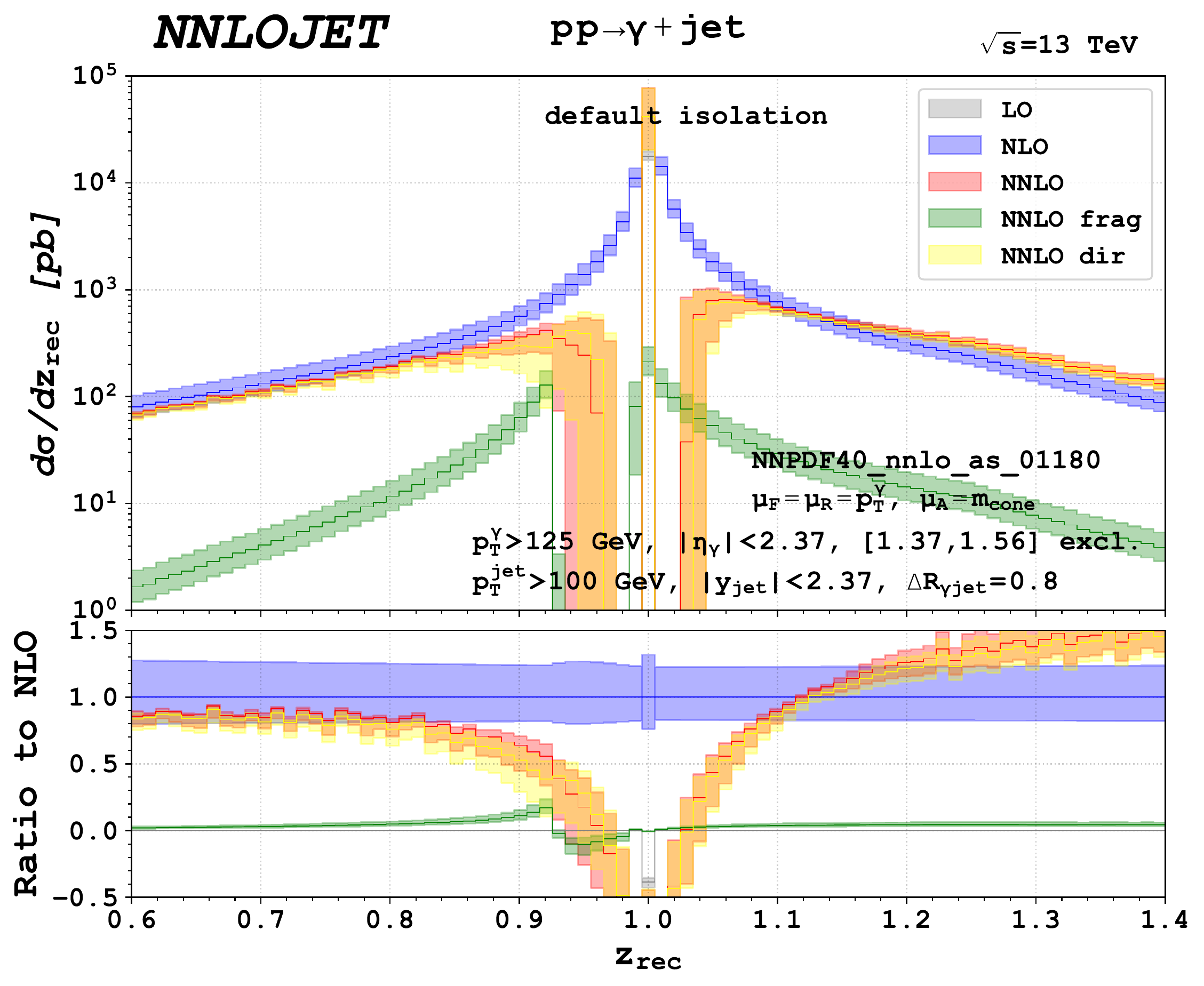}
\end{subfigure}
\hfill
\begin{subfigure}[b]{0.496\textwidth}
\centering
\includegraphics[width=\textwidth]{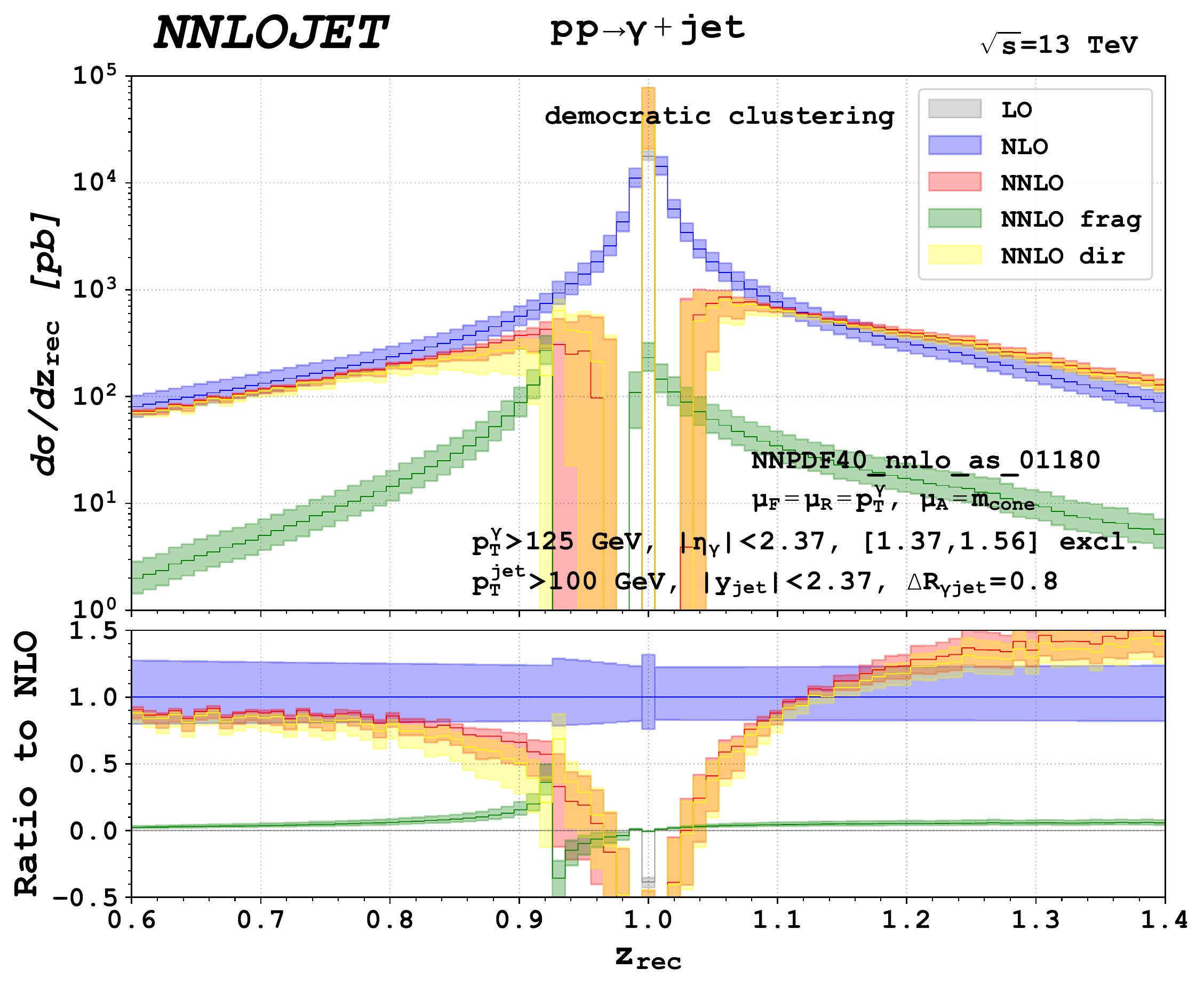}
\end{subfigure}
\vskip\baselineskip
\begin{subfigure}[b]{0.496\textwidth}   
\centering 
\includegraphics[width=\textwidth]{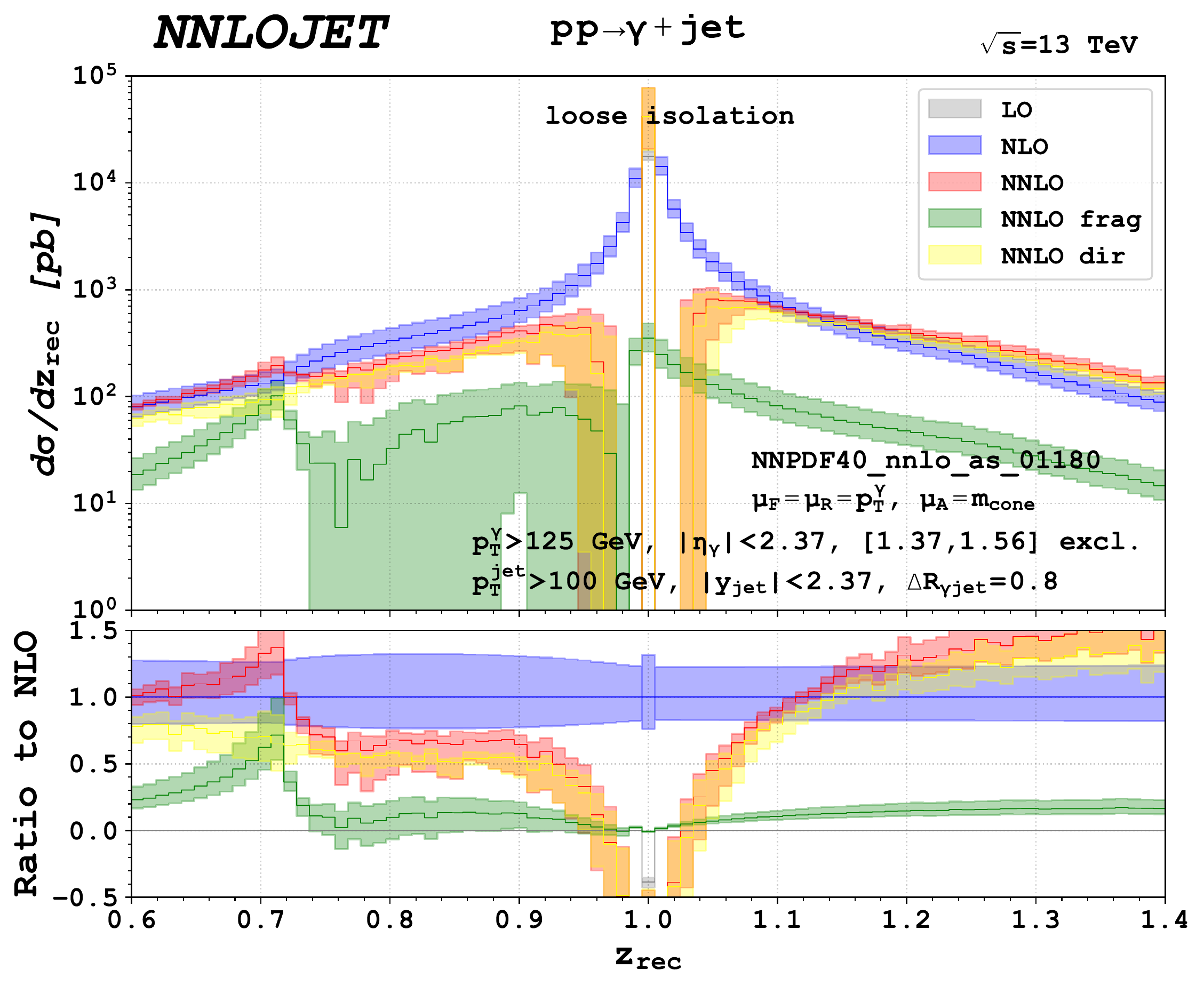}
\end{subfigure}
\hfill
\begin{subfigure}[b]{0.496\textwidth}   
\centering 
\includegraphics[width=\textwidth]{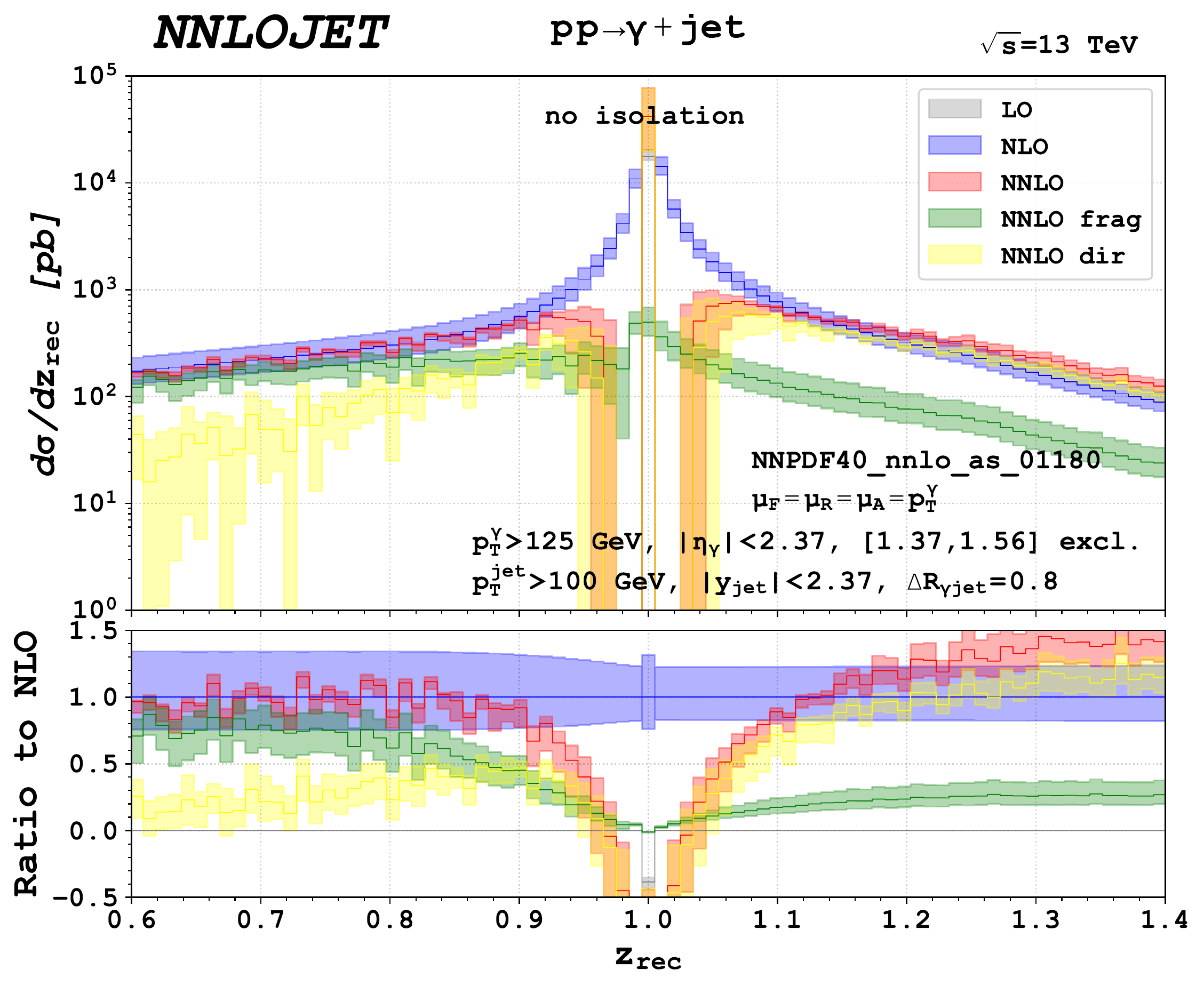}
\end{subfigure}
\caption{Decomposition of the cross section differential in $z_{\rm rec}$. Results obtained with the default isolation, a loose isolation, democratic clustering and without isolation are shown.}
\label{fig:bd_z_rec}
\end{figure}

Results for the $z_{\rm rec}$ distributions of the photon-plus-jet cross section are shown in Figure~\ref{fig:bd_z_rec}. For the predictions we used the set-up of the ATLAS 13\,\TeV study~\cite{ATLAS:2017xqp} and compared the default isolation, the fixed cone loose isolation, democratic clustering and the non-isolated cross section. For all set-ups we observe an increase of the NNLO cross section towards $z_{\rm rec} \approx 1$. Due to the Sudakov shoulder at $z_{\rm rec}=1$, the distributions show a poor perturbative convergence in the near vicinity of $z_{\rm rec}=1$. Considering the NNLO-to-NLO K-factors of the $z_{\rm rec}$ distributions, the region in which the fixed-order calculation fails to give a reliable description can be estimated to be $z_{\rm rec} \in [0.93, 1.06]$. The isolated distributions are dominated by direct photons over most of the kinematic range. In the region $z_{\rm rec} < 1$ the fragmentation contribution is increasing with $z_{\rm rec}$ only below the value $z_{\rm rec}^{\rm min}$ in \eqref{eq:z_rec_min}, corresponding to a kinematical region that requires the fragmenting photon cluster to be accompanied by at least two other partons and that is described only at NNLO and beyond. For the default isolation and for democratic clustering we have $z_{\rm rec}^{\rm min}=0.926$ and for the loose isolation $z_{\rm rec}^{\rm min}=0.714$. At $z_{\rm rec}^{\rm min}$ the fragmentation contribution drops, resulting in a kink of the full NNLO cross section, which is most prominent for the loose isolation. For $z_{\rm rec}^{\rm min}<z_{\rm rec} < 1$ the contribution from fragmentation is very small for the default isolation and for democratic clustering it even turns negative. Some modest amount of fragmentation contribution remains for 
the loose isolation prescription. 
\begin{figure}[!t]
\centering
\includegraphics[scale=0.40]{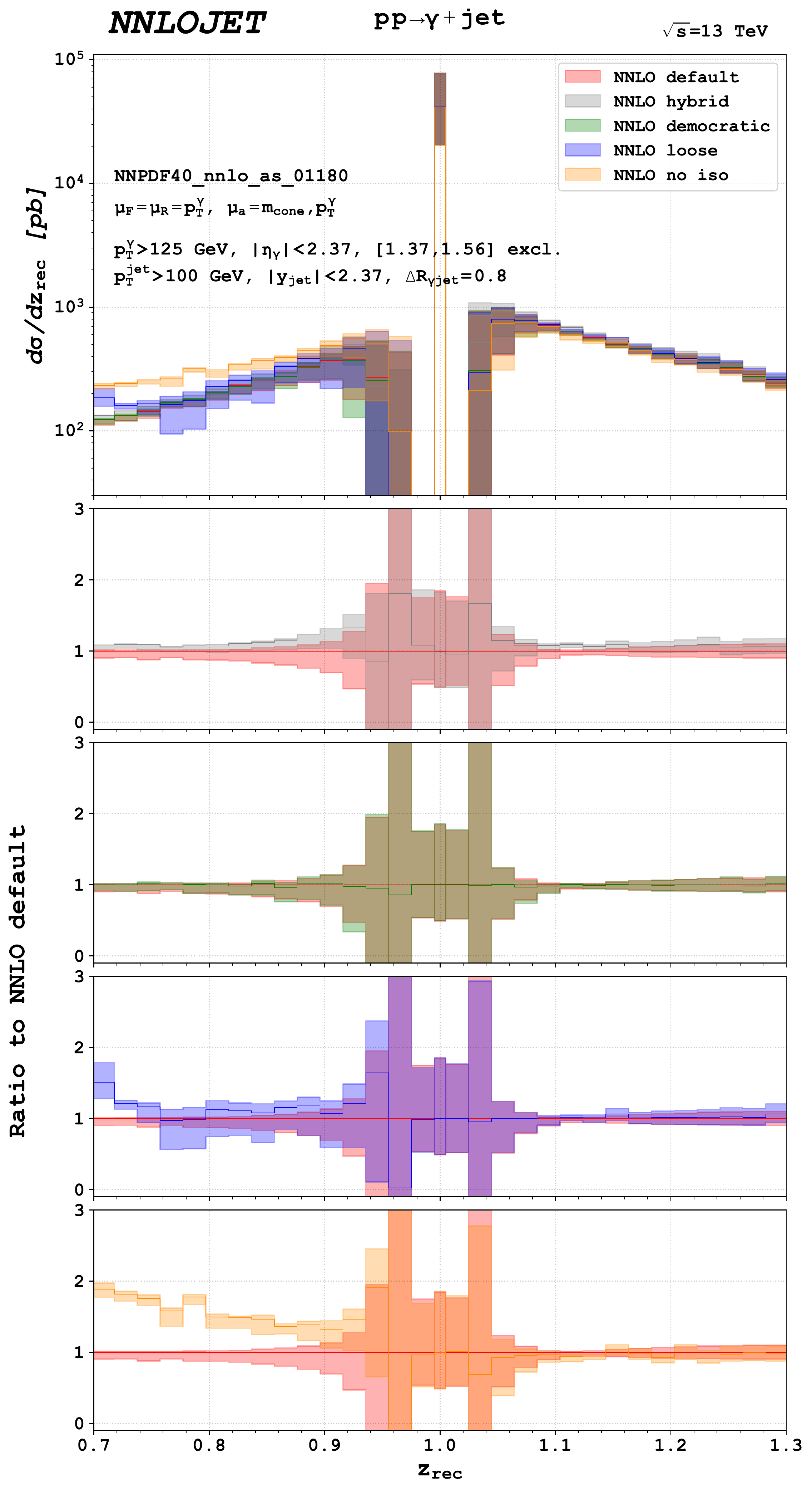}
\caption{Predictions for the photon-plus-jet cross section differential in $z_{\rm rec}$ using different isolation prescription.}
\label{fig:comp_zrec}
\end{figure}

A completely different picture emerges if no photon isolation is applied. In this case, the direct production processes dominate the $z_{\rm rec}$ distribution only for  $z_{\rm rec} > 1$. We observe a strong increase in the fragmentation contribution for $z_{\rm rec} < 1$. Already for $z_{\rm rec} < 0.89$, fragmentation processes dominate over direct photon production and for even smaller values of $z_{\rm rec}$ fragmentation processes contribute almost exclusively. Again, the immediate vicinity of $z_{\rm rec}=1$ 
can not be described by fixed-order perturbation theory due to the Sudakov shoulder behaviour. In contrast to the $z_{\rm rec}$ distributions for different photon isolation prescriptions, 
no further perturbative instability is observed around a $z_{\rm rec}^{\rm min}<1$  value. 

Figure~\ref{fig:comp_zrec} compares the $z_{\rm rec}$ distributions at NNLO obtained for the different photon isolation prescriptions, also including 
hybrid isolation. Substantial deviations outside the theory uncertainties
from the default ATLAS isolation are observed only if no isolation is applied.
While democratic clustering yields a distribution almost identical to the default isolation, loose and hybrid isolation predict a uniform but not significant enhancement of the distribution for $z_{\rm rec} < 1$. The substantial enhancement of the
non-isolated  $z_{\rm rec}$ distribution for $z_{\rm rec} < 1$ is entirely due to photon fragmentation processes. This observation
strongly suggests that a determination of the photon fragmentation functions at the LHC could potentially be attempted by a measurement of the  $z_{\rm rec}$
distribution in photon-plus-jet events without any isolation requirements on the photon. 
The sensitivity on the fragmentation functions could potentially be enhanced further by 
taking the difference or a ratio between isolation prescriptions and/or applying additional fiducial selection cuts, e.g.\ $|\cos\theta^*| \gtrsim 0.5$ or $\Delta\phi \gtrsim \tfrac{3\pi}{4}$, to enhance the relative contribution from the fragmentation processes. 

We are perfectly aware of the substantial experimental challenges  that a measurement of photon production cross sections without any isolation requirement entails.  Its  demonstrated sensitivity on the photon fragmentation functions presents a strong motivation to further investigate the prospects of such a measurement, and warrants 
a detailed assessment of potential experimental challenges.

\section{Conclusions}
\label{sec:conc}

In this paper, we used the antenna subtraction method to compute 
the NNLO QCD corrections to isolated photon and photon-plus-jet production, including contributions~\cite{Gehrmann:2022cih} from collinear photon radiation off 
partons and parton-to-photon fragmentation. 
Previous calculations at this order~\cite{Campbell:2017dqk,Campbell_2017,Chen_2020}  discarded these contributions by applying an idealised photon isolation based on a dynamical isolation cone~\cite{Frixione:1998jh}. Our newly derived results enable us for the first time to reproduce the photon isolation procedure used in the 
experimental measurements, which is based on a fixed-size cone around the photon direction, admitting a finite amount of hadronic energy. 
Our results are implemented in the \nnlojet parton-level event generator. A detailed 
comparison with experimental results from ATLAS, CMS and ALICE demonstrated the impact of the NNLO corrections for precision phenomenology. 

The mass factorisation scale in the photon fragmentation functions separates direct and fragmentation contributions; it reflects up to which scale an observable is 
inclusive in extra radiation accompanying the photon. We argue that in the case of isolated photons, this scale is related to the 
maximal invariant mass of photons and partons inside the isolation cone, which is typically much lower than the photon transverse momentum. Choosing a larger 
mass factorisation scale in the photon fragmentation functions would imply to cut out a substantial amount of out-of-cone parton radiation from the direct process at 
fixed order, which is added back in resummed (and thus attenuated) form through the fragmentation functions, thereby leading to an erroneous description of 
out-of-cone dynamics. With the invariant mass scale choice in the photon fragmentation functions, the fragmentation contributions are small and vanish with 
increasing photon transverse momentum. The smallness of the factorisation scale and the overall small impact of the fragmentation contributions justifies a power counting 
of $\alpha$ for the fragmentation functions, which is used throughout in our work. 

The newly derived NNLO corrections allow us to compute predictions for photon-plus-jet production without any photon isolation and to compare 
different isolation prescriptions. The application of the default ATLAS photon isolation leads to a decrease of the non-isolated cross section by only 30\%. 
Results obtained with a dynamical cone are close in 
normalisation to the default fixed-cone predictions, but deviate in the kinematical shape  especially at low photon transverse momenta. Loosening of the isolation 
parameters increases the cross sections, with the largest effects observed at low photon or jet transverse momentum. 

The democratic clustering procedure for photon isolation~\cite{Glover:1993xc, GehrmannDeRidder:1997gf} treats photons as any other partons in the jet clustering, 
identifying the photons only after the jet clustering by labelling all jets containing more than a predefined large fraction of electromagnetic energy as isolated photons. 
This isolation procedure will be particularly advantageous when electroweak corrections to photon observables are taken into account, since it eliminates 
all ambiguities relating to a discrimination in kinematical handling between identified and non-identified photons. It has however been proven impractical to be implemented 
in experimental analyses so far. We demonstrate that by using democratic clustering in the anti-$k_T$ algorithm with a cone size identical to the fixed cone isolation 
and properly adjusted energy fraction threshold, the predicted cross sections are nearly identical to those obtained for fixed cone isolation. These results may 
help to enable measurements applying democratic clustering especially for electroweak precision observables. 

We introduce the observable  $z_{\rm rec}$ derived from photon-plus-jet final states, which allows us to identify kinematical regions that are 
potentially sensitive on photon fragmentation contributions and the form of the fragmentation functions. 
The highest sensitivity to parton-to-photon fragmentation functions is observed if no 
photon isolation is applied, and we discuss possible hadron collider measurements of these functions. 

Our newly derived results eliminate the systematic uncertainty resulting from the mismatch of photon isolation prescription between experiment and 
theory for isolated photon final states at hadron colliders. They will thus enable a broad range of precision phenomenology studies with these observables, including 
electroweak precision studies, constraints on parton distributions, and the determination of  photon fragmentation functions.

\section*{Acknowledgements}
The authors thank Juan Cruz-Martinez, James Currie, Rhorry Gauld, Aude Gehrmann-De Ridder,
 Imre Majer, Matteo Marcoli, Jonathan Mo, Thomas Morgan, Jan Niehues, Jo\~ao Pires, Adrian Rodriguez Garcia, 
  Giovanni Stagnitto, Duncan Walker, Steven Wells and James Whitehead for useful 
discussions and their many contributions to the \nnlojet code. We would also like to thank Thomas Becher and Joey Huston for interesting discussions on 
photon isolation and its implications.  
This work has received funding from the Swiss National Science Foundation (SNF) under contract 200020-204200, from the European Research Council (ERC) under the European Union's Horizon 2020 research and innovation programme grant agreement 101019620 (ERC Advanced Grant TOPUP), from the UK Science and Technology Facilities Council (STFC) through grant ST/T001011/1 and from the Deutsche Forschungsgemeinschaft (DFG, German Research Foundation) under grant 396021762-TRR 257.

\bibliographystyle{JHEP}
\bibliography{bib_GJ}

\providecommand{\href}[2]{#2}\begingroup\raggedright\begin{thebibliography}{10}

\bibitem{Athens-Athens-Brookhaven-CERN:1979oxx}
{\scshape Athens-Athens-Brookhaven-CERN} collaboration, M.~Diakonou et~al.,
  \emph{{Direct Production of High $p_T$ Single Photons at the CERN
  Intersecting Storage Rings}},
  \href{http://dx.doi.org/10.1016/0370-2693(79)90985-7}{\emph{Phys. Lett. B}
  {\bf 87} (1979) 292--296}.

\bibitem{Anassontzis:1982gm}
E.~Anassontzis et~al., \emph{{High p(t) Direct Photon Production in p p
  Collisions}}, \href{http://dx.doi.org/10.1007/BF01572345}{\emph{Z. Phys. C}
  {\bf 13} (1982) 277--289}.

\bibitem{CMOR:1989qzc}
{\scshape CMOR} collaboration, A.~L.~S. Angelis et~al., \emph{{Direct Photon
  Production at the {CERN} {ISR}}},
  \href{http://dx.doi.org/10.1016/0550-3213(89)90305-2}{\emph{Nucl. Phys. B}
  {\bf 327} (1989) 541--568}.

\bibitem{UA1:1988zam}
{\scshape UA1} collaboration, C.~Albajar et~al., \emph{{Direct Photon
  Production at the CERN Proton - anti-Proton Collider}},
  \href{http://dx.doi.org/10.1016/0370-2693(88)90968-9}{\emph{Phys. Lett. B}
  {\bf 209} (1988) 385--396}.

\bibitem{UA2:1991wce}
{\scshape UA2} collaboration, J.~Alitti et~al., \emph{{A Measurement of the
  direct photon production cross-section at the CERN anti-p p collider}},
  \href{http://dx.doi.org/10.1016/0370-2693(91)90503-I}{\emph{Phys. Lett. B}
  {\bf 263} (1991) 544--550}.

\bibitem{WA70:1987vvj}
{\scshape WA70} collaboration, M.~Bonesini et~al., \emph{{Production of High
  Transverse Momentum Prompt Photons and Neutral Pions in Proton Proton
  Interactions at 280-GeV/c}},
  \href{http://dx.doi.org/10.1007/BF01584385}{\emph{Z. Phys. C} {\bf 38} (1988)
  371}.

\bibitem{FermilabE706:2004emk}
{\scshape Fermilab E706} collaboration, L.~Apanasevich et~al.,
  \emph{{Measurement of direct photon production at Tevatron fixed target
  energies}}, \href{http://dx.doi.org/10.1103/PhysRevD.70.092009}{\emph{Phys.
  Rev. D} {\bf 70} (2004) 092009},
  [\href{http://arxiv.org/abs/hep-ex/0407011}{{\tt hep-ex/0407011}}].

\bibitem{D0:2005ofv}
{\scshape D0} collaboration, V.~M. Abazov et~al., \emph{{Measurement of the
  isolated photon cross section in $p \bar{p}$ collisions at $\sqrt{s}$ =
  1.96-TeV}},
  \href{http://dx.doi.org/10.1016/j.physletb.2006.04.048}{\emph{Phys. Lett. B}
  {\bf 639} (2006) 151--158}, [\href{http://arxiv.org/abs/hep-ex/0511054}{{\tt
  hep-ex/0511054}}]. [Erratum: Phys.Lett.B 658, 285--289 (2008)].

\bibitem{D0:2008chx}
{\scshape D0} collaboration, V.~M. Abazov et~al., \emph{{Measurement of the
  Differential Cross-Section for the Production of an Isolated Photon with
  Associated Jet in $p \bar{p}$ Collisions at $\sqrt{s}$ = 1.96-TeV}},
  \href{http://dx.doi.org/10.1016/j.physletb.2008.06.076}{\emph{Phys. Lett. B}
  {\bf 666} (2008) 435--445}, [\href{http://arxiv.org/abs/0804.1107}{{\tt
  0804.1107}}].

\bibitem{D0:2013lra}
{\scshape D0} collaboration, V.~M. Abazov et~al., \emph{{Measurement of the
  differential cross section of photon plus jet production in $p\bar{p}$
  collisions at $\sqrt{s}=1.96$ TeV}},
  \href{http://dx.doi.org/10.1103/PhysRevD.88.072008}{\emph{Phys. Rev. D} {\bf
  88} (2013) 072008}, [\href{http://arxiv.org/abs/1308.2708}{{\tt 1308.2708}}].

\bibitem{CDF:2017cuc}
{\scshape CDF} collaboration, T.~Aaltonen et~al., \emph{{Measurement of the
  inclusive-isolated prompt-photon cross section in $p\bar{p}$ collisions using
  the full CDF data set}},
  \href{http://dx.doi.org/10.1103/PhysRevD.96.092003}{\emph{Phys. Rev. D} {\bf
  96} (2017) 092003}, [\href{http://arxiv.org/abs/1703.00599}{{\tt
  1703.00599}}].

\bibitem{ATLAS:2017nah}
{\scshape ATLAS} collaboration, M.~Aaboud et~al., \emph{{Measurement of the
  cross section for inclusive isolated-photon production in $pp$ collisions at
  $\sqrt s=13$ TeV using the ATLAS detector}},
  \href{http://dx.doi.org/10.1016/j.physletb.2017.04.072}{\emph{Phys. Lett. B}
  {\bf 770} (2017) 473--493}, [\href{http://arxiv.org/abs/1701.06882}{{\tt
  1701.06882}}].

\bibitem{ATLAS:2019buk}
{\scshape ATLAS} collaboration, G.~Aad et~al., \emph{{Measurement of the
  inclusive isolated-photon cross section in $pp$ collisions at $\sqrt{s}=13$
  TeV using 36 fb$^{-1}$ of ATLAS data}},
  \href{http://dx.doi.org/10.1007/JHEP10(2019)203}{\emph{JHEP} {\bf 10} (2019)
  203}, [\href{http://arxiv.org/abs/1908.02746}{{\tt 1908.02746}}].

\bibitem{ATLAS:2017xqp}
{\scshape ATLAS} collaboration, M.~Aaboud et~al., \emph{{Measurement of the
  cross section for isolated-photon plus jet production in $pp$ collisions at
  $\sqrt s=13$ TeV using the ATLAS detector}},
  \href{http://dx.doi.org/10.1016/j.physletb.2018.03.035}{\emph{Phys. Lett. B}
  {\bf 780} (2018) 578--602}, [\href{http://arxiv.org/abs/1801.00112}{{\tt
  1801.00112}}].

\bibitem{CMS:2018qao}
{\scshape CMS} collaboration, A.~M. Sirunyan et~al., \emph{{Measurement of
  differential cross sections for inclusive isolated-photon and photon+jets
  production in proton-proton collisions at $\sqrt{s} =$ 13 TeV}},
  \href{http://dx.doi.org/10.1140/epjc/s10052-018-6482-9}{\emph{Eur. Phys. J.
  C} {\bf 79} (2019) 20}, [\href{http://arxiv.org/abs/1807.00782}{{\tt
  1807.00782}}].

\bibitem{CMS:2019jlq}
{\scshape CMS} collaboration, A.~M. Sirunyan et~al., \emph{{Measurements of
  triple-differential cross sections for inclusive isolated-photon+jet events
  in pp collisions at $\sqrt{s} = 8\,\text {TeV} $}},
  \href{http://dx.doi.org/10.1140/epjc/s10052-019-7451-7}{\emph{Eur. Phys. J.
  C} {\bf 79} (2019) 969}, [\href{http://arxiv.org/abs/1907.08155}{{\tt
  1907.08155}}].

\bibitem{ALICE:2019rtd}
{\scshape ALICE} collaboration, S.~Acharya et~al., \emph{{Measurement of the
  inclusive isolated photon production cross section in pp collisions at
  $\sqrt{s}=7$ TeV}},
  \href{http://dx.doi.org/10.1140/epjc/s10052-019-7389-9}{\emph{Eur. Phys. J.
  C} {\bf 79} (2019) 896}, [\href{http://arxiv.org/abs/1906.01371}{{\tt
  1906.01371}}].

\bibitem{Lindert:2017olm}
J.~M. Lindert et~al., \emph{{Precise predictions for $V+$ jets dark matter
  backgrounds}},
  \href{http://dx.doi.org/10.1140/epjc/s10052-017-5389-1}{\emph{Eur. Phys. J.
  C} {\bf 77} (2017) 829}, [\href{http://arxiv.org/abs/1705.04664}{{\tt
  1705.04664}}].

\bibitem{Halzen:1978rx}
F.~Halzen and D.~M. Scott, \emph{{Testing QCD in the Hadroproduction of Real
  and Virtual Photons}},
  \href{http://dx.doi.org/10.1103/PhysRevLett.40.1117}{\emph{Phys. Rev. Lett.}
  {\bf 40} (1978) 1117}.

\bibitem{Ruckl:1978mx}
R.~Rückl, S.~J. Brodsky and J.~F. Gunion, \emph{{The Production of Real
  Photons at Large Transverse Momentum in $pp$ Collisions}},
  \href{http://dx.doi.org/10.1103/PhysRevD.18.2469}{\emph{Phys. Rev.} {\bf D18}
  (1978) 2469--2483}.

\bibitem{Harriman:1990hi}
P.~N. Harriman, A.~D. Martin, W.~J. Stirling and R.~G. Roberts, \emph{{Parton
  Distributions Extracted From Data on Deep Inelastic Lepton Scattering, Prompt
  Photon Production and the {Drell-Yan} Process}},
  \href{http://dx.doi.org/10.1103/PhysRevD.42.798}{\emph{Phys. Rev. D} {\bf 42}
  (1990) 798--810}.

\bibitem{Vogelsang:1995bg}
W.~Vogelsang and A.~Vogt, \emph{{Constraints on the proton'$s$ gluon
  distribution from prompt photon production}},
  \href{http://dx.doi.org/10.1016/0550-3213(95)00424-Q}{\emph{Nucl. Phys.} {\bf
  B453} (1995) 334--354}, [\href{http://arxiv.org/abs/hep-ph/9505404}{{\tt
  hep-ph/9505404}}].

\bibitem{dEnterria:2012kvo}
D.~d'Enterria and J.~Rojo, \emph{{Quantitative constraints on the gluon
  distribution function in the proton from collider isolated-photon data}},
  \href{http://dx.doi.org/10.1016/j.nuclphysb.2012.03.003}{\emph{Nucl. Phys.}
  {\bf B860} (2012) 311--338}, [\href{http://arxiv.org/abs/1202.1762}{{\tt
  1202.1762}}].

\bibitem{Carminati:2012mm}
L.~Carminati, G.~Costa, D.~D'Enterria, I.~Koletsou, G.~Marchiori, J.~Rojo
  et~al., \emph{{Sensitivity of the LHC isolated-gamma+jet data to the parton
  distribution functions of the proton}},
  \href{http://dx.doi.org/10.1209/0295-5075/101/61002}{\emph{EPL} {\bf 101}
  (2013) 61002}, [\href{http://arxiv.org/abs/1212.5511}{{\tt 1212.5511}}].

\bibitem{Campbell:2018wfu}
J.~M. Campbell, J.~Rojo, E.~Slade and C.~Williams, \emph{{Direct photon
  production and PDF fits reloaded}},
  \href{http://dx.doi.org/10.1140/epjc/s10052-018-5944-4}{\emph{Eur. Phys. J.}
  {\bf C78} (2018) 470}, [\href{http://arxiv.org/abs/1802.03021}{{\tt
  1802.03021}}].

\bibitem{Koller:1978kq}
K.~Koller, T.~F. Walsh and P.~M. Zerwas, \emph{{Testing {QCD}: Direct Photons
  in $e^+ e^-$ Collisions}},
  \href{http://dx.doi.org/10.1007/BF01474661}{\emph{Z. Phys. C} {\bf 2} (1979)
  197}.

\bibitem{Laermann:1982jr}
E.~Laermann, T.~F. Walsh, I.~Schmitt and P.~M. Zerwas, \emph{{Direct Photons in
  $e^+ e^-$ Annihilation}},
  \href{http://dx.doi.org/10.1016/0550-3213(82)90162-6}{\emph{Nucl. Phys.} {\bf
  B207} (1982) 205--232}.

\bibitem{Frixione:1998jh}
S.~Frixione, \emph{{Isolated photons in perturbative QCD}},
  \href{http://dx.doi.org/10.1016/S0370-2693(98)00454-7}{\emph{Phys. Lett.}
  {\bf B429} (1998) 369--374}, [\href{http://arxiv.org/abs/hep-ph/9801442}{{\tt
  hep-ph/9801442}}].

\bibitem{Aurenche:1987fs}
P.~Aurenche, R.~Baier, M.~Fontannaz and D.~Schiff, \emph{{Prompt Photon
  Production at Large $p_T$ Scheme Invariant QCD Predictions and Comparison
  with Experiment}},
  \href{http://dx.doi.org/10.1016/0550-3213(88)90553-6}{\emph{Nucl. Phys.} {\bf
  B297} (1988) 661--696}.

\bibitem{Baer:1990ra}
H.~Baer, J.~Ohnemus and J.~F. Owens, \emph{{A Next-to-leading Logarithm
  Calculation of Direct Photon Production}},
  \href{http://dx.doi.org/10.1103/PhysRevD.42.61}{\emph{Phys. Rev.} {\bf D42}
  (1990) 61--71}.

\bibitem{Aurenche:1992yc}
P.~Aurenche, P.~Chiappetta, M.~Fontannaz, J.~P. Guillet and E.~Pilon,
  \emph{{Next-to-leading order bremsstrahlung contribution to prompt photon
  production}},
  \href{http://dx.doi.org/10.1016/0550-3213(93)90615-V}{\emph{Nucl. Phys.} {\bf
  B399} (1993) 34--62}.

\bibitem{Gordon:1993qc}
L.~E. Gordon and W.~Vogelsang, \emph{{Polarized and unpolarized prompt photon
  production beyond the leading order}},
  \href{http://dx.doi.org/10.1103/PhysRevD.48.3136}{\emph{Phys. Rev.} {\bf D48}
  (1993) 3136--3159}.

\bibitem{Gluck:1994iz}
M.~Glück, L.~E. Gordon, E.~Reya and W.~Vogelsang, \emph{{High $p_T$ photon
  production at $p \bar{p}$ collider}},
  \href{http://dx.doi.org/10.1103/PhysRevLett.73.388}{\emph{Phys. Rev. Lett.}
  {\bf 73} (1994) 388--391}.

\bibitem{Catani:2002ny}
S.~Catani, M.~Fontannaz, J.~P. Guillet and E.~Pilon, \emph{{Cross-section of
  isolated prompt photons in hadron hadron collisions}},
  \href{http://dx.doi.org/10.1088/1126-6708/2002/05/028}{\emph{JHEP} {\bf 05}
  (2002) 028}, [\href{http://arxiv.org/abs/hep-ph/0204023}{{\tt
  hep-ph/0204023}}].

\bibitem{Catani:2013oma}
S.~Catani, M.~Fontannaz, J.~P. Guillet and E.~Pilon, \emph{{Isolating Prompt
  Photons with Narrow Cones}},
  \href{http://dx.doi.org/10.1007/JHEP09(2013)007}{\emph{JHEP} {\bf 09} (2013)
  007}, [\href{http://arxiv.org/abs/1306.6498}{{\tt 1306.6498}}].

\bibitem{Aurenche:2006vj}
P.~Aurenche, M.~Fontannaz, J.-P. Guillet, E.~Pilon and M.~Werlen, \emph{{A New
  critical study of photon production in hadronic collisions}},
  \href{http://dx.doi.org/10.1103/PhysRevD.73.094007}{\emph{Phys. Rev. D} {\bf
  73} (2006) 094007}, [\href{http://arxiv.org/abs/hep-ph/0602133}{{\tt
  hep-ph/0602133}}].

\bibitem{Owens:1986mp}
J.~F. Owens, \emph{{Large Momentum Transfer Production of Direct Photons, Jets,
  and Particles}},
  \href{http://dx.doi.org/10.1103/RevModPhys.59.465}{\emph{Rev. Mod. Phys.}
  {\bf 59} (1987) 465}.

\bibitem{Gluck:1992zx}
M.~Glück, E.~Reya and A.~Vogt, \emph{{Parton fragmentation into photons beyond
  the leading order}}, \href{http://dx.doi.org/10.1103/PhysRevD.51.1427,
  10.1103/PhysRevD.48.116}{\emph{Phys. Rev.} {\bf D48} (1993) 116}. [Erratum:
  Phys. Rev.D51,1427(1995)].

\bibitem{Bourhis:1997yu}
L.~Bourhis, M.~Fontannaz and J.~P. Guillet, \emph{{Quarks and gluon
  fragmentation functions into photons}},
  \href{http://dx.doi.org/10.1007/s100520050158}{\emph{Eur. Phys. J.} {\bf C2}
  (1998) 529--537}, [\href{http://arxiv.org/abs/hep-ph/9704447}{{\tt
  hep-ph/9704447}}].

\bibitem{Buskulic:1995au}
{\scshape ALEPH} collaboration, D.~Buskulic et~al., \emph{{First measurement of
  the quark to photon fragmentation function}},
  \href{http://dx.doi.org/10.1007/BF02907417}{\emph{Z. Phys.} {\bf C69} (1996)
  365--378}.

\bibitem{Ackerstaff:1997nha}
{\scshape OPAL} collaboration, K.~Ackerstaff et~al., \emph{{Measurement of the
  quark to photon fragmentation function through the inclusive production of
  prompt photons in hadronic Z0 decays}},
  \href{http://dx.doi.org/10.1007/s100520050122}{\emph{Eur. Phys. J.} {\bf C2}
  (1998) 39--48}, [\href{http://arxiv.org/abs/hep-ex/9708020}{{\tt
  hep-ex/9708020}}].

\bibitem{GehrmannDeRidder:1997gf}
A.~Gehrmann-De~Ridder and E.~W.~N. Glover, \emph{{A Complete O ($\alpha
  \alpha_s$) calculation of the photon + 1 jet rate in $e^+ e^-$
  annihilation}},
  \href{http://dx.doi.org/10.1016/S0550-3213(97)00818-3}{\emph{Nucl. Phys.}
  {\bf B517} (1998) 269--323}, [\href{http://arxiv.org/abs/hep-ph/9707224}{{\tt
  hep-ph/9707224}}].

\bibitem{GehrmannDeRidder:1998ba}
A.~Gehrmann-De~Ridder and E.~W.~N. Glover, \emph{{Final state photon production
  at LEP}}, \href{http://dx.doi.org/10.1007/s100520050382,
  10.1007/s100529800958}{\emph{Eur. Phys. J.} {\bf C7} (1999) 29--48},
  [\href{http://arxiv.org/abs/hep-ph/9806316}{{\tt hep-ph/9806316}}].

\bibitem{Campbell_2017}
J.~M. Campbell, R.~K. Ellis and C.~Williams, \emph{{Direct Photon Production at
  Next-to\textendash{}Next-to-Leading Order}},
  \href{http://dx.doi.org/10.1103/PhysRevLett.118.222001}{\emph{Phys. Rev.
  Lett.} {\bf 118} (2017) 222001}, [\href{http://arxiv.org/abs/1612.04333}{{\tt
  1612.04333}}]. [Erratum: Phys.Rev.Lett. 124 (2020) 259901].

\bibitem{Campbell:2017dqk}
J.~M. Campbell, R.~K. Ellis and C.~Williams, \emph{{Driving missing data at the
  LHC: NNLO predictions for the ratio of $\gamma+j$ and $Z+j$}},
  \href{http://dx.doi.org/10.1103/PhysRevD.96.014037}{\emph{Phys. Rev.} {\bf
  D96} (2017) 014037}, [\href{http://arxiv.org/abs/1703.10109}{{\tt
  1703.10109}}].

\bibitem{Chen_2020}
X.~Chen, T.~Gehrmann, N.~Glover, M.~H\"ofer and A.~Huss, \emph{{Isolated photon
  and photon+jet production at NNLO QCD accuracy}},
  \href{http://dx.doi.org/10.1007/JHEP04(2020)166}{\emph{JHEP} {\bf 04} (2020)
  166}, [\href{http://arxiv.org/abs/1904.01044}{{\tt 1904.01044}}].

\bibitem{Siegert:2016bre}
F.~Siegert, \emph{{A practical guide to event generation for prompt photon
  production with Sherpa}},
  \href{http://dx.doi.org/10.1088/1361-6471/aa5f29}{\emph{J. Phys.} {\bf G44}
  (2017) 044007}, [\href{http://arxiv.org/abs/1611.07226}{{\tt 1611.07226}}].

\bibitem{Czakon:2021ohs}
M.~L. Czakon, T.~Generet, A.~Mitov and R.~Poncelet, \emph{{B-hadron production
  in NNLO QCD: application to LHC t$ \overline{t} $ events with leptonic
  decays}}, \href{http://dx.doi.org/10.1007/JHEP10(2021)216}{\emph{JHEP} {\bf
  10} (2021) 216}, [\href{http://arxiv.org/abs/2102.08267}{{\tt 2102.08267}}].

\bibitem{Gehrmann:2022cih}
T.~Gehrmann and R.~Sch\"urmann, \emph{{Photon fragmentation in the antenna
  subtraction formalism}},
  \href{http://dx.doi.org/10.1007/JHEP04(2022)031}{\emph{JHEP} {\bf 04} (2022)
  031}, [\href{http://arxiv.org/abs/2201.06982}{{\tt 2201.06982}}].

\bibitem{GehrmannDeRidder:2005cm}
A.~Gehrmann-De~Ridder, T.~Gehrmann and E.~W.~N. Glover, \emph{{Antenna
  subtraction at NNLO}},
  \href{http://dx.doi.org/10.1088/1126-6708/2005/09/056}{\emph{JHEP} {\bf 09}
  (2005) 056}, [\href{http://arxiv.org/abs/hep-ph/0505111}{{\tt
  hep-ph/0505111}}].

\bibitem{Daleo:2006xa}
A.~Daleo, T.~Gehrmann and D.~Maitre, \emph{{Antenna subtraction with hadronic
  initial states}},
  \href{http://dx.doi.org/10.1088/1126-6708/2007/04/016}{\emph{JHEP} {\bf 04}
  (2007) 016}, [\href{http://arxiv.org/abs/hep-ph/0612257}{{\tt
  hep-ph/0612257}}].

\bibitem{Currie:2013vh}
J.~Currie, E.~W.~N. Glover and S.~Wells, \emph{{Infrared Structure at NNLO
  Using Antenna Subtraction}},
  \href{http://dx.doi.org/10.1007/JHEP04(2013)066}{\emph{JHEP} {\bf 04} (2013)
  066}, [\href{http://arxiv.org/abs/1301.4693}{{\tt 1301.4693}}].

\bibitem{Glover:1993xc}
E.~W.~N. Glover and A.~G. Morgan, \emph{{Measuring the photon fragmentation
  function at LEP}}, \href{http://dx.doi.org/10.1007/BF01560245}{\emph{Z.
  Phys.} {\bf C62} (1994) 311--322}.

\bibitem{Hall:2018jub}
E.~Hall and J.~Thaler, \emph{{Photon isolation and jet substructure}},
  \href{http://dx.doi.org/10.1007/JHEP09(2018)164}{\emph{JHEP} {\bf 09} (2018)
  164}, [\href{http://arxiv.org/abs/1805.11622}{{\tt 1805.11622}}].

\bibitem{Anastasiou:2001sv}
C.~Anastasiou, E.~W.~N. Glover, C.~Oleari and M.~E. Tejeda-Yeomans, \emph{{Two
  loop QCD corrections to massless quark gluon scattering}},
  \href{http://dx.doi.org/10.1016/S0550-3213(01)00195-X}{\emph{Nucl. Phys. B}
  {\bf 605} (2001) 486--516}, [\href{http://arxiv.org/abs/hep-ph/0101304}{{\tt
  hep-ph/0101304}}].

\bibitem{Bern:2003ck}
Z.~Bern, A.~De~Freitas and L.~J. Dixon, \emph{{Two loop helicity amplitudes for
  quark gluon scattering in QCD and gluino gluon scattering in supersymmetric
  Yang-Mills theory}},
  \href{http://dx.doi.org/10.1088/1126-6708/2003/06/028}{\emph{JHEP} {\bf 06}
  (2003) 028}, [\href{http://arxiv.org/abs/hep-ph/0304168}{{\tt
  hep-ph/0304168}}]. [Erratum: JHEP 04, 112 (2014)].

\bibitem{Bern:1994fz}
Z.~Bern, L.~J. Dixon and D.~A. Kosower, \emph{{One loop corrections to two
  quark three gluon amplitudes}},
  \href{http://dx.doi.org/10.1016/0550-3213(94)00542-M}{\emph{Nucl. Phys. B}
  {\bf 437} (1995) 259--304}, [\href{http://arxiv.org/abs/hep-ph/9409393}{{\tt
  hep-ph/9409393}}].

\bibitem{Signer:1995np}
A.~Signer, \emph{{One loop corrections to five parton amplitudes with external
  photons}}, \href{http://dx.doi.org/10.1016/0370-2693(95)00905-Z}{\emph{Phys.
  Lett. B} {\bf 357} (1995) 204--210},
  [\href{http://arxiv.org/abs/hep-ph/9507442}{{\tt hep-ph/9507442}}].

\bibitem{Signer:1995a}
A.~Signer, \emph{{Helicity method for next-to-leading order corrections in
  QCD}}.
\newblock PhD thesis, {ETH Z\"{u}rich}, 1995.

\bibitem{DelDuca:1999pa}
V.~Del~Duca, W.~B. Kilgore and F.~Maltoni, \emph{{Multiphoton amplitudes for
  next-to-leading order QCD}},
  \href{http://dx.doi.org/10.1016/S0550-3213(99)00663-X}{\emph{Nucl. Phys. B}
  {\bf 566} (2000) 252--274}, [\href{http://arxiv.org/abs/hep-ph/9910253}{{\tt
  hep-ph/9910253}}].

\bibitem{Balsiger:2018ezi}
M.~Balsiger, T.~Becher and D.~Y. Shao, \emph{{Non-global logarithms in jet and
  isolation cone cross sections}},
  \href{http://dx.doi.org/10.1007/JHEP08(2018)104}{\emph{JHEP} {\bf 08} (2018)
  104}, [\href{http://arxiv.org/abs/1803.07045}{{\tt 1803.07045}}].

\bibitem{Ebert:2019zkb}
M.~A. Ebert and F.~J. Tackmann, \emph{{Impact of isolation and fiducial cuts on
  q$_{T}$ and N-jettiness subtractions}},
  \href{http://dx.doi.org/10.1007/JHEP03(2020)158}{\emph{JHEP} {\bf 03} (2020)
  158}, [\href{http://arxiv.org/abs/1911.08486}{{\tt 1911.08486}}].

\bibitem{NNPDF:2021njg}
{\scshape NNPDF} collaboration, R.~D. Ball et~al., \emph{{The path to proton
  structure at 1\% accuracy}},
  \href{http://dx.doi.org/10.1140/epjc/s10052-022-10328-7}{\emph{Eur. Phys. J.
  C} {\bf 82} (2022) 428}, [\href{http://arxiv.org/abs/2109.02653}{{\tt
  2109.02653}}].

\bibitem{NNPDF:2017mvq}
{\scshape NNPDF} collaboration, R.~D. Ball et~al., \emph{{Parton distributions
  from high-precision collider data}},
  \href{http://dx.doi.org/10.1140/epjc/s10052-017-5199-5}{\emph{Eur. Phys. J.
  C} {\bf 77} (2017) 663}, [\href{http://arxiv.org/abs/1706.00428}{{\tt
  1706.00428}}].

\bibitem{Cacciari:2008gp}
M.~Cacciari, G.~P. Salam and G.~Soyez, \emph{{The anti-$k_t$ jet clustering
  algorithm}},
  \href{http://dx.doi.org/10.1088/1126-6708/2008/04/063}{\emph{JHEP} {\bf 04}
  (2008) 063}, [\href{http://arxiv.org/abs/0802.1189}{{\tt 0802.1189}}].

\bibitem{Becher:2012xr}
T.~Becher, C.~Lorentzen and M.~D. Schwartz, \emph{{Precision Direct Photon and
  $W$-Boson Spectra at High $p_T$ and Comparison to LHC Data}},
  \href{http://dx.doi.org/10.1103/PhysRevD.86.054026}{\emph{Phys. Rev. D} {\bf
  86} (2012) 054026}, [\href{http://arxiv.org/abs/1206.6115}{{\tt 1206.6115}}].

\bibitem{Becher:2013zua}
T.~Becher and X.~Garcia~i Tormo, \emph{{Electroweak Sudakov effects in $W, Z$
  and $\gamma$ production at large transverse momentum}},
  \href{http://dx.doi.org/10.1103/PhysRevD.88.013009}{\emph{Phys. Rev. D} {\bf
  88} (2013) 013009}, [\href{http://arxiv.org/abs/1305.4202}{{\tt 1305.4202}}].

\bibitem{Catani:1997xc}
S.~Catani and B.~R. Webber, \emph{{Infrared safe but infinite: Soft gluon
  divergences inside the physical region}},
  \href{http://dx.doi.org/10.1088/1126-6708/1997/10/005}{\emph{JHEP} {\bf 10}
  (1997) 005}, [\href{http://arxiv.org/abs/hep-ph/9710333}{{\tt
  hep-ph/9710333}}].

\bibitem{Cacciari:2008gn}
M.~Cacciari, G.~P. Salam and G.~Soyez, \emph{{The Catchment Area of Jets}},
  \href{http://dx.doi.org/10.1088/1126-6708/2008/04/005}{\emph{JHEP} {\bf 04}
  (2008) 005}, [\href{http://arxiv.org/abs/0802.1188}{{\tt 0802.1188}}].

\bibitem{Gehrmann-DeRidder:2006zbx}
A.~Gehrmann-De~Ridder, T.~Gehrmann and E.~Poulsen, \emph{{Isolated photons in
  deep inelastic scattering}},
  \href{http://dx.doi.org/10.1103/PhysRevLett.96.132002}{\emph{Phys. Rev.
  Lett.} {\bf 96} (2006) 132002},
  [\href{http://arxiv.org/abs/hep-ph/0601073}{{\tt hep-ph/0601073}}].

\bibitem{Gehrmann-DeRidder:2006lpc}
A.~Gehrmann-De~Ridder, T.~Gehrmann and E.~Poulsen, \emph{{Measuring the Photon
  Fragmentation Function at HERA}},
  \href{http://dx.doi.org/10.1140/epjc/s2006-02574-x}{\emph{Eur. Phys. J. C}
  {\bf 47} (2006) 395--411}, [\href{http://arxiv.org/abs/hep-ph/0604030}{{\tt
  hep-ph/0604030}}].

\bibitem{Belghobsi:2009hx}
Z.~Belghobsi, M.~Fontannaz, J.~P. Guillet, G.~Heinrich, E.~Pilon and M.~Werlen,
  \emph{{Photon - Jet Correlations and Constraints on Fragmentation
  Functions}}, \href{http://dx.doi.org/10.1103/PhysRevD.79.114024}{\emph{Phys.
  Rev. D} {\bf 79} (2009) 114024}, [\href{http://arxiv.org/abs/0903.4834}{{\tt
  0903.4834}}].

\bibitem{Kaufmann:2016nux}
T.~Kaufmann, A.~Mukherjee and W.~Vogelsang, \emph{{Access to Photon
  Fragmentation Functions in Hadronic Jet Production}},
  \href{http://dx.doi.org/10.1103/PhysRevD.93.114021}{\emph{Phys. Rev. D} {\bf
  93} (2016) 114021}, [\href{http://arxiv.org/abs/1604.07175}{{\tt
  1604.07175}}].

\end{thebibliography}\endgroup

\end{document}